\documentclass[letterpaper, 10 pt, conference]{ieeeconf}  

\IEEEoverridecommandlockouts                              

\usepackage{graphicx}
\usepackage{url}  
\usepackage{pifont}
\usepackage{xcolor}
\usepackage{hyperref}

\usepackage{amsmath}
\DeclareMathOperator*{\argmax}{argmax}

\title{\LARGE \bf
A Comprehensive  Survey with Critical Analysis \\ for Deepfake Speech Detection 
}

\author{Lam~Pham$^{1*}$, 
        Phat~Lam$^{2*}$,
        Dat~Tran$^{3*}$,
        Hieu~Tang$^{4}$,
        Tin~Nguyen$^{5}$, \\
        Alexander~Schindler$^{6}$,         
        Florian~Skopik$^{7}$,
        Alexander~Polonsky$^{8}$,
        Canh~Vu$^{9**}$
\thanks{L. Pham, A. Schindler, and F. Skopik are with Austrian Institute of Technology, Vienna, Austria.}%
\thanks{P. Lam and T. Nguyen are with HCM University of Technology,  Ho Chi Minh city, Vietnam}%
\thanks{H. Tang is with University of Technology of Troyes, France}%
\thanks{D. Tran is with FPT University, Ho Chi Minh city, Vietnam}
\thanks{A. Polonsky is with BLOOM Social Analytics, France}%
\thanks{C. Vu is with Laboratoire Roberval, Université de technologie de Compiègne, France}%
\thanks{(*) Main and equal contribution into the paper.}
\thanks{(**) Corresponding author.}
}

\begin{document}

\maketitle
\thispagestyle{empty}
\pagestyle{empty}

\begin{abstract}

Thanks to advancements in deep learning, speech generation systems now power a variety of real-world applications, such as text-to-speech for individuals with speech disorders, voice chatbots in call centers, cross-linguistic speech translation, etc. While these systems can autonomously generate human-like speech and replicate specific voices, they also pose risks when misused for malicious purposes. This motivates the research community to develop models for detecting synthesized speech (e.g., fake speech) generated by deep-learning-based models, referred to as the Deepfake Speech Detection task. As the Deepfake Speech Detection task has emerged in recent years, there are not many survey papers proposed for this task. Additionally, existing surveys for the Deepfake Speech Detection task tend to summarize techniques used to construct a Deepfake Speech Detection system rather than providing a thorough analysis. This gap motivated us to conduct a comprehensive survey, providing a critical analysis of the challenges and developments in Deepfake Speech Detection. Our survey is innovatively structured, offering an in-depth analysis of current challenge competitions, public datasets, and the deep-learning techniques that provide enhanced solutions to address existing challenges in the field. From our analysis, we propose hypotheses on leveraging and combining specific deep learning techniques to improve the effectiveness of Deepfake Speech Detection systems. Beyond conducting a survey, we perform extensive experiments to validate these hypotheses and propose a highly competitive model for the task of Deepfake Speech Detection. Given the analysis and the experimental results, we finally indicate potential and promising research directions for the Deepfake Speech Detection task.

\indent \textit{Items}--- Deepfake speech detection (DSD), challenge competition, ensemble, audio embedding, pre-trained model.
\end{abstract}
\section{INTRODUCTION}
\label{intro}

\begin{figure*}[h]
    \centering
    \includegraphics[width =1.0\linewidth]{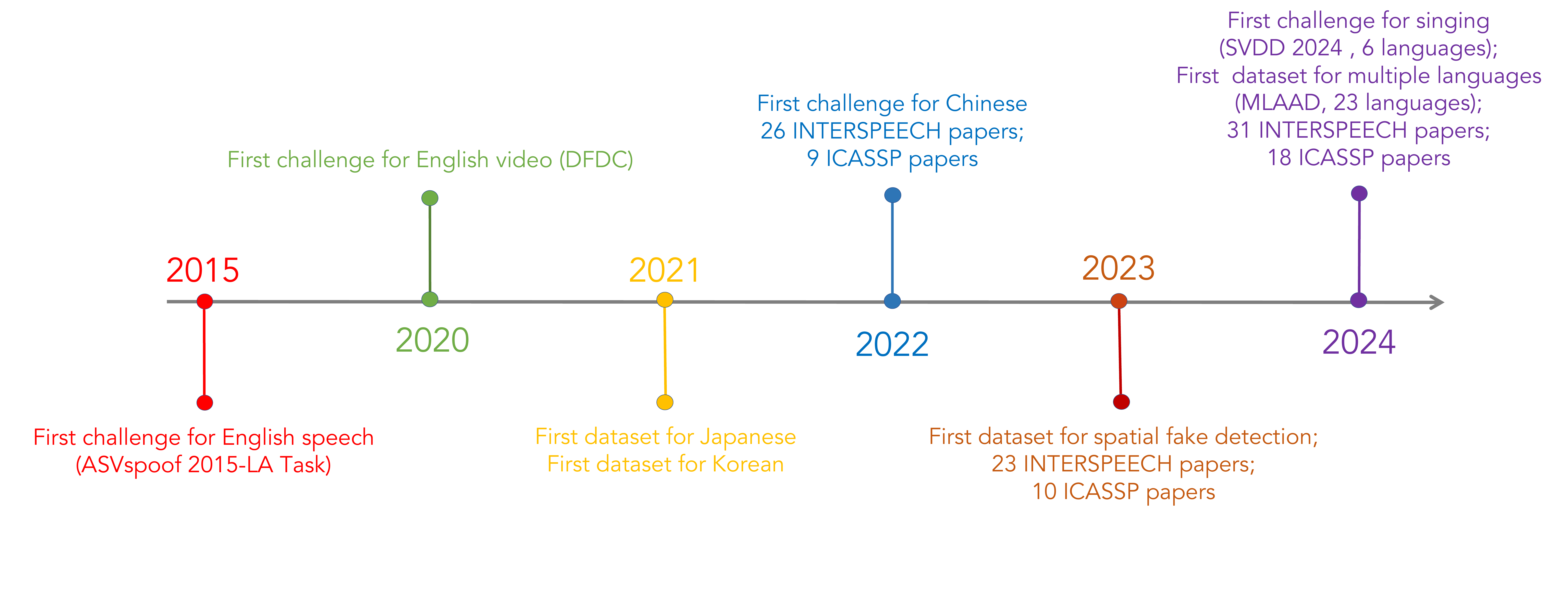}
   	\vspace{-1.6cm}
	\caption{The timeline of Deepfake Speech Detection (DSD) task}
    \label{F0}
\end{figure*}

\begin{table*}[t]
    \caption{The main factors analyzed in survey papers} 
  	\vspace{-0.2cm}
    \centering
    \scalebox{0.85}{
    \begin{tabular}{|c|c|c|c|c|c|c|c|c|c|c|c|} 
        \hline 
        \textbf{Papers} &\textbf{Years} &\textbf{Audio/Video} &\textbf{Challenge} &\textbf{Public} &\textbf{Data}&\textbf{Feature} &\textbf{Classification} &\textbf{Loss} &\textbf{Training}  &\textbf{Proposed}&\textbf{Continue}\\
                        &               &                     &\textbf{Competitions}  &\textbf{Datasets}  &\textbf{Augmentation}  &\textbf{Extraction}  &\textbf{Models}  &\textbf{Functions}  &\textbf{Strategies}  &\textbf{Models}  &\textbf{Updating}\\
        \hline 
         \hline 
           \cite{survey2021} &2021 &Yes/Yes  &No   &Yes  &No &No  &Yes &No  &No  &No   &No\\  
        
        \hline 
           \cite{survey_11} &2023 &Yes/Yes &No   &No  &No &Yes  &Yes &No  &No  &No   &No\\   
        \hline 
           \cite{survey_12} &2023 &Yes/Yes  &No   &No  &No &Yes  &Yes &Yes  &No  &No   &No\\   
            \hline     
           \cite{survey_patel} &2023 &Yes/Yes  &No  &Yes  &No &No   &Yes &Yes  &Yes  &No  &No\\
       \hline        
           \cite{survey_01} &2023 &Yes/No  &Yes  &Yes  &No &Yes   &Yes &Yes  &Yes  &Yes  &No\\
        \hline       
           \cite{survey2023_usa} &2023 &Yes/Yes  &No  &Yes  &No &No   &Yes &No  &No  &No  &No\\
        \hline   \cite{survey2024video} &2024 &Yes/Yes  &No   &Yes  &No &No  &Yes &No  &No  &No   &No\\
       \hline 
           \cite{survey_14} &2024 &Yes/Yes  &No   &Yes  &No &Yes  &Yes &No  &No  &No   &No\\  
        \hline
           \cite{survey_02} &2024 &Yes/No  &No   &Yes  &Yes &Yes  &Yes &Yes  &Yes  &No   &No\\
       
       \hline
       \hline 
          \textbf{Our Survey}    &\textbf{2024} &\textbf{Yes/No}  &\textbf{Yes}  &\textbf{Yes}  &\textbf{Yes} &\textbf{Yes}  &\textbf{Yes} &\textbf{Yes}  &\textbf{Yes}  &\textbf{Yes}  &\textbf{Yes}\\
       \hline     
    \end{tabular}
    }
    \vspace{-0.3cm}
    \label{survey} 
\end{table*}

In recent years, remarkable advancements in deep learning techniques and neural networks have revolutionized the field of generative AI. Today, core communication mediums such as audio, images, video, and text can be automatically generated and applied across various domains, including chatbot systems (e.g., ChatGPT), film production~\cite{survey_03}, code generation~\cite{survey_04}, and audio synthesis~\cite{survey_05_TTS, survey_06_VC}, etc.
However, AI-synthesized data could pose a serious threat to social security when there is an increasing number of crimes related to leveraging the synthesized data~\cite{survey_07_crime}.
To address this concern, the tasks, which are proposed for detecting synthesized data (e.g. fake data) generated from deep-learning-based methods, referred to as deepfake detection, have drawn much attention from the research community recently.

Focusing on human speech, this paper provides a comprehensive survey for the task of Deepfake Speech Detection (DSD).
To this end, the milestones presenting the development progress of the DSD task are first presented in Fig.~\ref{F0}.
As the figure shows, the earliest public dataset and challenge proposed for the DSD task was introduced in 2015, focusing exclusively on the English language. 
Then, the first challenge for video deepfake detection (DFDC~\cite{dfdc_ds}) was introduced in 2020.
In subsequent years, datasets for the DSD task in Japanese~\cite{wake_ds}, Korean~\cite{wake_ds}, and Chinese~\cite{add22} were introduced in 2021 and 2022, respectively.
Recently, in 2024, multilingual datasets for the DSD task have been published, including MLAAD~\cite{MAILABS_ds} for conversational speech and SVDD~\cite{svdd2024} for singing.
%
Fig.~\ref{F0} also highlights a growing number of papers, datasets, and challenge competitions for the DSD task from 2021 to the present.
This trend indicates that the DSD task has recently gained prominence and has attracted significant interest from the research community.

To further understand the DSD task, we summarized recent survey papers related to the DSD task in Table~\ref{survey}.
As shown in the table, most of these surveys focus on detecting general fake data (e.g., images, videos, audio, or text), with audio or human speech typically being addressed only as a subsection or a part of the broader discussion~\cite{survey_14, survey_11, survey_12}.
Therefore, the main techniques, existing concerns, and potential research for the DSD task have not been comprehensively analyzed in these papers.
Among the survey papers, only two survey papers of~\cite{survey_01} and~\cite{survey_02} focus on the DSD task. However, as conventional surveys, these papers primarily summarize the technologies used to construct a DSD system such as datasets, feature extraction, classification model, loss functions, rather than providing a comprehensive analysis and highlighting existing concerns.
For instance, while challenge competitions proposed for the DSD task are very important in advancing the research community, their importance and various aspects have not been thoroughly analyzed (e.g., the number of research teams participating in these competitions and their results are interesting to analyze).
Although this information reflects the level of interest in DSD within the research community, it has not been addressed in any existing survey papers.
The second concern is related to public datasets proposed for the DSD task.
In particular, the current survey papers do not adequately analyze the imbalance among (1) the number of utterances, (2) the AI-synthesized speech systems used to generate fake speech; and (3) the original/real human speech resource used to generate fake speech utterances. 
These key factors are essential in creating a high-quality DSD dataset for evaluating DSD models.
Additionally, survey papers are at risk of becoming outdated as new datasets, techniques, and models continue to emerge. However, current surveys do not offer solutions for regularly updating essential information, such as details about challenge competitions, public datasets, and the top-performing models on specific datasets.
\begin{table*}[t]
    \caption{The challenge competitions proposed for Deepfake Speech Detection} 
        	\vspace{-0.2cm}
    \centering
    \scalebox{0.95}{
    \begin{tabular}{|l|c|c|c|c|c|c|c|c|c|} 
        \hline 
        \textbf{Challenge Competitions} &\textbf{Years} &\textbf{Data Types} &\textbf{Languages}   & \textbf{Public Labels}  & \textbf{Audio} &\textbf{Visual} &\textbf{Team No.} &\textbf{Top-1}  \\ 
        &&&\textbf{(Number)} &\textbf{(train}\&\textbf{dev/test)}&&&&\textbf{System} \\ 
       \hline
       \hline     
        ASVspoof 2015~\cite{ASV15}      &2015 &Speech &English  &Yes/Yes &Yes &No   &16 &Ensemble Model\\ 
               \hline
        ASVspoof 2019 (LA Task)~\cite{ASV19}      &2019 &Speech &English  &Yes/Yes &Yes &No   &48  &Ensemble Model\\
               \hline
        DFDC~\cite{dfdc_ds}   &2020 &Speech &English &Yes/Yes &Yes &Yes  &\textbf{2114} &Ensemble Model \\
        \hline
        FTC~\cite{FTC_ch} &2020 &Speech &English &No/No  &Yes &No &n/a &n/a  \\
        \hline        
        ASVspoof 2021 (LA Task)~\cite{ASV21}      &2021 &Speech &English  &Yes/Yes &Yes &No   &41 &Ensemble Model\\
        ASVspoof 2021 (DF Task)~\cite{ASV21}      &2021 &Speech &English  &Yes/Yes &Yes &No   &33 &Ensemble Model\\
                       \hline
        ADD 2022 Track 1~\cite{add22}            &2022 &Speech &Chinese  &Yes/Yes &Yes  &No &48 & Single Model \\
        ADD 2022 Track 2~\cite{add22}            &2022 &Speech &Chinese  &Yes/Yes &Yes  &No &27 &Single Model \\
        ADD 2022 Track 3.2~\cite{add22}          &2022 &Speech &Chinese  &Yes/Yes &Yes  &No &33 &Single Model \\
               \hline
        ADD 2023 Track 1.2~\cite{add23}           &2023 &Speech &Chinese  &No/No &Yes  &No &49 &Ensemble Model\\
        ADD 2023 Track 2~\cite{add23}           &2023 &Speech &Chinese  &No/No &Yes  &No &16 &Single Model\\

       \hline
        AV-Deepfake1M~\cite{cai2023av, 1mdeepfake_ch}         &2024 &Speech  &English &Yes/No  &Yes &Yes  &n/a &n/a  \\
        \hline 
        ASVspoof 2024~\cite{ASV24}          &2024 &Speech &English  &Yes/No &Yes &No &53 &Ensemble Model\\
        \hline
        SVDD 2024~\cite{sing, svdd2024} &2024 &Singing &\textbf{Multilanguages (6)} &Yes/No  &Yes &No &47 &Ensemble Model  \\
        \hline
    \end{tabular}
    }
    \vspace{-0.3cm}
    \label{table:T1} 
\end{table*}
Regarding technologies used to construct a DSD model such as feature extraction, classification model, or loss functions, current survey papers mainly summarize and then present conclusions rather than conducting experiments to provide strong evidence and validation.

The above concerns about the existing survey papers for the DSD task motivate and inspire us to provide a much more comprehensive survey in this paper.
By addressing these concerns, we make the following main contributions:
\begin{itemize}
    \item We provide a comprehensive analysis and then indicate concerns related to three main topics: The current challenge competition, the published datasets, and the deep-learning-based techniques used to develop a DSD system. 
    Each topic consists of three main parts: `Analysis', `Discussion', and `Contribution'. 
    The `Analysis' summarizes concrete information about the topic. 
    The `Discussion' indicates concerns in each topic. 
    Finally, the `Contribution' provides our suggestion and solution to further improve each topic.
    \item To solve the out-of-date issue of a survey paper, we set up a Github repository to update further challenge competitions, public datasets, and top-performance systems. New versions of the paper are also continually updated on `https://arxiv.org'.      
    \item More than a survey, we conduct extensive experiments to verify assumptions from the comprehensive analysis (i.e., different types of data augmentation, multiple input features, multiple network architectures, cross-dataset and cross-language evaluation, etc.), achieving a competitive DSD model. 
    Given the analysis and experimental results, we indicate potential research directions for the DSD task.
\end{itemize}
The remainder of this paper is structured as follows: Section~\ref{challenges} discusses  challenge competitions for the DSD task. 
Section~\ref{dataset} deeply analyses the public and benchmark datasets proposed for the DSD task. 
In Section~\ref{systems}, we summarize the key techniques for constructing the main components of a DSD system, including data augmentation, feature extraction, classification models, and loss functions
Section~\ref{evidence} presents extensive experiments that validate the techniques described in Section~\ref{systems}. 
Building on the analysis and results from the previous sections, Section~\ref{potential} outlines our proposed research directions in the DSD task. Finally, Section~\ref{conclusion} concludes the paper.

\section{Challenge Competitions Proposed for Deepfake Speech Detection}
\label{challenges}

\textbf{Analysis:} Challenge competitions for the DSD task play a crucial role in motivating the research community. 
These competitions not only introduce new benchmark datasets but also host workshops where research teams can discuss their ideas and share their motivations. 
This environment encourages the community to publish more datasets and develop new techniques to address the DSD challenges.
To analyze DSD challenge competitions, we first summarize all challenges in Table~\ref{table:T1}. 
Importantly, we will continually update information about future DSD challenge competitions in our GitHub repository\footnote{https://github.com/AI-ResearchGroup/A-Comprehensive-Survey-with-Critical-Analysis-for-Deepfake-Speech-Detection}.
%

\begin{figure}[t]
    \centering
    \includegraphics[width =1.0\linewidth]{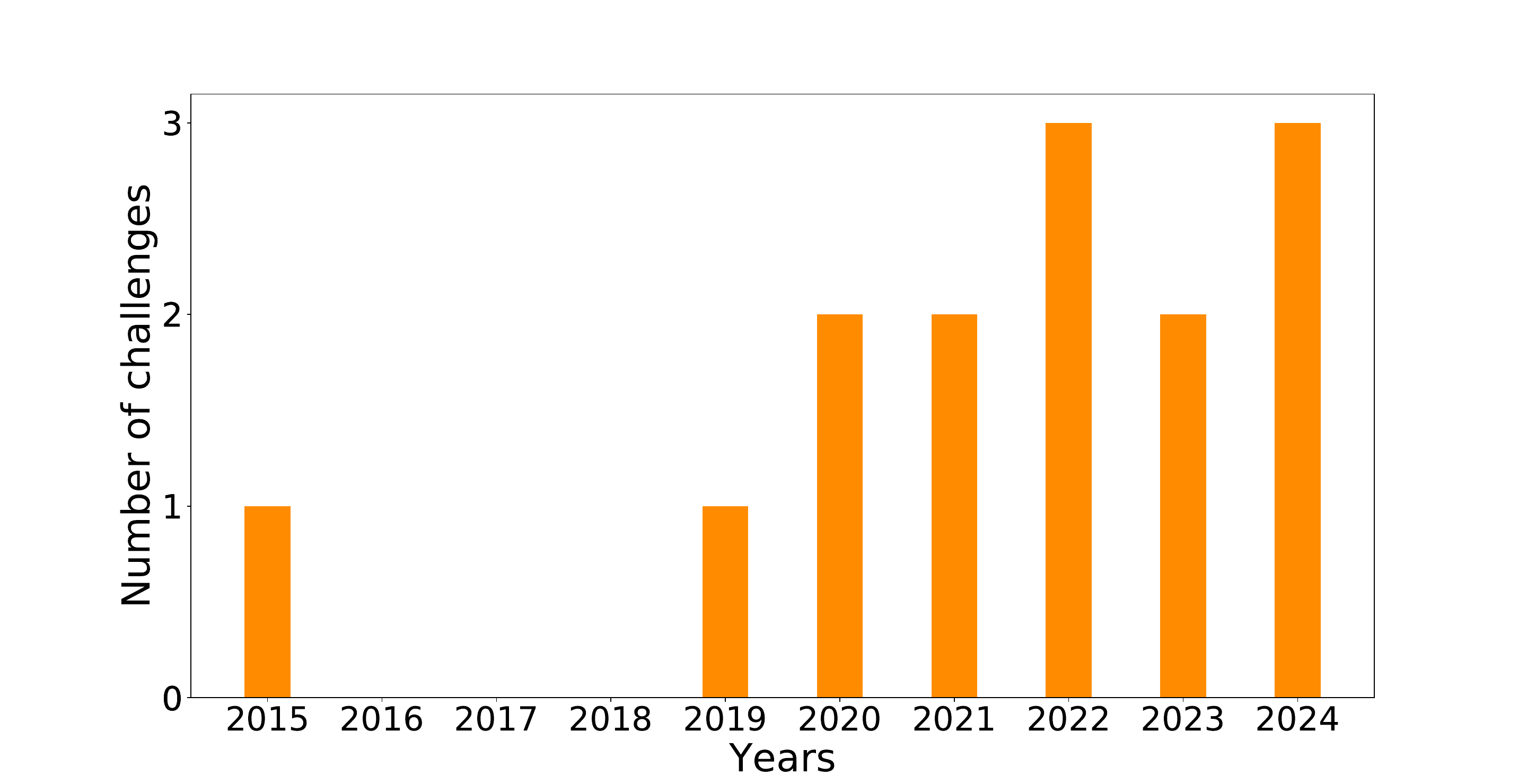}
	\caption{The number of competitions proposed for DSD task from 2015}
    \label{F1}
\end{figure}

As Table~\ref{table:T1} shows, most challenge competitions focus on detecting fake speech in a conversation except for the SVDD 2024 challenge~\cite{sing} for the fake singing detection.
All challenge competitions for fake speech detection in a conversation have been proposed for a single language (i.e., While ADD 2022 and ADD 2023 are for Chinese, the others are proposed for English).
%
Regarding the number of DSD challenge competitions, Fig.~\ref{F1} shows that there has been an increase in recent years. This trend indicates that the DSD task has gained attention from the research community, particularly due to the rise of advanced deep learning systems capable of generating highly realistic human-like speech, which poses significant security risks.
%
DSD challenge competitions, which explore fake speech in a conversation, can be separated into two groups.
The first group is proposed for only audio~\cite{ASV15, ASV17, ASV19, ASV21, ASV24, FTC_ch, add22, add23}.
Meanwhile, the second group is for video in which a fake video is identified by fake audio, fake image, or both fake audio and image~\cite{dfdc_ds, 1mdeepfake_ch}.
This indicates that DSD is not only treated as an individual task independently but also considered as a sub-task in multimodal systems.
It is also evident that the second group, which focuses on fake video detection, has attracted significantly more research teams (e.g., 2,114 teams in the DFDC challenge~\cite{dfdc_ds}) compared to the first group (e.g., the largest team count was 74 in the ASVspoof 2021 challenge~\cite{ASV21}). This provides an insight that fake video detection is a more compelling task, drawing greater interest and participation from research teams. 
%
Regarding top-1 systems in these challenge competitions, they leveraged the ensemble techniques which combine a wide range of input features or multiple models (i.e., most submitted systems mainly use deep learning based models).

\begin{table*}[t]
    \caption{Public and Benchmark Datasets Proposed for Deepfake Speech Detection} 
        	\vspace{-0.2cm}
    \centering
    \scalebox{0.72}{
    \begin{tabular}{|l|c|c|c|c|c|c|c|c|c|} 
       \hline  
       \textbf{Datasets}          &\textbf{Years}  &\textbf{Languages}  &\textbf{Speakers} &\textbf{Utt. No.} &\textbf{Fake speech}  &\textbf{Speech}  &\textbf{Real Speech} &\textbf{Utt. length (s)} &\textbf{Evaluation} \\  
					     &                &                    &(Male/Female)     &(Real/Fake)      &\textbf{Generators}                            &\textbf{Condition}                    &\textbf{Resources}                           &                               &\textbf{Metrics}    \\
       \hline     
       \hline     
       ASVspoof 2015~\cite{ASV15}(audio)            
       &2015            
       &English             
       &45/61           
       &16,651/246,500   
       &10                          
       &Clean               
       &Speaker Volunteers          
       &1 to 2                          
       &EER\\
       \hline
       
       FoR~\cite{for_ds}(audio)                     &2019            
       &English             
       &140                
       &-/195541         
       &7                           
       &Clean               
       &Kaggle~\cite{for_source}   
       &2.35                           
       &Acc\\
       \hline
       
       ASVspoof 2019 (LA task)~\cite{ASV19}(audio)  &2019            
       &English             
       &46/61           
       &12,483/108,978          
       &19                          
       &Clean     
       &Speaker Volunteers          
       &n/a                            
       &EER \\
       \hline
       
       DFDC~\cite{dfdc_ds}(video)           
       &2020            
       &English             
       &3426              
       &128,154/104,500
       &1                           
       &Clean \& Noisy                   
       &Speaker Volunteers                 
       &68.8                           
       &Pre., Rec. \\
       \hline
       
       ASVspoof 2021 (LA task)~\cite{ASV21}(audio)  &2021        
       &English             
       &21/27           
       &18,452/163,114                
       &13                          
       &Clean \& Noisy      
       &Speaker Volunteers          
       & n/a                             
       &EER \\
       \hline
       
       ASVspoof 2021 (DF task)~\cite{ASV21}(audio)  &2021        
       &English            
       &21/27           
       &\textbf{22,617/589,212}                 
       &\textbf{100+}                        
       &Clean \& Noisy      
       &Speaker Volunteers          
       & n/a                            
       &EER \\
       \hline
  
       WaveFake~\cite{wake_ds}(audio)               &2021            
       &English, 
       &0/2             
       &-/117,985         
       &6                          
       &Clean               
       &LJSPEECH~\cite{ljspeech_ds},    
       &6/4.8                    
       &EER \\
       & &Japanese & & & & &JSUT~\cite{jsut_ds}& & \\
       \hline

       KoDF~\cite{kokd_ds}(video)           &2021            &Korean              &198/205         &62,116/175,776   &2                           &Clean               &Speaker Volunteers         &90/15 (real/fake)            &Acc, AuC  \\
       \hline
       ADD 2022~\cite{add22} &2022 &Chinese & 40/40& 3012/24072&2&Clean&AISHELL-3 \cite{aishell-3}&1 to 10&EER \\
       \hline

       FakeAVCeleb~\cite{khalid2021fakeavceleb}(video) &2022  &English             &250/250         &570/25,000            &2                           &Clean \& Noisy       &Vox-Celeb2~\cite{voxceleb2} &7  &AuC\\
       \hline

       In-the-Wild~\cite{intwi_ds}(video)           
       &2022 &English             
       &58                
       &19963/11816              
       &0                          
       &Clean \& Noisy                   &Self-collected              &4.3                       &EER \\
       \hline
       LAV-DF~\cite{cai2022you}(video)      &2022            &English             &153               &36,431/99,873    &1                           &Clean \& Noisy       &Vox-Celeb2~\cite{voxceleb2}   & 3 to 20                         & AP \\      
       \hline
       
       Voc.v~\cite{VoC_ds}(audio)                   
       &2023            
       &English             
       &46/61           
       &14,250/41,280             
       &5                           
       &Clean \& Noisy       
       &ASVspoof 2019           
       &n/a        
       &EER \\
       \hline

       PartialSpoof~\cite{pasp_ds}(audio)         
       &2023            
       &English             
       &46/61           
       &12,483/108,978   
       &19                          
       &Clean \& Noisy       
       &ASVspoof 2019               
       &0.2 to 6.4  
       &EER \\
       \hline
       
       LibriSeVoc~\cite{f_ds_01}(audio)             
       &2023            
       &English             
       &n/a                 
       &13,201/79,206     
       &6                           
       &Clean \& Noisy       
       &Librispeech                    
       &5 to 34        
       &EER\\      
       
       \hline
       AV-Deepfake1M~\cite{cai2023av, 1mdeepfake_ch}(video)   
       &2023 
       &English      
       &2,068              
       &\textbf{286,721/860,039}  
       &2                           
       &Clean \& Noisy        
       &Voxceleb2~\cite{voxceleb2}                     
       &5 to 35        
       &Acc,  AuC  \\
       \hline

       CFAD~\cite{cfad2023}(audio)                   
       &2024            
       &Chinese             
       &1023           
       &-/374,000             
       &11                           
       &Clean \& Noisy       
       &AISHELL1-3~\cite{aishell1,aishell3}            
       &n/a        
       &EER \\
       
       & & & & & &
       \& Codecs & MAGICDATA~\cite{magicdata}&  & \\
       \hline

        MLAAD~\cite{muller2024mlaad}(audio)         
        &2024            
        &\textbf{Multilanguages (23)}      
        &n/a                  
        &-/76,000             
        &54                          
        &Clean \& Noisy                     
        &M-AILABS~\cite{MAILABS_ds}     
        &n/a        
        &Acc \\
       \hline

    ASVspoof 2024~\cite{ASV24}(audio)     
    &2024            
    &English      
    &n/a                  
    &\textbf{188,819/815,262}             
    &28        
    &Clean \& Noisy      
    &MLS~\cite{mls}    
    &n/a        
    &EER \\
    \hline     
        
   SVDD2024~\cite{svdd2024}(audio)  
   &2024    
   &\textbf{Mutilanguages (6)}      
   &59                  
   &12,169/72,235             
   &48        
   &      
   Clean 
   & Mandarin,   
   & n/a        
   &EER \\
   & & & & & & &Japanese & &  \\

       \hline
    \end{tabular}
    }
    \vspace{-0.3cm}
    \label{table:T2} 
\end{table*}

\textbf{Discussion}: Given the recent analysis of challenge competitions proposed for the DSD task, some concerns can be indicated. 
Firstly, the DSD task has drawn attention from the research community and is now recognized as one of the critical components in a complex system of deepfake detection.
However, most current challenge competitions are limited to single languages, such as Chinese or English,  and primarily focus on detecting fake speech within conversations.
Secondly, some challenge competitions have not published datasets for different reasons. For example, FTC~\cite{FTC_ch} was organized by the US government, and the top-performing systems are used by the US government. Similarly, ADD 2023~\cite{add23} only provides the dataset for the teams that attended during the competition.
These limitations hinder research motivation and further development once the challenges conclude.
%
Third, it is recognized that fake speech utterances are mainly generated from deep-learning-based speech generation systems.
Therefore, if selected deep-learning-based speech generators are not general or up-to-date, this significantly affects the effectiveness and quality of the challenge competition. 
This highlights the need for collaboration between two tasks of deep-learning-based speech generation and detection within the same challenge competition.
Competitions like ASVspoof 2024~\cite{ASV24} and ADD 2022~\cite{add22} have addressed this by not only published datasets but also presented a two-phase or two-track challenge in which the first phase/track is for Deepfake Speech Generation and the second one is for Deepfake Speech Detection.
%
Finally, regarding techniques used in these competitions, ensemble models have become widely leveraged to enhance performance in many challenge competitions, enabling research teams to develop top-performing systems. However, this approach has several drawbacks, including limited interpretability, increased system complexity, high training costs, and concerns related to power consumption and green AI. Therefore, different aspects of using deep-learning-based models such as using a single model, low complexity, or real-time inference can be regarded as main constraints in challenge competitions for the DSD task in the future. For example, the DCASE challenge Task 1~\cite{dcase2022} for Sound Scene Classification requires the submitted systems to obey two constraints: (1) not larger than 128 K parameters and (2) not larger than 30 MMAC units.

\textbf{Our contribution:} Given the analysis and the discussion about the DSD challenge competitions above, our work further motivates the research community by:
\begin{itemize}
    \item We present and highlight the important role of DSD challenge competitions. We then provide a comprehensive analysis and indicate the existing concerns.       
    \item We continue updating new challenge competitions in the future by creating a Github project\footnote{https://github.com/AI-ResearchGroup/A-Comprehensive-Survey-with-Critical-Analysis-for-Deepfake-Speech-Detection}. 
    The GitHub repository serves as a reference for up-to-date information on challenge competitions and current concerns. In other words, it provides a summary of challenge competitions related to the DSD task, ensuring that this survey paper stays updated.
\end{itemize}

\begin{figure}[t]
    \centering
    \includegraphics[width =1.0\linewidth]{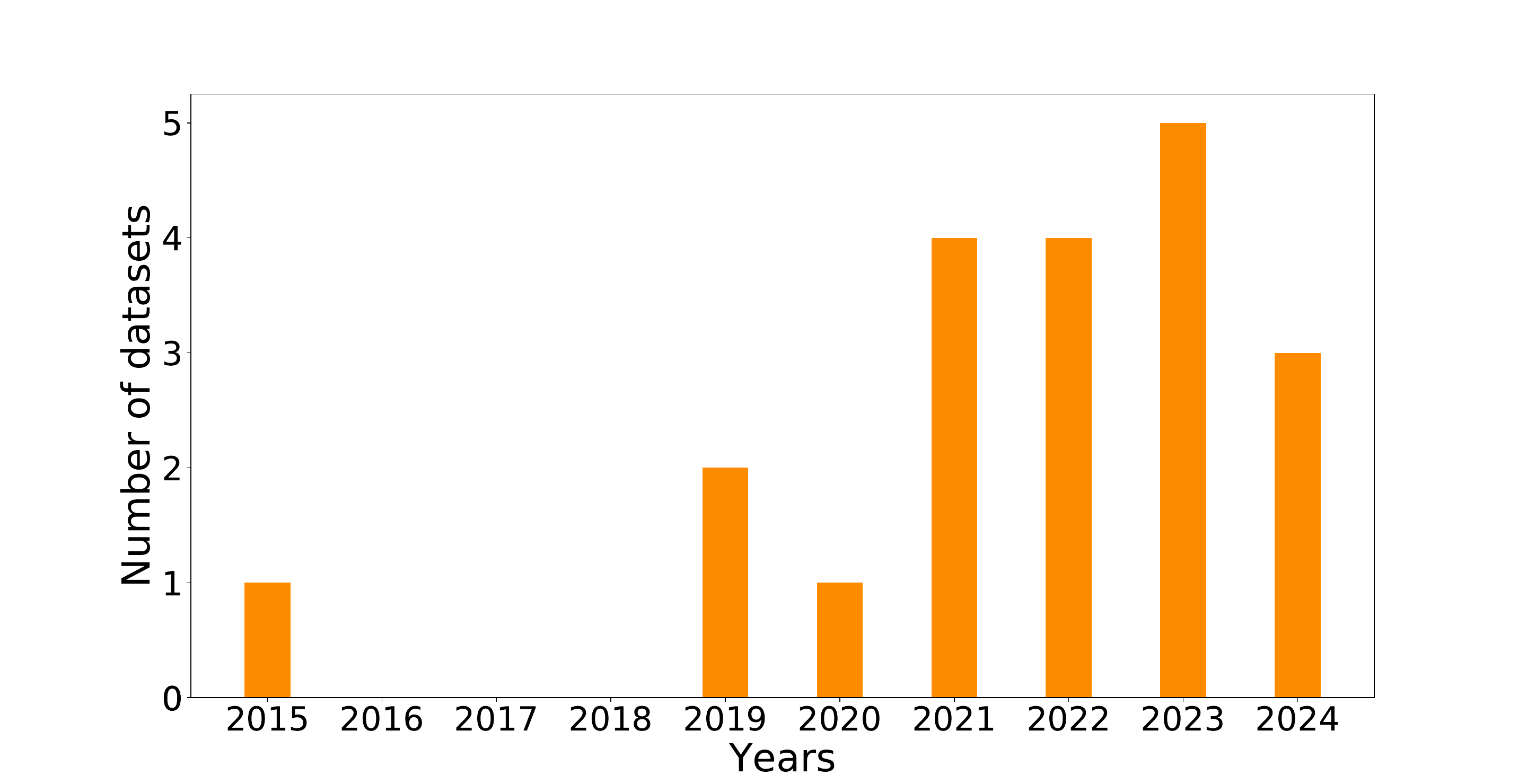}
	\caption{The number of public datasets proposed for DSD task from 2015}
    \label{F2}
\end{figure}

\section{Public Datasets Proposed For Deepfake Speech Detection}
\label{dataset}

\textbf{Analysis:} 
Public datasets proposed for the DSD task, including those introduced through challenge competitions, play a crucial role in motivating the research community to develop and evaluate DSD systems. In this section, we present a summary of the public and benchmark datasets for the DSD task, as shown in Table~\ref{table:T2}. These datasets have been introduced through various challenge competitions and published papers.

As illustrated in Fig.\ref{F2}, the number of public datasets for the DSD task has grown significantly in recent years. Most of these datasets include both clean and noisy speech. Notably, nearly all datasets have been designed for English, with WaveFake~\cite{wake_ds}, KoDF~\cite{kodf_s1}, and ADD 2022~\cite{add22} being the exceptions, focusing on Japanese, Korean, and Chinese languages, respectively. 
Recently, the first multilingual datasets for the DSD task were introduced in~\cite{muller2024mlaad} and~\cite{svdd2024}. 
The MLAAD dataset~\cite{MAILABS_ds} provides fake speech in conversations generated in 23 widely spoken languages.
Meanwhile, the SVDD dataset~\cite{svdd2024} was proposed for deepfake singing detection with six different languages (i.e., the Chinese songs are the majority).

\begin{table*}[t]
    \caption{Deepfake Speech Generation Systems Used in Public DSD Datasets \\ (TTS: Text to Speech, VC: Voice Conversion, AT: Adversarial attach using Malafide or Malocopula)} 
        	\vspace{-0.2cm}
    \centering
    \scalebox{0.85}{
    \begin{tabular}{|l|c|c|l|} 
        \hline 
        \textbf{Datasets} &\textbf{Year} &\textbf{No. of TTS/VC/AT} &\textbf{Deepfake Speech Generation Systems}   \\  
       \hline
        ASVspoof 2015~\cite{ASV15}   &2015   &7 VC, 3 TTS & VC-01~\cite{asv15_s1_01, asv15_s1_02}, VC-02~\cite{asv15_s2}, TTS-01~\cite{asv15_s3}, TTS-02~\cite{asv15_s3}, VC-03~\cite{asv15_s5}, \\
        & & & VC-04~\cite{asv15_s6}, VC-05~\cite{asv15_s6}, VC-06~\cite{asv15_s8}, VC-07~\cite{asv15_s9}, TTS-03~\cite{asv15_s10} \\
        \hline
        FoR~\cite{for_ds}          &2019     &7 TTS   & Deep Voice 3, Amazon AWS Polly, Baidu TTS, Google Traditional TTS, \\
        &&&Google Cloud TTS, Google Wavenet TTS, Microsoft Azure TTS\\
        \hline
        ASVspoof 2019 (LA task)~\cite{ASV19} &2019  & 8 VC, 11 TTS &TTS-01~\cite{A1}, TTS-02~\cite{A1, A2}, TTS-03~\cite{A3}, TTS-04~\cite{A4}, VC-01~\cite{A5}, VC-02~\cite{A6}, \\
        &&& TTS-05~\cite{A3, A7}, TTS-06~\cite{A1, A8}, TTS-07~\cite{A9_1, A9_2}, TTS-08~\cite{A10_1, A10_2}, TTS-09~\cite{A10_1, A10_2, A11}, \\
        &&& TTS-10~\cite{A12}, VC-03+TTS~\cite{A13_1}, VC-04+TTS~\cite{A14_1, A14_2}, VC-05+TTS~\cite{A14_1, A14_2}, TTS-11~\cite{A4}, \\
        &&& VC-06~\cite{A17_1, A17_2}, VC-07~\cite{A18_1, A18_2, A18_3}, VC-08~\cite{A6}\\
        \hline
         DFDC~\cite{dfdc_ds} &2020  & 1 TTS & TTS Skins voice conversion~\cite{dfdc_s1}\\
       \hline        
        KoDF~\cite{kokd_ds} &2021 &2 TTS & ATFHP~\cite{kodf_s1} and Wav2Lip~\cite{kodf_s2} \\
       \hline 
        ASVspoof 2021 (LA task)~\cite{ASV21} &2021 & 13 TTS/VC & Reuse ASVspoof 2019  \\
          \hline
        ASVspoof 2021 (DF task)~\cite{ASV21} &2021 & 100 TTS/VC   & Vocoders \cite{asv_2021_journal} \\
        

          \hline
        WaveFake~\cite{wake_ds}  &2021   &6 TTS   &MelGAN~\cite{wake_s1}, FB-MelGAN~\cite{wake_s1}, HiFi-GAN~\cite{wake_s3}, WaveGlow~\cite{wake_s4}, PWG~\cite{wake_s2}, MB-MelGAN~\cite{wake_s1}\\
        
          \hline

        FakeAVCeleb~\cite{khalid2021fakeavceleb}  &2022 &  2 TTS &SV2TTS~\cite{FakeAVCeleb_s1, FakeAVCeleb_s2}  \\
        
        \hline
          In-the-Wild~\cite{intwi_ds}   &2022    & n/a & n/a \\
               \hline
             LAV-DF~\cite{cai2022you} &2022  & 1 TTS &SV2TTS~\cite{la_vd_s1} \\
      \hline   
        Voc.v~\cite{VoC_ds}        &2023   &  5 TTS &   HiFi-GAN~\cite{wake_s3}, MB-MelGAN~\cite{wake_s1}, WaveGlow~\cite{wake_s4}, PWG~\cite{wake_s2},  Hn-NSF~\cite{voc_s1}\\                
        \hline
        
        PartialSpoof~\cite{pasp_ds} &2023  &   21 TTS/VC    & Reuse ASVspoof 2019 \\
           \hline
       LibriSeVoc~\cite{f_ds_01} &2023  &   6 TTS/VC    & WaveNet~\cite{A12}, WaveRNN~\cite{wavernn}, MelGAN~\cite{wake_s1},
       Parallel WaveGAn~\cite{paralel_wavegan},
       WaveGrad~\cite{wave_grad},
       DiffWave~\cite{diff_wave}\\
           \hline
       
      AV-Deepfake1M~\cite{cai2023av, 1mdeepfake_ch} &2023 &  2 TTS  &VITS~\cite{av_deepfake_1M_tts_s2}, YoursTTS~\cite{av_deepfake_1M_tts_s1} \\
            \hline
         CFAD~\cite{cfad2023}        &2024   &  11 TTS &  STRAIGHT~\cite{straight}, Griffin-Lim~\cite{griffin_lim}, LPCNet~\cite{lpcnet}, WaveNet~\cite{A12}, PWG~\cite{wake_s2}, HiFi-GAN\cite{hifigan},\\
         &&&
         MB-MelGAN~\cite{wake_s1},
         MelGAN~\cite{wake_s1},
         WORLD~\cite{world},
         FastSpeech~\cite{fast_speech},
         Tacotron-HifiGAN~\cite{tacotron}
         \\
         \hline
      MLAAD~\cite{muller2024mlaad} &2024 &54 TTS & Bark, Capacitron, FastPitch, GlowTTS, Griffin Lim, Jenny, NeuralHMM, Overflow, \\
      &&& Parler TTS,  Speech5, Tacotron DDC, Tacotron2, Tacotron2 DCA, Tacotron2 DH, Tcotron2-DDC, \\
      &&& Tortoise, VITS, VITS Neon, VITS-MMS, XTTS v1.1, XTTS v2  \\
       \hline 
       ASVspoof 2024~\cite{ASV24}&2024 &15 TTS, 6 VC, 7 AT &TTS-01~\cite{asv24_tts1}, TTS-02~\cite{asv24_tts2}, TTS-03~\cite{asv24_tts3},  TTS-04~\cite{asv24_tts4}, TTS-05~\cite{asv24_tts5}, TTS-06\cite{asv24_tts6}, TTS-07\cite{asv24_tts7}, \\
       &&& TTS-08(self-develop), VC-01\cite{asv24_vc1}, TTS-09\cite{asv24_tts9}, VC-02~\cite{asv24_vc2}, VC-03(self-develop), TTS-10~\cite{asv24_tts10}, \\
       &&& AT-01 (Malafide+TTS-10~\cite{asv24_tts10}), TTS-11~\cite{asv24_tts11}, AT-02(self-Develop), TTS-12~\cite{asv24_tts12}, TTS-13~\cite{asv24_tts13}, \\
       &&& AT-03(Malafide+TTS~\cite{asv24_tts_xx}), VC-04(self-develop), VC-05~\cite{asv24_vc5}[24], VC-06(add noise), \\
       &&& AT-04(Malacopula+VC-06), TTS-14~\cite{asv24_tts14}, TTS-15~\cite{asv24_tts15}, AT-05(Malacopula+AT-01), \\
       &&& AT-06(Malacopula+TTS-13~\cite{asv24_tts13}), AT-07(Malacopula+VC-05~\cite{asv24_vc5}) \\
       \hline
    \end{tabular}
    }
    \vspace{-0.3cm}
    \label{table:T3} 
\end{table*}

Most deepfake datasets are generated from one of three generator techniques: Text-to-Speech (TTS), Voice Conversion (VC), and Adversarial Attacks (AT), as shown in Table~\ref{table:T3}.
Notably, ASVspoof 2024~\cite{ASV24} is the first dataset that uses AT systems to generate fake speech.
While TTS systems generate fake speech from text, VC systems generate fake speech from real speech (e.g., audio).
To mimic the target speakers, TTS and VC systems attempt to explore the audio embeddings extracted from the target speakers.
These audio embeddings are treated as a part of the feature map in the entire network architecture in TTS and VC systems.
Regarding AT systems, they mainly apply Malafide~\cite{Malafide} and Malocopula~\cite{Malacopula} methods to generate fake speech.
Both Malafide~\cite{Malafide} and Malocopula~\cite{Malacopula} methods involve  leveraging filter banks.
Malafide~\cite{Malafide} applies multiple techniques of linear time-invariant (LTI), non-causal filter, and the coefficients (e.g., tap weights) to create TTS/VC-based fake speech that mimics the target speaker.
Meanwhile, Malocopula~\cite{Malacopula} combines both linear filter and non-linear filter (e.g., one-dimensional convolutional layer) to replicate the target speaker's voice.

To compare among DSD datasets, we analyze three different aspects: (1) the number of fake utterances; (2) the AI-synthesized speech systems used to generate fake speech; and (3) the original/real human speech resource used to generate fake speech utterances.
As Table~\ref{table:T2} shows, most datasets present lower than 300,000 utterances of fake speech, except  ASVspoof 2021 (DF Task)~\cite{ASV21}, ASVspoof 2024~\cite{ASV24}, and 
AV-Deepfake1M dataset~\cite{cai2023av, 1mdeepfake_ch} with 589212, 815262, and 860039 fake samples, respectively.
Although DFDC~\cite{dfdc_ds,dfdc_s1} and AV-Deepfake1M dataset~\cite{cai2023av, 1mdeepfake_ch} present a large number of fake data, this was proposed for video in which audio may not be fake.
Additionally, these fake utterances were generated from only a few deep-learning-based speech-generation systems.
Indeed, two TTS models of VITS~\cite{av_deepfake_1M_tts_s2}, YoursTTS~\cite{av_deepfake_1M_tts_s1} and one TTS model~\cite{dfdc_s1} were used to generate fake speech in DFDC~\cite{dfdc_ds} and  AV-Deepfake1M dataset~\cite{cai2023av, 1mdeepfake_ch} datasets, respectively. 
On the other hand, the ASVspoof 2021 (DF Eva) dataset~\cite{ASV21} contains 589212 fake utterances, generated using over 100 voice conversion (VC) and text-to-speech (TTS) systems. 
To catch up with state-of-the-art deepfake speech generators, Table~\ref{table:T3} presents the architectures and resources of deepfake speech generators.
The table indicates that the ASVspoofing series show up-to-date and diverse deepfake speech generators compared to the others.
In terms of the original human speech resources, most DSD datasets are based on recordings from a limited number of speaker volunteers. 
For example, although the ASVspoof 2021 (DF Eva) dataset~\cite{ASV21} used 100 VC and TTS systems to create fake utterances, the real speech resource is from 107 speaker volunteers.
%
%
Some DSD datasets of AV-Deepfake1M~\cite{cai2023av, 1mdeepfake_ch}, CFAD~\cite{cfad2023} leveraged the large and available human speech datasets to generate fake utterances such as Voxceleb2~\cite{voxceleb2}, AISHELLI-3~\cite{aishell-3}, MAGICDATA~\cite{magicdata}.
However, these datasets use a limited number of speech generators (e.g., 2 TTS and 11 TTS for AV-Deepfake1M~\cite{cai2023av, 1mdeepfake_ch} and CFAD~\cite{cfad2023}, respectively). 

%
\begin{figure*}[t]
    \centering
    \includegraphics[width =0.9\linewidth, height = 0.15\linewidth]{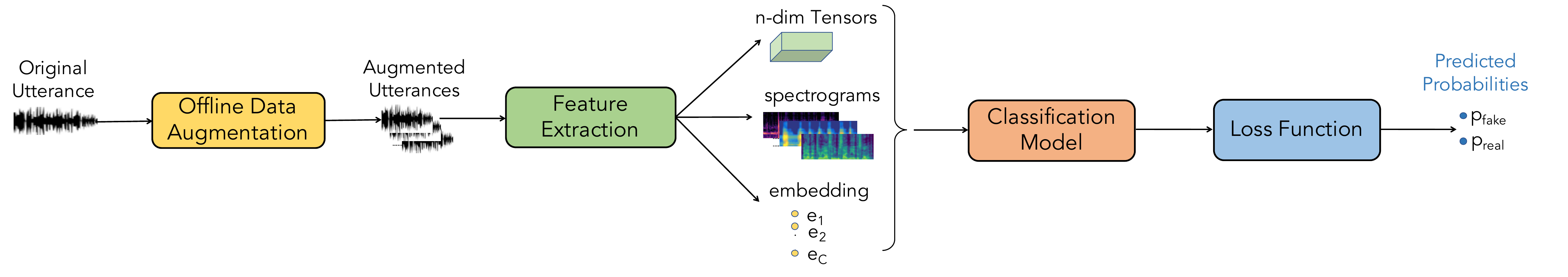}
	\caption{The high-level architecture of Deepfake Speech Detection (DSD) systems}
    \label{F4}
\end{figure*}
Regarding metrics evaluation, all datasets proposed for the DSD task come together with a baseline and metrics for the evaluation.
Regarding the baseline systems, all baselines leveraged convolutional neural network (CNN) based architectures.
These baselines are evaluated mainly by the Equal Error Rate (EER) metric. 
Some datasets such as KoDF~\cite{kokd_ds}, AV-Deepfake1M~\cite{cai2023av, 1mdeepfake_ch}, MLAAD~\cite{muller2024mlaad}, FoR~\cite{for_ds} used Accuracy (Acc.) and Area Under The Curve  (AUC) metrics instead of EER.

\textbf{Discussion}: Given the analysis of benchmark datasets proposed for the DSD task, some existing issues can be outlined. These include the limited number of datasets available for multiple languages and the imbalance of several aspects within existing datasets.

Firstly, more public and benchmark datasets have been proposed for the DSD task. 
However, there is only one multilingual dataset currently. The lack of multilingual datasets for DSD tasks presents several challenges for current model development and evaluation such as performance degradation on cross-language settings that leads to a limited applicability in real-world applications. This motivates the research community to propose more datasets for multiple languages to enhance model's capability in real-life settings.
%
\begin{figure}[t]
    \centering
    \includegraphics[width =0.8\linewidth]{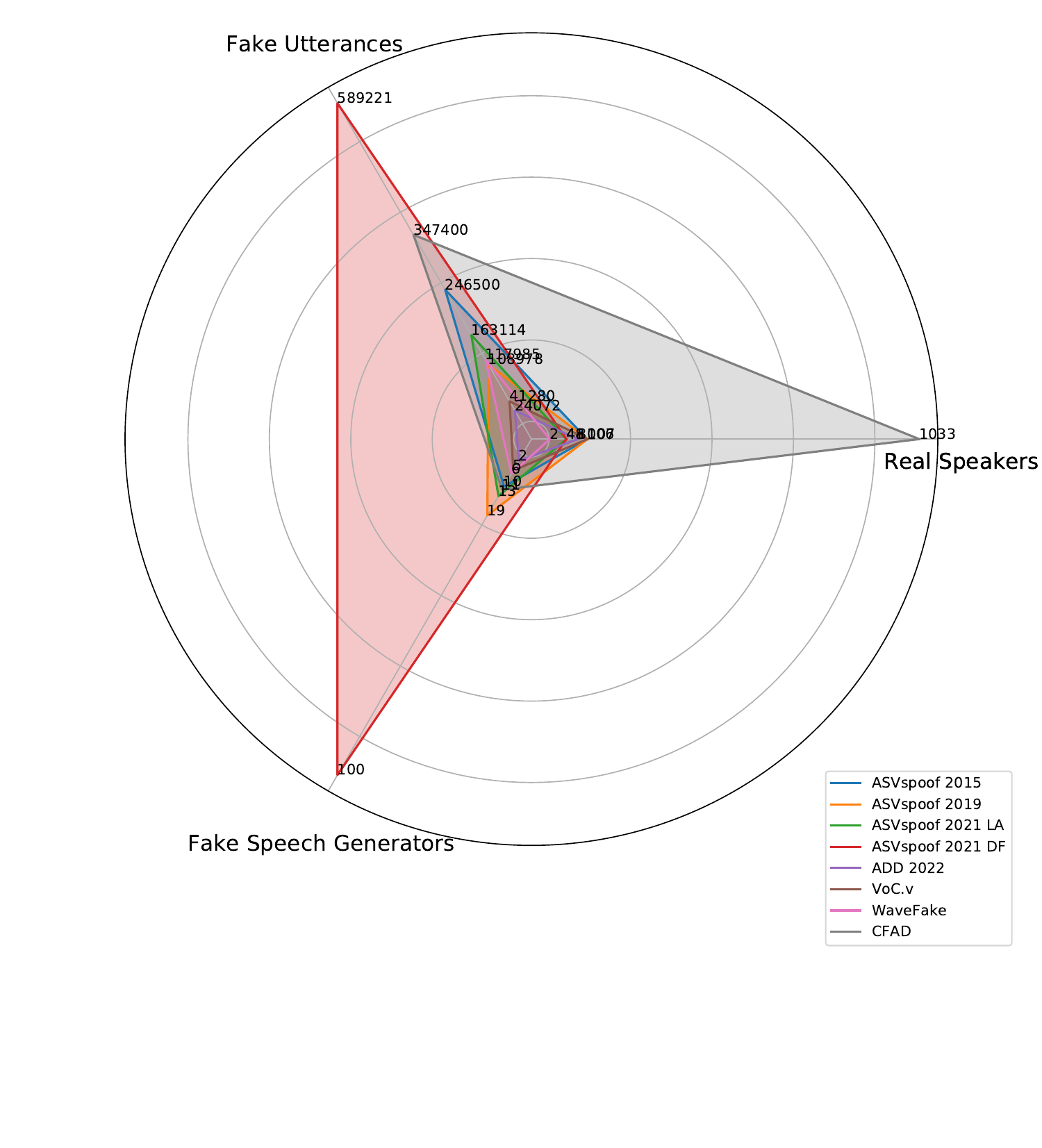}
       	\vspace{-1.5cm}
	\caption{The imbalance among the fake speech utterances, the fake speech generators, and the real speaker volunteers in benchmark DSD datasets}
    \label{F3}
\end{figure}
Secondly, another limitation of currently available datasets is that they focus on a limited number of DSD use cases. In particular, two use cases should be clearly distinguished: 1) detecting deepfakes \textit{without} access to the original voice, and 2) detecting deepfakes \textit{with} access to the original voice. The current datasets are designed for addressing the former but not the latter use case as they lack authentic-cloned speech pairing. Another highly relevant use case that should be addressed in the future is \textit{partially} deepfake speech whereby just a part of the speech is being replaced by a synthetic component.
Thirdly, we highlight an imbalance among DSD datasets regarding three aspects: (1) the number of fake utterances; (2) the AI-synthesized speech systems used to generate fake speech; and (3) the original/real human speech resource used to generate fake speech utterances.
The imbalance can be clearly described in Fig.~\ref{F3} regarding DSD datasets using speaker volunteers.
\begin{itemize}
    \item \textbf{The number of utterances:} The quantity of utterances within the datasets is not uniform. Some datasets may contain a large number of samples, while others have significantly fewer. A small number of real or fake utterances within datasets (e.g., FakeAVCeleb \cite{khalid2021fakeavceleb}, ADD~\cite{add22}) limits the model's exposure to a wide variety of speech patterns and scenarios, affecting the detection robustness and generalization on new, unseen data. Additionally, a controlled ratio between real and fake samples created within datasets (e.g., ASVspoof 2024~\cite{ASV24}, ASVsproof 2021~\cite{ASV21}) also ensure diversity of fake techniques and avoid overfitting on the fake data, especially if the fake samples are generated using similar techniques. Therefore, maintaining a moderately controlled ratio between real and fake utterances, along with a diverse range of these utterances, is essential for future dataset development.
\begin{table*}[t]
    \caption{Individual DSD Systems Exploring Raw Audio} 
        	\vspace{-0.2cm}
    \centering
    \scalebox{0.85}{
    \begin{tabular}{|c|l|l|l|l|l|l|} 
        \hline 
        \textbf{Systems} &\textbf{Years} &\textbf{Datasets} &\textbf{Features}  &\textbf{Data Augmentation}  &\textbf{Models} &\textbf{Loss Functions}  \\  
                         &               &                  &                   &(\textbf{Distoration/Compression})          &                & \\ 

       \hline
       
        ~\cite{m01} 
        &2021 
        &ASVspoof 2021 (LA Task) 
        &Raw Audio 
        &\underline{Comp.}: MP3, ACC, OGG 
        &RawNet2 
        &Focal loss\\
        \hline
        
        ~\cite{m04} 
        &2021 
        &ASVspoof 2021 (LA\&DF Tasks) 
        &Raw Audio 
        &\underline{Comp.}: G.723, G.726, 
        &RawNet2 
        & Cross Entropy (CE)\\ 
        & & & &GSM, opus,  speex, mp2, & & \\
        & & & & ogg, tta, wma, acc, ra & & \\
        \hline
        
        ~\cite{m05} 
        &2021 
        &ASVspoof 2019 (LA Task) 
        &Raw Audio 
        &\underline{Dis.}: Channel Drop, 
        &SinC+CRNN &MSE Loss\\
        
        & & & &Frequency masking & &\\
        \hline
        
       ~\cite{m06} 
       &2021 
       &ASVspoof 2021 (LA Task)  
       &Raw Audio 
       &\underline{Comp.}: mp3, mp2, m4a, m4r,   
       & RawNet2 
       & OC-Softmax \\  
       & & & &opus, ogg, mov, PCM µ-law, & & \\
       & & & &PCM a-law, speex, ilbc,  & & \\
       & & & & G.729, GSM, G.722, AMR & & \\
       \hline
       
       ~\cite{m07} 
       &2021 
       &ASVspoof 2021 (LA\&DF Tasks)  
       &Raw Audio 
       &\underline{Dis.}: Time-wise,  
       &RawNet2 
       &Cross Entropy  \\
       &&&&Silence Strimming && \\
        \hline

       ~\cite{m08} 
       &2021 
       &ASVspoof 2021 (LA\&DF Tasks)  
       & Raw Audio 
       &n/a  
       &\underline{Encoder}: SinC+Residual 
       &WCE Loss \\
       & & & & &\underline{Decoder}: Graph Attention Network & \\
       \hline

       ~\cite{m09} 
       &2021 
       &ASVspoof 2021 (LA\&DF Tasks)  
       &Raw Audio 
       &\underline{Dis.}: Mixup, FIR filters  
       &Sinc+CNN 
       &WCE Loss  \\
        \hline  
        
       ~\cite{m12} 
       &2021 
       &ASVspoof 2021 (LA Task)  
       &Raw Audio 
       &\underline{Comp.}: G.711-alaw,G.722,  
       &SinC+RawNet2 
       &AM-softmax \\
       
       & & & &GSM-FR, and G.729& &\\
        \hline

        ~\cite{intwi_ds} 
        &2022 
        &ASVspoof 2019 (LA Task) 
        &Raw Audio 
        & n/a 
        &RawNet2, RawNet-GAT, CRNNSpoof 
        &Cross Entropy \\
        & &In The Wild & & & & \\
        \hline

        ~\cite{m35} 
        &2022 
        &ASVspoof 2019 (LA Task)  
        & Raw Audio 
        &n/a  
        &\underline{Encoder}: RawNet2 
        & WCE Loss \\
        & & & & & \underline{Decoder}: Graph Attention Neural Network & \\
        \hline
        
        ~\cite{m34} 
        &2022 
        &ASVspoof 2021 (LA\&DF Tasks)  
        &Raw Audio 
        &\underline{Dis.}: RawBoost~\cite{rawboost}  
        &\underline{Encoder}: Sinc+CNN, Wave2Vec2.0+CNN  
        &WCE loss \\
        & & & & &\underline{Decoder}: Graph Attention network &  \\
        \hline

       ~\cite{m17} 
       &2023 
       &ASVspoof 2019 (LA Task), 
       &Raw Audio 
       &\underline{Dis.}: Stereo speech  
       &\underline{Encoder}: SinC+ResNet 
       & AM-softmax \\
       & &ASVspoof 2021 (LA\&DF Tasks) & & &\underline{Decoder}: Graph Attention network & \\
        \hline

        ~\cite{m25} 
        &2023 
        &ASVspoof 2019 (LA Task)  
        &Raw Audio 
        &n/a 
        &\underline{Encoder}: Wav2vec2.0~\cite{wav2vec20}, HuBERT~\cite{Hubert} 
        &Cross Entropy\\
        & & & & &\underline{Decoder}: LCNN-LSTM-Graph Attention & \\
        \hline

        ~\cite{m43} 
        &2023 
        &ADD 2023 
        & Raw Audio 
        &\underline{Dis.}: Add noise, mix utterance  
        &\underline{Encoder}: Wav2Vec2.0  
        &Cross Entropy \\
        & & & & &\underline{Decoder}: ECAPA-TDNN& \\
        \hline

        ~\cite{m53} 
        &2022 
        &ASVspoof 2019 (LA Task), 
        & Raw Audio 
        &n/a  
        &\underline{Encoder}: ECAPA-TDNN, RawNet  
        &Cross Entropy,  \\
        & & & & &\underline{Decoder}: Linear layers & Triplet loss, \\
        &&&&&&AM-Softmax\\
        \hline

        ~\cite{m49} 
        &2023 
        &ADD 2023 
        & Raw Audio
        &\underline{Dis.}:Add noise, vibration, mixup 
        &\underline{Encoder}: Wav2Vec2.0 &A-Softmax,\\
        &&&&&\underline{Decoder}:CNN-Transformer &Triplet loss,\\
        &&&&& &Adversial loss\\
        \hline

        ~\cite{m37} 
        &2023 
        &ASVspoof 2019 (LA Task),  
        & Raw Audio 
        &n/a  
        &\underline{Encoder}: Wav2Vec2.0~\cite{wav2vec20} 
        & Triplet, BCE,   \\
        & &WaveFake,      & & &\underline{Decoder}: LCNN-Transformer & Adversarial loss\\
        & &FakeAVCeleb   & & & &\\
        \hline

       ~\cite{m15} 
       &2024 
       &ASVspoof 2019 (LA Task), 
       &Raw Audio 
       &n/a 
       &SincNet/LEAF+ResNet 
       &Cross Entropy \\
        & &ASVspoof 2021 (LA\&DF Tasks), & & & & \\
        & &In The Wild~\cite{intwi_ds} & & & & \\
        \hline
        
        ~\cite{m15} 
        &2024 
        &ASVspoof 2021 (LA\&DF Tasks) 
        &Raw Audio 
        &n/a 
        & \underline{Encoder}:  EnCodec~\cite{Encode_01}, AudioDec~\cite{Encode_02},  
        &Cross Entropy \\
        
        & & & & & AudioMAE~\cite{Encode_03}, HuBERT~\cite{Hubert},  & \\
        & & & & & WavLM~\cite{Wavlm}, Whisper~\cite{whisper} & \\
        & & & & &\underline{Decoder}: ResNet & \\
        \hline   
        
        ~\cite{m20} 
        &2024 
        &ASVspoof 2019 (LA Task),       
        &Raw Audio 
        &\underline{Dis.}: Add noise, overlapping  
        &\underline{Encoder}: WavLM~\cite{Wavlm}, 
        &Cross Entropy \\
        & &ASVspoof 2021 (LA\&DF Tasks) & & &\underline{Decoder}: Multi-Fusion Attentive & \\
         \hline
         
        ~\cite{m32} 
        &2024 
        &ASVspoof 2019 (LA Task),  
        &Raw Audio &n/a &\underline{Encoder}: Wav2vec2.0~\cite{wav2vec20}, BEATS~\cite{beats},  
        &n/a \\
        & &ASVspoof 2021 (LA Task),  & & & LationCLAP~\cite{clap}, AudioCLIP~\cite{audioClib}, & \\
        & &In The Wild & & &  \underline{Decoder}: Similarity Score Measurement & \\ 
         \hline  
    \end{tabular}
    }
    \label{S1} 
\end{table*}
    \item \textbf{Deepfake speech generation systems:} The variety of deep-learning-based systems used to generate deepfake speech is another area of concern. As Table~\ref{table:T3} shows, some of datasets such as MLAAD~\cite{muller2024mlaad}, ASVspoof 2021(DF task)~\cite{ASV21}, ASVspoof 2024~\cite{ASV24} present more than 20 systems (e.g., TTS, VC, or AT systems).
    Among these datasets, ASVspoof 2021 (DF Task)~\cite{ASV21} and ASVspoof 2024~\cite{ASV24} present diverse TTS, VC, and AT systems.
    In particular, while more than 100 TTS and VC are for ASVspoof 2021 (DF Task)~\cite{ASV21}, 28 TTS, VC, and AT are used in ASVspoof 2024~\cite{ASV24}.
    Although MLAAD~\cite{muller2024mlaad} has been the unique multiple-language dataset currently, fake speech in this dataset was only generated from 54 TTS systems. Overall, some datasets may predominantly feature speech synthesized by a few specific deep-learning-based generators or techniques, while others might include a broader range. 
    Datasets generated from a limited number of deep-learning-based generators possibly lead to over-specialization, reducing the model’s ability to detect deepfakes generated by other systems and affecting the performance in real-world scenarios. 
    Therefore, this imbalance motivates the research community to create more diverse datasets that include a wide range of AI-synthesized speech methods.
\begin{table*}[t]
    \caption{Individual DSD Systems Exploring Spectrogram Based Features} 
        	\vspace{-0.2cm}
    \centering
    \scalebox{0.85}{
    \begin{tabular}{|l|l|l|l|l|l|l|} 
        \hline 
        \textbf{Systems} &\textbf{Years} &\textbf{Datasets}   &\textbf{Data Augmentation}   &\textbf{Features}          &\textbf{Models} &\textbf{Loss Functions} \\  
                         &               &                    &(\textbf{Distoration/Compression})                             &                           &                & \\ 
        
       \hline
               ~\cite{m48} &2020 &ASVspoof 2019 (LA Task)  &\underline{Dis.}: Add noise, reverberation,       &LFCC                       &ResNet   &LMC loss, \\
               &&&FreqAugment & & &Cross Entropy \\

       \hline
        ~\cite{m01} &2021 &ASVspoof 2021 (LA Task)  &\underline{Comp.}: MP3, ACC, OGG        &LFCC                       &LCNN   &Focal loss, \\
                    & &                         &                             &MEL                        &TDNN   &Focal, Cross Entropy \\
        \hline
        ~\cite{m50} &2021 &ASVspoof 2021 (LA Task)  &n/a        &LFB, SPEC, LFCC                       &LCNN, LCNN-LSTM   &Cross Entropy, MSE \\
        \hline
       ~\cite{m02} &2021  &ASVspoof 2021 (LA Task)  &\underline{Comp.}: MP3, ACC, &LFCC  &ECAPA-TDNN &Focal loss\\
                   &&&landlie, cellular, VoiP &  && \\
        \hline   
     ~\cite{m04} &2021 &ASVspoof 2021 (LA\&DF Tasks) &\underline{Comp.}: G.723, G.726, GSM,  &CQT & LCNN & Cross Entropy\\
                 &&&opus, speex, mp2, ogg,   &CQCC, LFCC &GMM &\\
                &&& tta, wma, acc, ra&LFCC &GMM, LCNN & \\
        \hline
       ~\cite{m06} &2021 &ASVspoof 2021 (LA Task) & \underline{Comp.}: G.723, G.726, GSM & PSCC, LFCC,    &Resnet18, TDNN &OC-Softmax \\  
                   &&&opus, speex, mp2, ogg, &DCT-DFT, LLFB&& \\
                   &&&tta, wma, acc, ra & &&\\
       
       \hline
       ~\cite{m07} &2021 &ASVspoof 2021 (LA\&DF Tasks)  &\underline{Dis.}: Time-wise, &CQT & ResNet, CNN, LSTM &Cross Entropy \\
       &&&Silence Strimming &&& \\
        \hline   
       ~\cite{m09} &2021 &ASVspoof 2021 (LA\&DF Tasks)  &\underline{Dis.}: Mixup, FIR filters & MSTFT  & ResNet, LCNN  &Central loss \\
        \hline
       ~\cite{m10} &2021 &ASVspoof 2019, 2021 (LA Task) &n/a &LFCCs, logLFBs,  &Squeeze CNN &Cross Entropy,\\
                   & &&&GM-LFBs,   & &A-Softmax loss \\
                   & &&&Textrograms& &MLC loss \\
        \hline
      ~\cite{m11} &2021&ASVspoof 2021 (LA\&DF Tasks)  &\underline{Comp.}: MP3, AAC, &LFCCs  &ECAPA-TDNN, ResNet  &OC-Softmax, \\
                  & &                              &Landlie, cellular; &&&P2SGrad losses \\
                  & &                              &\underline{Dis.}: device impulse &&& \\ 
        \hline
      ~\cite{m12} &2021 &ASVspoof 2021 (LA Task)  &\underline{Comp.}: G.711-alaw, G.722, &LFCCs &LCNN &AM-softmax \\
                  &&&GSM-FR, and G.729 &&& \\
       \hline
       ~\cite{m31} &2021 &ASVspoof 2019 (LA Tasks)  &n/a &LFCC &ResNet &OC-Softmax \\
        \hline
        ~\cite{m33} &2021 &ASVspoof 2019 (LA Tasks)  &n/a &LFCC &LSTM-SECNN &MSE loss \\
        \hline
        ~\cite{m46} &2021 &ASVspoof 2019 (LA Tasks)  &\underline{Dis.}: SpecAug &log-Mel &ResNet &n/a \\
         \hline
       ~\cite{intwi_ds} &2022 &ASVspoof 2019 (LA Task),   &n/a  &CQT, log-STFT        &LCNN, CNN-LSTM, Inception, &Cross Entropy \\
                         & &In the Wild     &   &MEL            &ResNet, Transformer & \\      
                         
        \hline
        ~\cite{m45} &2022 &ADD 2022  &\underline{Dis.}: Add noise/music/babele, &LFCC &ResNet &Focal loss \\
                    & &          &Reverb, Modify Volume, SpecAug; & & & \\
                    & &          &\underline{Comp.}: MP3, OGG, AAC, OPUS &&& \\
        \hline
       ~\cite{m22} &2023 &ASVspoof 2019 (LA Task), &n/a  &LFCC &LCNN-LSTM &Cross Entropy, \\
               & &WaveFake, FakeAVCeleb  &&&&Adversarial loss, \\
               &&&&&&Triplet loss \\
         \hline
          ~\cite{m24} &2023 &ASVspoof 2019 (LA Task)   &\underline{Comp.}: FLAC &MEL &Finetune SSAT Transformer &Cross Entropy \\
         \hline 
        ~\cite{m49} &2023 &ASVspoof 2019 (LA Task)   &n/a &STFT+F0 sub-bands  &SENet34 &A-Softmax,\\
        &&&&&&KL loss\\
        \hline
                ~\cite{m52} &2023 &ASVspoof 2019 (LA Task)   &n/a &LFCC, CQT  &Teacher-Student &OC-Softmax,\\
        &&&&&(ResNet, LCNN) &MSE loss\\
          \hline
       ~\cite{m15} &2024 &ASVspoof 2019 (LA Task), &n/a &CQT, MEL,  &ResNet &Cross Entropy \\
                                             &&&  &logSpec, LFCC &  & \\                              
       \hline  
       ~\cite{m16} &2024 &ASVspoof 2019 (LA Task),  &\underline{Dis.}: SpecAugment  &FBank  &ECAPA-TDNN &AM-Softmax \\
                   &     &ASVspoof 2021 (LA\&DF Tasks) &&&& \\
        \hline       
        ~\cite{m19} &2024 &ASVspoof 2019 (LA Task),  &\underline{Dis.}: RawBoost~\cite{rawboost} &log-MEL &\underline{Encoder}: CNN, ResNet,  &Cross Entropy, \\
                    & &ASVspoof 2021 (LA\&DF Tasks) &&&SE-ResNet &Contrastive loss \\
                    & & &&&\underline{Decoder}: GAN networks~\cite{Decoder_01_GAN} & \\
       \hline  
       ~\cite{m21} &2024 &ASVspoof 2019 (LA Task) &\underline{Dis.}: Oversampling  &STFT &\underline{Encoder}: Transformer &Cross Entropy \\
                           & &   &&&\underline{Decoder}: Transformer& \\

         \hline       
       ~\cite{m23} &2024 &ASVspoof 2019 (LA Task), &\underline{Dis.}: RawBoost~\cite{rawboost} &MEL &Finetune Wav2Vec2.0 &Cross Entropy,  \\
          & & ASVspoof 2021 (LA\&DF Tasks),   &&&(XLSR-53~\cite{wav2vec20})&Contrastive loss \\
          & & FakeAVCeleb, WaveFake  &&&& \\
         \hline
        ~\cite{m26} 
        &2024 
        &ASVspoof 2019 (LA Task)  &\underline{Comp.}: aac, flac, mp3, m4a  
        &LFCC 
        &\underline{Encoder}: Transformer 
        &OC-Softmax \\
                    
        & & ASVspoof 2021 (DF Task)  &wma, ogg, wa & &\underline{Decoder}: Transformer& \\
        & &  & \underline{Dis.}: Speed perturbation, SpecAug& & & \\
        \hline
     \end{tabular}
     }
    \label{S2} 
\end{table*}
    \item \textbf{Real human speech
resource:} The source of real voice plays a crucial role in shaping the effectiveness, generalization, and ethical aspects of deepfake detection models. 
As highlighted in Table~\ref{table:T3}, there are two main sources for building DSD datasets: voice samples from volunteer speakers or from existing datasets. 
Voice samples from volunteers offer greater control over diversity (if managed thoroughly) and address ethical concerns, as they are collected with explicitly informed consent. However, this approach can be resource-intensive in terms of time and cost and may not scale efficiently. 
In contrast, utilizing existing human speech datasets offers better accessibility and scalability.
However, it may introduce biases toward certain groups, such as public figures, reducing diversity in real-world applications and especially raising significant ethical issues. 
These problems suggest other balanced approaches to build DSD datasets that consider both diversity and scalability in the future.
\end{itemize}


Based on the above discussions and statistic information in Fig.~\ref{F3}, it can be concluded that ASVspoof 2019 (LA task)~\cite{ASV19}, ASVspoof 2021 (LA \& DF tasks)~\cite{ASV21}, ASVspoof 2024~\cite{ASV24} are among the most balanced datasets at the writing time. 
Additionally, the MLAAD~\cite{muller2024mlaad} is the largest and most suitable DSD dataset for evaluating cross-languages. The discussions on existing datasets for the DSD task underscore the importance of future efforts by the research community to release comprehensive, multilingual, and balanced datasets.
Also, Fig.~\ref{F3} emphasizes the significant costs and workload involved in creating such datasets, while ensuring compliance with essential security protocols for speaker volunteers.

\textbf{Our contribution:} Given the analysis and the discussion above, we make the following main contributions: 
\begin{itemize}
    \item We focus on the important role of public datasets proposed for the DSD task, providing a comprehensive analysis and indicating the existing issues. The analysis shows different aspects that are not mentioned in the other surveys: (1) We report the original resource of real human speech; (2) We provide an overview of deep learning-based systems used to generate fake speech; (3) The survey is not only for fake speech but also for fake video; (4) We highlight imbalances and other concerns in current public DSD datasets, along with their impact on model performance and practical applicability.
    
    \item In line with the evolution of challenge competitions, we will continue to update new DSD datasets via our GitHub repository\footnote{https://github.com/AI-ResearchGroup/A-Comprehensive-Survey-with-Critical-Analysis-for-Deepfake-Speech-Detection} in the future. This ensures the ongoing relevance of the survey and provides an up-to-date resource for DSD datasets.
\end{itemize}

\section{Overview on Proposed Systems For Deepfake Speech Detection}
\label{systems}

To conduct a comprehensive analysis of DSD systems, we first review state-of-the-art research papers addressing the DSD task. Notably, a large number of the selected papers are from high-reputation journals and conferences such as INTERSPEECH (48 papers) and ICASSP (29 papers) in recent years. Then, we categorize these DSD systems into three groups based on input type, as detailed in Tables \ref{S1}, \ref{S2}, and \ref{S3}. The first group, shown in Table \ref{S1}, consists of DSD systems that directly process audio utterances using single models. These models are based on a single machine learning algorithm or one specific network architecture. In the second group (Table \ref{S2}), audio utterances are first transformed into spectrograms, representing temporal-frequency features. After this transformation, a single model is applied to analyze the data.
The final group, shown in Table \ref{S3}, features a diverse range of ensemble models that utilize various input features and combine multiple models.
%
\begin{table*}[t]
    \caption{DSD Systems Leveraging Ensemble Techniques To Enhance The Performance} 
        	\vspace{-0.2cm}
    \centering
    \scalebox{0.75}{
    \begin{tabular}{|l|l|l|l|l|l|l|l|}        
        \hline 
        \textbf{Systems} &\textbf{Years} &\textbf{Datasets} & \textbf{Features}  &\textbf{Data Augmentation}  &\textbf{Models}   &\textbf{Loss Functions} &\textbf{Ensemble Methods}\\  
                         &               &                  &                    &(\textbf{Distoration/Compression})    &   & & \\ 

       \hline
        ~\cite{m39} &2019 &ASVspoof 2019 (LA Task),   &LFCC, CQT, FFT &n/a &LCNN  &A-Softmax &Multiple inputs\\
      \hline      
         ~\cite{m40} &2021 &ASVspoof 2019 (LA Task)  &Raw Audio &\underline{Dis.}: Mixup &ResNet  &Cross Entropy & Multiple branches\\
        \hline
         ~\cite{m51} &2021 &ASVspoof 2019 (LA Task)  &LSB, SPEC, LFCC &n/a &LCNN, LCNN-LSTM  &Cross Entropy, & Multiple inputs, models\\
         & & & & &&MSE for P2SGrad &\\
        \hline
        ~\cite{m02} &2021 &ASVspoof 2021 (LA\&DF Tasks)  &LFCC &\underline{Comp.}: MP3, ACC, landlie,  & Variants of ECAPA-TDNN  &OC-Softmax & Multiple models \\
        & & & & cellular, VoiP & & & \\
        \hline        
       ~\cite{m03}  &2021 &ASVspoof 2021 (LA\&DF Tasks)  &LFCC &\underline{Dis.}: Reverberation, add noise, &ResNet, MLP, SWA[18] &large margin cosine, & Multiple models  \\
       &&&& \underline{Comp.}: mp3, mp4 &&Cross Entropy & \\
       \hline
       ~\cite{m01} &2021 &ASVspoof 2021 (LA Task)     &LFCC, MFCC, draw &\underline{Comp.}: MP3, ACC, OGG &TDNN, RawNet2 &Focal loss &Multiple inputs, models\\
        \hline       
        ~\cite{m04} &2021 &ASVspoof 2021 (LA\&DF Tasks) &Draw, CQCC, LFCC &\underline{Comp.}: G.723, G.726,  &GMM, LCNN &Cross Entropy &Multiple inputs, models \\
        &&&& GSM, opus, speex, mp2, ogg, &&& \\
        &&&& tta, wma, acc, ra &&& \\
        \hline       
       ~\cite{m06} &2021 &ASVspoof 2021 (LA Task)  & Raw, PSCC, LFCC,  & \underline{Comp.}: TODO set 1+2  &ResNet18, GMM,  &OC-Softmax &Multiple inputs, models \\
       &&& DCT-DFT, LLFB &&TDNN, RawNet2&&\\
        \hline   
        
       ~\cite{m09} &2021 &ASVspoof 2021 (LA Task)  & MSTFT & \underline{Dis.}: Mixup, FIR filters  &Resnet18, LCNN, Sinc+CNN &Central loss &Multiple inputs, models \\
        \hline
       ~\cite{m31} &2021 &ASVspoof 2019 (LA Tasks)  &LFCC &n/a &ResNet  &OC-Softmax & Multiple branches\\
       \hline
       ~\cite{m29} &2022  &ASVspoof 2021 (LA\&DF Tasks)  &LFCC &\underline{Comp.}: G.711-alaw, G.711-$\mu$law &GMM-MobileNet  &Cross Entropy & Multiple branches\\
       \hline
       ~\cite{m44} &2022 &ASVspoof 2021 (LA Task)  & CQT, MEL & \underline{Dis.}: Mixup, Frequency Masking  &BC-ResNet, FreqCNN &n/a &Multiple inputs, models \\
       \hline
       ~\cite{m30} &2022  &ASVspoof 2019 (LA Tasks)  &LFCC &n/a &ResNet, LSTM  &OC-Softmax loss & Multiple branches\\
       \hline
       ~\cite{m47} &2022 &ASVspoof 2019, 2021 (LA Task) &Log-Mel    &\underline{Dis.}: Add music, noise, speech &ResNet  &A-Softmax & Multiple models\\
                    &      &          &                                & Reverb, pitch shift, SpecAug &&&\\

        \hline
              ~\cite{m41} &2023  &ASVspoof 2019, 2021 (LA Task) &Raw Audio    &\underline{Dis.}: Mixup, SpecAug &ResNet  &Cross Entropy & Multiple branches\\
        \hline
               ~\cite{m42} &2023 &ADD 2023  &Raw Audio, Log-Mel &\underline{Dis.}: Add noise, room inpulse, &ResNet  &Cross Entropy, & Multiple branches\\
                           & &          &                     & mixup, speed shifting,  &&KL loss &\\
                           & & & &frequency masking && & \\
        \hline
         ~\cite{m25} &2023  &ASVspoof 2019 (LA Task)  &Wav2vec, Duration,  &n/a &LCNN-LSTM-GAP &Cross Entropy &Multiple inputs\\
                    &&& Pronunciation &&&Cross Entropy& \\
        \hline
        ~\cite{m21} &2024 &ASVspoof 2019 (LA Task) &STFT phase, magnitude &\underline{Dis.}: Oversampling  &Transformer &Entropy & Multiple inputs\\
        \hline
        
         ~\cite{m28} &2024 &ASVspoof 2019 (LA Task),  &LFCC, MPE &n/a &LCNN &Cross Entropy &Multiple inputs\\
                           &&In The Wild & & & & & \\
        
         \hline
       ~\cite{m38} &2024 &ASVspoof 2019 (LA Tasks)  &Raw Audio &\underline{Dis.}: Noise, Reverb, SpecAug,  &\underline{Encoders}:     &Cross Entropy, &Multiple models \\
                    & &ASVspoof 2021 (LA Task)  &          &Drop Frequencies         &Wav2vec-XLSR-ASR,   &MSE for P2SGrad & \\
                    & &In-the-wild, MLAAD-EN                &&                &Wav2vec-XLSR-SER & &\\                
        \hline       
       ~\cite{m15} &2024 &ASVspoof 2019 (LA Task),   &Raw Audio &n/a & \underline{Encoders}: XLS-R,  &Cross Entropy &Multiple inputs, models\\
                   & &ASVspoof 2021 (LA\&DF Tasks) &&&Hubert, WavLM & & \\
                   & & &&&\underline{Decoder}: ResNet && \\       
       \hline

       

    \end{tabular}
    }
    \vspace{-0.3cm}
    \label{S3} 
\end{table*}
Given the summary of DSD systems in Table~\ref{S1}, \ref{S2}, \ref{S3}, we describe the high-level architecture of DSD systems as shown in Fig.~\ref{F4}.
From Fig~\ref{F4}, we then identify and analyze four main components that directly impact the DSD system performance: (1) Offline data augmentation, (2) Feature extraction, (3) Classification model, and (4) Loss function and Training strategy.

\subsection{Offline data augmentation}
\label{off_data_aug}
\textbf{Analysis:} Data augmentation involves generating variations of the original data to increase the size of DSD datasets, which enhances the robustness and generalization capabilities of machine learning models. Since this step is applied to original audio utterances before the training process, it can be referred to as offline data augmentation.
As shown in Tables \ref{S1}, \ref{S2}, and \ref{S3}, offline data augmentation methods can be separated into two main groups, referred to as compression and distortion.
The compression methods involve compress and decompress algorithms, mainly using audio codec techniques. 
A codec, short for `coder-decoder', is a software used to compress and decompress digital audio. Among these methods, MP3, AAC, OGG, G.7XX, and Opus formats are commonly applied. Codec data augmentation helps simulate these real-world conditions through various compression schemes (e.g., phone calls, music streaming, or online video playback on applications such as Facebook, WhatsApp, etc.). 
Since different codecs use various compression and decompression algorithms, they impact audio-related factors such as signal-to-noise ratio (SNR), high-frequency formants, energy loss, sample rate, bit depth, and bitrate in distinct ways. 
This suggests that if there are subtle differences between real and fake speech in these aspects, generating diverse audio utterances using different codecs can be an effective approach for distinguishing between them.
%
%

Codec methods can be divided into three main categories based on the quality of audio data: uncompressed format, lossless compressed format, and lossy compressed format.
Audio files with uncompressed formats such as WAV, AIFF, or PCM are large and contain all audio information recorded from an audio device.
The lossless compressed formats such as FLAC, WMA, or ALAC only reduce unnecessary features of audio data and retain the almost original audio data.
Meanwhile, lossy compressed formats such as MP3 or AAC significantly reduce audio features such as sample rate or bit depth to achieve low-volume audio files, which is suitable for streaming-based applications with real-time requirements.

The second distortion method tends to modify the raw audio by adding reverberation, background, and music in~\cite{m03, m38, m47} or using techniques of time-wise, silence streaming in~\cite{m07} without affecting the audio quality such as sample rate, bit depth, or bit rate. 
The distortion method enforces classification models to learn distinct features between fake and real speech while these features are mixed by different noise resources. 
Notably, conventional data augmentation methods, such as pitch shifting and time stretching, which are commonly applied to raw audio in tasks like Acoustic Scene Classification~\cite{hu2020device}, Speech Emotion Detection~\cite{PHAM2023120608}, and Speech Separation~\cite{ALEX2023102949}, have not been applied popularly to the DSD task~\cite{m42,m47}.
%

\begin{figure}[t]
    \centering
    \includegraphics[width =1.0\linewidth]{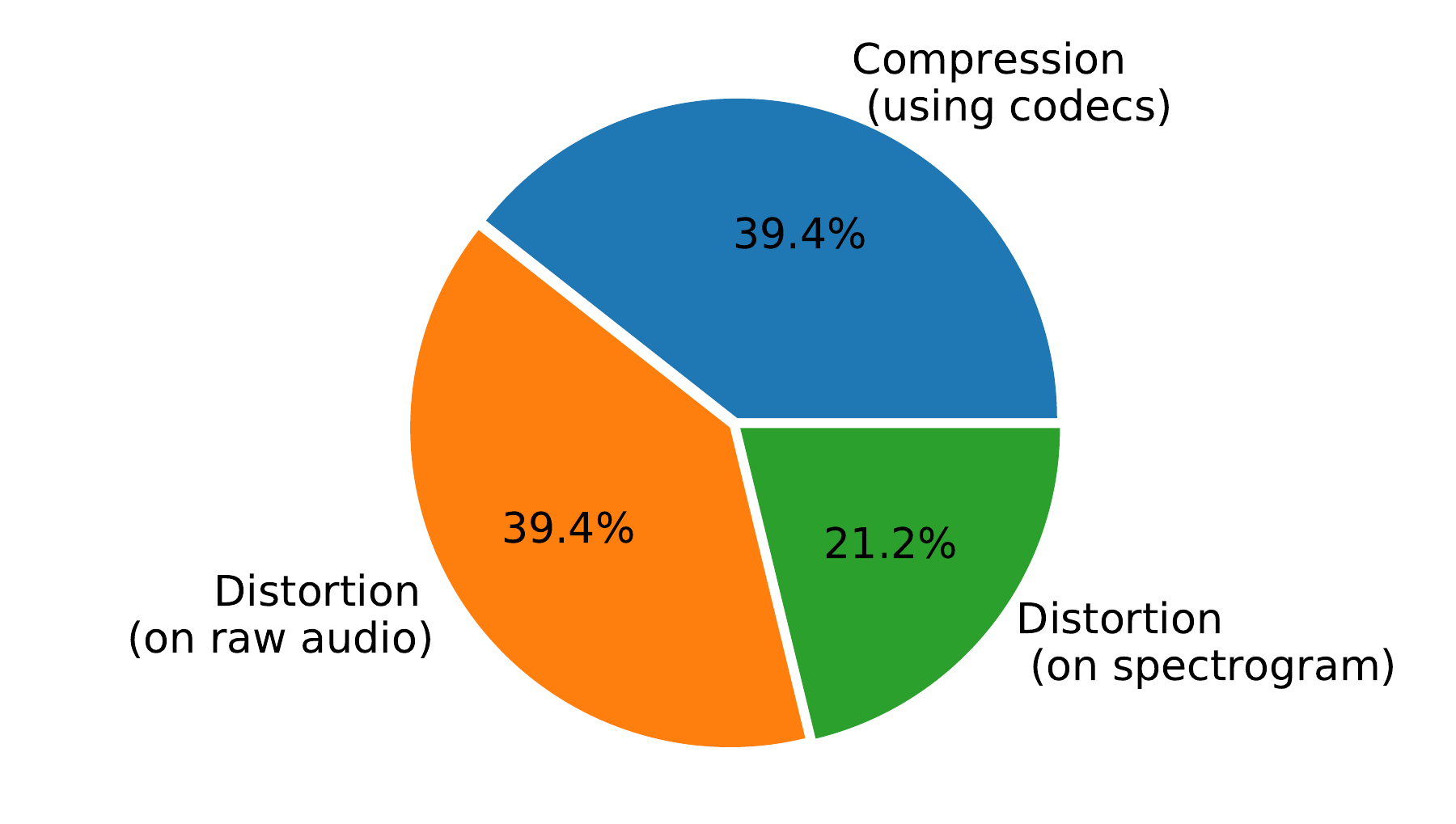}
        \vspace{-1.0cm}
	\caption{The statistics of data augmentation methods obtained from Table~\ref{S1},~\ref{S2},~\ref{S3} }
    \label{F11}
\end{figure}

\textbf{Discussion:} Although compression methods and distortion methods present different approaches to generate more audio data, none of the papers has compared, analyzed, and indicated if one of the approaches is superior in the DSD task.
Indeed, the statistical information in Fig~\ref{F11} indicates that the number of state-of-the-art DSD systems using offline distortion augmentation and offline compression augmentation are equal. 

Regarding codec-based data augmentation, little research has examined the differences among codec methods to identify which are most suitable for the DSD task in certain real-life scenarios. 
Indeed, social networks such as Facebook, Instagram, or YouTube and Internet-based communication tools such as WhatsApp, and WeChat (VoIP call) utilize specific and relevant codec methods.
For example, YouTube shares audio with MP3 formats, while VoIP calls normally use G.722 audio format as the standard.
However, many proposed DSD systems have been evaluated on current and benchmark datasets with WAV files, which do not accurately reflect the codec-specific conditions of real-life DSD applications. 

In speech-relevant tasks such as speaker recognition, speaker emotion detection, etc., some distortion data augmentations of Mixup~\cite{mixup2} or SpecAugment~\cite{spec_aug}, which are inspired from the computer vision domain, are widely used. 
These data augmentation methods focus on synthesizing new spectrograms in various manners (e.g., merging, masking), which might not accurately reflect artifacts of the audio signal. 
Additionally, these data augmentation methods are applied to batches of spectrograms, referred to as online data augmentation.
As shown in Fig.~\ref{F11}, Mixup~\cite{mixup2} or SpecAugment~\cite{spec_aug} are also used in a wide range of DSD systems.
However, none of the papers has analyzed or compared the efficiency between offline data augmentation and online data augmentation.

\textbf{Our contribution:} Given the analysis and the existing concerns above, we make the following main contributions: 
\begin{itemize}
    \item To evaluate the role and the effect of the online and offline data augmentation methods, we conducted extensive experiments in this paper. Based on our findings, we identify data augmentation techniques that are compatible with DSD systems. In particular, we compare the performance of codec-based methods with the Mixup~\cite{mixup2} and SpecAugment~\cite{spec_aug}.
    \item On our GitHub repository, we regularly update codec-based methods and other data augmentation techniques featured in the latest research.    
    We also released code for implementing codec-based methods and other data augmentation methods in this GitHub repository, which are used to conduct our experiments in this paper.      
\end{itemize}

\subsection{Feature extraction}
\label{fea_ext}
\textbf{Analysis:} As shown in Fig.~\ref{F4}, feature extraction methods can be categorized into two main groups: non-parameter and trainable-parameter methods. 

In \textbf{non-parameter feature extraction}, a raw audio utterance (e.g., a 1-D tensor) is first transformed into a time-frequency spectral features (e.g., a 2-D tensor) using various transformation
ranging from spectral coefficients (e.g., MFCC~\cite{ASV15_result_sum, m01}, LFCC~\cite{m06,m30,m31}, CQCC~\cite{m04}, etc) to spectrogram-based representations such as STFT-spectrogram~\cite{m21,m09}, CQT-spectrogram~\cite{m04,m07}, etc. 
Once the time-frequency spectrograms are generated, some DSD systems directly use them for training with classification models~\cite{m07}, while other systems use several approaches to enhance feature quality before applying a classification model.
%
The first approach involves applying auditory filter banks such as Mel~\cite{m02, m15}, Linear Filter~\cite{m01, m11, m12} (LF), etc,  to capture the relationships between frequency bands.
Then a Discrete Fourier Transform (DFT) is applied to analyze the relationship across temporal dimension before the features are fed into a model for the training process~\cite{m01, m02, m11, m12}. Notably, the output of Mel, LF, or DFT operations remains a 2-D tensor (similar to a spectrogram), representing both temporal and spectral features.
In the second approach, spectrograms are fed into pre-trained models, such as XLS-R~\cite{xls-r}, Hubert~\cite{Hubert}, WavLM~\cite{Wavlm}, or Whisper~\cite{whisper}, to extract embeddings. These embeddings are the output feature maps from a specific layer of the pre-trained model~\cite{m15}. Typically, the embeddings form a 1-D tensor, similar to a vector, where each dimension of the vector is treated as an independent value.
In this approach, the choice of spectrogram depends on the one used to train the pre-trained models. Typically, the Mel-spectrogram is preferred, as most pre-trained models use it as input for training upstream tasks~\cite{Wavlm, Hubert, whisper}. In general, non-parameter feature extraction leverages various spectrogram transformations, auditory filters, auditory statistics, and pre-trained models to generate distinct features (e.g., 1-D audio embeddings, 2-D spectrograms) of audio input.

%
%

%
\textbf{Trainable-parameter feature extraction} involves extracting audio features by applying trainable network layers.
In particular, systems proposed in~\cite{m15, m05, m08} applied SincNet layers~\cite{sincnet}, LEAF layers~\cite{leaf}, FBanks~\cite{m16} to learn and extract features from raw audio. These techniques construct learnable filterbanks
or approximate the standard filtering process. For example, SincNet and LEAF layers keep the role of adaptive and bandpass filters to capture frequency features between two pre-defined cut-off frequencies. The outputs of these trainable layers are the feature maps that are then fed into the next parts of detection systems.
In other words, trainable feature extraction includes trainable network layers as a part of entire network architectures that directly train and learn features from raw audio without the spectrogram transformation steps. 


\textbf{Discussion:} 
By allowing learnable temporal-spatial features during the training process, trainable-parameter feature extraction is compatible with end-to-end systems and shows effectiveness in distinguishing artifacts in fake speech. However, as most proposed systems using trainable features were evaluated on single datasets rather than cross-dataset settings, this possibly leads to challenges in generalization since learned feature sets perform well under specific conditions but fail in unseen fake speech in real-world environments.
\begin{figure}[t]
    \centering
    \includegraphics[width =1.0\linewidth]{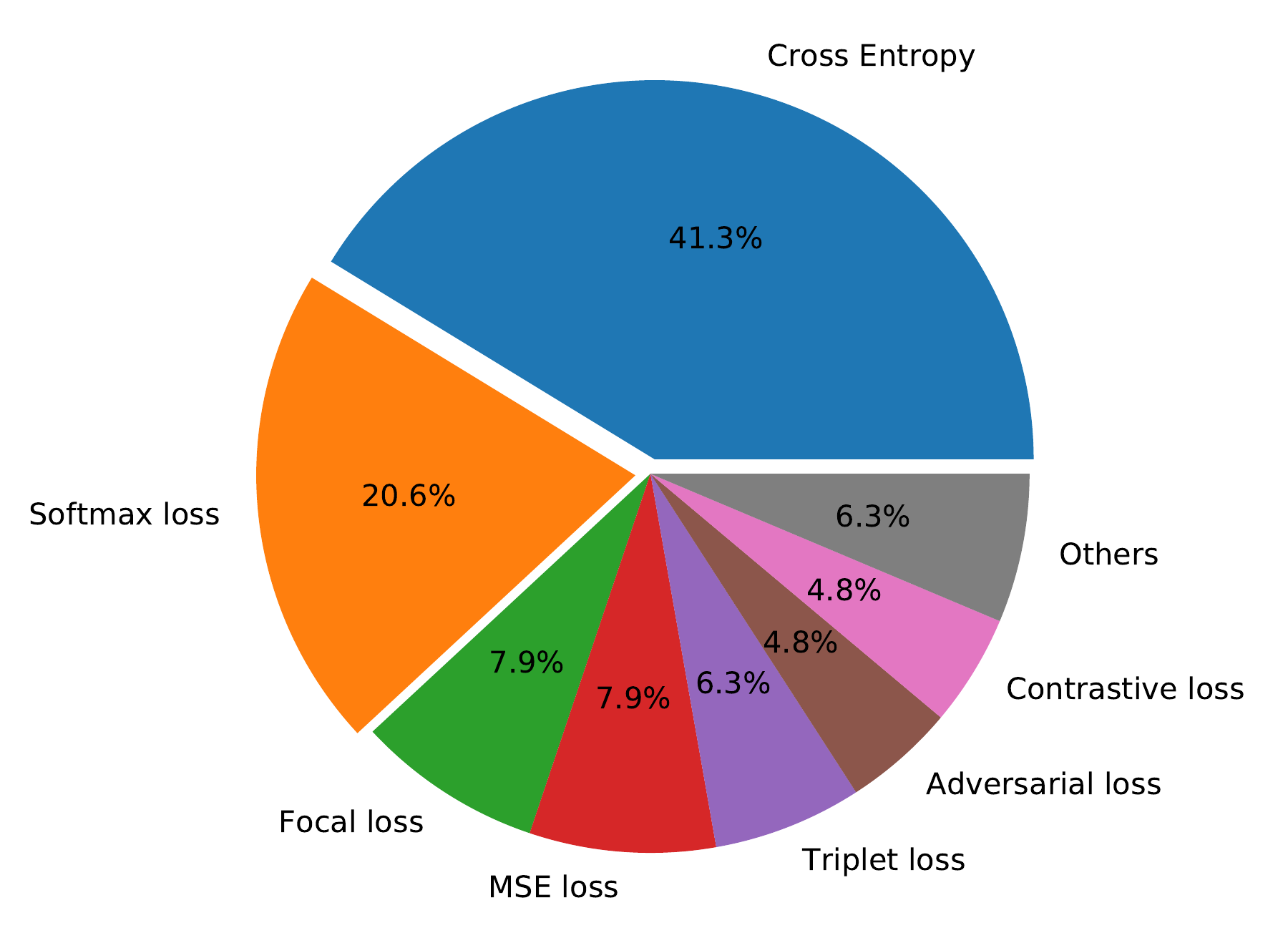}
     	\vspace{-0.5cm}
	\caption{The statistics of loss functions obtained from Table~\ref{S1},~\ref{S2},~\ref{S3}}
    \label{F10}
\end{figure}
Regarding feature extraction using audio embeddings from pre-trained models, although these pre-trained models are effective for many audio tasks, using them for deepfake detection presents several challenges. Firstly, as pre-trained models are initially trained for upstream tasks such as speech-to-text, speaker identification, emotion detection, etc, that focus on different aspects (i.e.,  speech-to-text or emotion detection), the audio melody and harmony (i.e., emotion detection), or distinct frequencies (i.e., speaker identification), embeddings can fail to capture subtle artifacts specific to synthesized speech. Secondly, audio deepfakes are generated to closely mimic real speech, they often  have the same formants, pitch, and rhythm as real audio, especially when generated by advanced deep-learning-based speech generation systems. 
Additionally, the use of pre-trained models can add complexity due to their large network architectures.
%

For systems using spectrograms such as CQT, MEL, GAM, etc., each spectrogram is designed to capture specific frequency ranges. These spectrograms focus on different central frequencies, which allows them to highlight distinct features of an audio signal. However, human speech contains a wide range of formants - characteristics of sound determined by factors such as language, accent, vocal tract shape, and vocal fold behavior. Therefore, relying on only one type of spectrogram may miss important features, leading to incomplete or insufficient representations of the speech signal that are useful for deepfake detection.
To address this, DSD systems have begun to use ensembles of multiple spectrogram inputs~\cite{m01, m02, m04, m06, m09}. 
By leveraging the unique strengths of each spectrogram type, this approach aims to enhance detection accuracy and has shown significant improvements in model performance. Many top-performing systems in recent competitions have demonstrated the effectiveness of using ensembles to boost overall system robustness. However, ensemble models present several limitations, including reduced interpretability, increased system complexity, and higher training costs. 

\begin{figure}[t]
    \centering
    \includegraphics[width =1.0\linewidth]{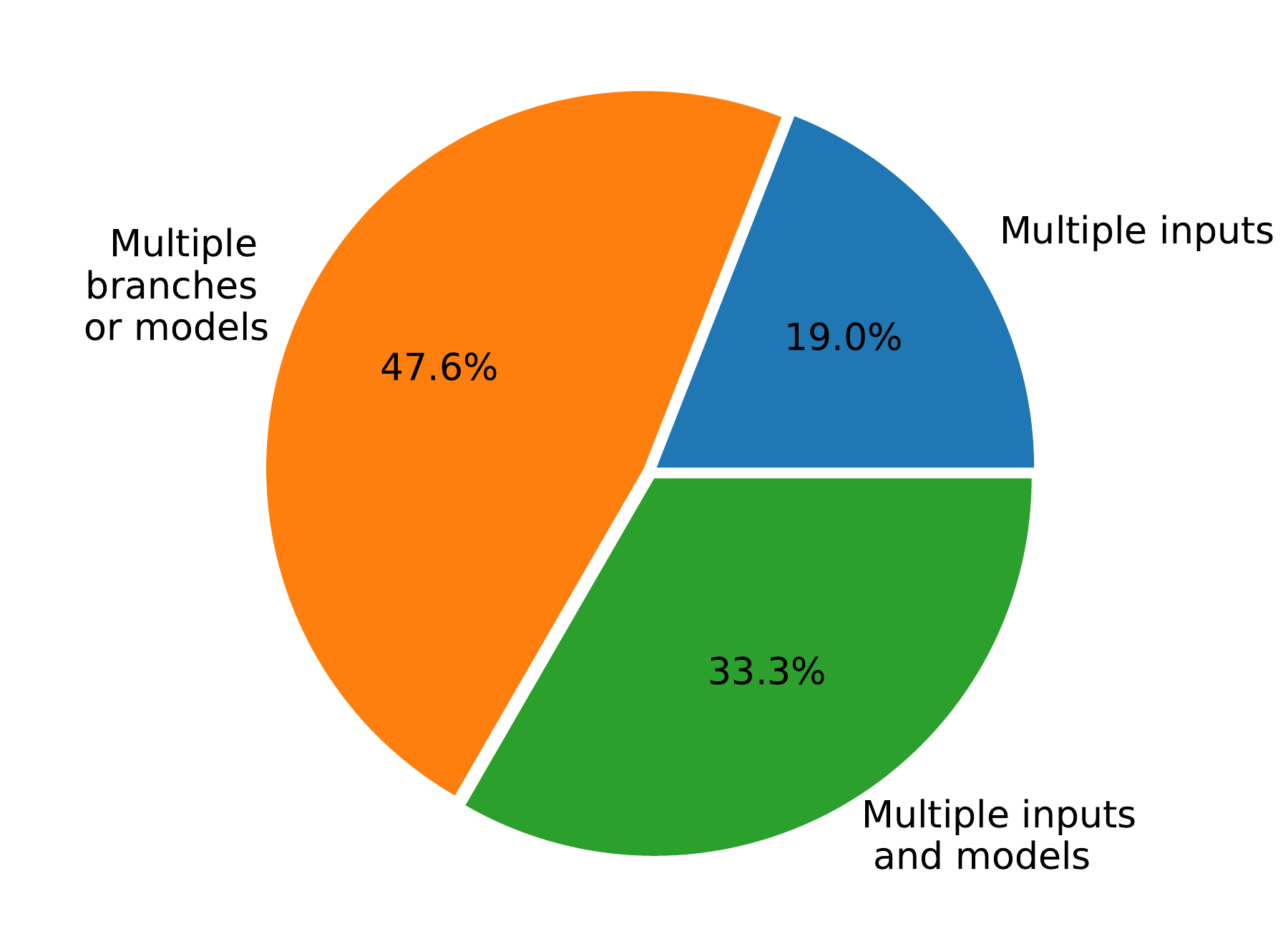}
         	\vspace{-0.5cm}
	\caption{The statistics of ensemble methods obtained from Table~\ref{S1},~\ref{S2},~\ref{S3}}
    \label{F12}
\end{figure}

\textbf{Our contribution:} Given the analysis and the discussion above, we make the following main contributions: 
\begin{itemize}
    \item We presented the commonly used feature extraction methods in DSD systems, highlighting their characteristics and potential challenges associated with each approach.
    \item In the next section, we conduct extensive experiments of various feature extraction methods to evaluate the most effective approach for the DSD task. Additionally, we explore different feature ensembles to determine the optimal combinations for enhancing performance. 
    \item In our GitHub, we release code for different spectrogram transformations using Librosa toolbox~\cite{librosa_tool}.      
\end{itemize}
\subsection{Classification models}
\label{model}
\textbf{Analysis:} Early models proposed for DSD task approached conventional machine learning algorithms.
For example, 9 over 16 submitted systems in ASVspoofing 2015 challenge~\cite{ASV15_result_sum} extract MFCC feature (i.e. Systems A, B, E, G, H, I, N, O, and P in~\cite{ASV15_result_sum}).
Then, various machine learning-based models such as Mahalanobis distance measurement, Gaussian-based model (GMM), Support vector machine-based models (SVM, SVM-RBF), or fusion models (GMM and SVM) are used to explore MFCC features. 
%
%
However, recent DSD systems as shown in Table~\ref{S1},~\ref{S2},~\ref{S3} present a wide range of neural network architectures due to the powerful deep learning techniques.
Recently proposed deep neural networks for the DSD task can be separated into four main approaches.
The first approach, which focuses on exploring spatial features, leverages convolutional-based network architectures (CNN).
Among the CNN-based networks, Resnet, LCNN, and RawNet architectures are widely used.
ResNet and LCNN are used to explore spectrogram-based features such as LFCC~\cite{m04}, CQT~\cite{m09}, and MEL~\cite{m15}.
Meanwhile, RawNet architectures are normally combined with SincNet layer~\cite{sincnet} to learn raw audio~\cite{m01, m04, m06, m07, m12, m15}.
The second approach, which focuses on exploring the temporal features, presents recurrent neural network (RNN) based architectures.
For example, LSTM-based networks, TDNN, or ECAPA-TDNN are proposed in~\cite{m07, m01},~\cite{m02, m06} and~\cite{m16}, respectively.
As shown in Table~\ref{S1},~\ref{S2},~\ref{S3}, RNN-based networks have not been popularly applied for the DSD task compared to the CNN-based architectures. The third approach involves combining both convolutional layers and recurrent layers to explore both temporal and spatial features, referred to as hybrid network architectures. In particular, recurrent network-based layers such as LSTM, GRU are combined with CNN-based layers to perform convolutional-recurrent neural network (CRNN) architectures~\cite{m01, m16, m22}.
%

Recently, encoder-decoder based network architectures have been popularly used for the DSD task. 
Indeed, along with conventional encoder and decoder in transformer-based architectures~\cite{m21, m26}, various encoder architectures such as XLSR-53~\cite{m23}, WavLM~\cite{m20}, CNN or ResNet~\cite{m19} are explored.
Decoder architectures also show diverse using GAN-based architecture~\cite{m19}, Multi-feature attention~\cite{m20}, Graph Attention Network~\cite{m08,m34,m35}, etc.

To further enhance the DSD performance, the DSD research community leverages a wide range of ensemble models.
These ensemble models can be separated into three main approaches which are marked in the final column in Table~\ref{S3}.
In the first approach (Multiple inputs), multiple input features are explored~\cite{m39, m25, m21}.
This approach is inspired by the idea that multiple features contain different and distinct features between fake and real utterances.
Given different features, each feature is trained by the same classification model (i.e., the individual model shares the same network architecture but presents different training parameters after the training process).
For example, while~\cite{m21} explores the magnitude and phase features of STFT spectrogram, different features of Wav2Vec embeddings, duration, and pronunciation are explored in~\cite{m25}.
Similarly, multiple spectrograms such as LFCC, CQT, and STFT are trained by one classification model of CNN~\cite{lampham_01}.
Finally, the scores obtained from individual models are fused to achieve the final and best result.
The second approach (Multiple branches or models) leverages different network architectures that explore one type of input feature~\cite{m02, m03, m29, m31, m40, m41, m42, m47, m38}.
This approach is inspired by the idea that different network architectures are likely to capture distinct properties from the input feature.
For example,~\cite{m03} proposed multiple branches of GMM-DNN and ResNet to explore the LFCC spectrogram.
Similarly,~\cite{m02} explores the raw audio by different variants of ECAPA-TDNN.
The final approach (Multiple inputs, models) leverages both multiple input features and different network architectures.
For example,~\cite{m01} explore raw audio by RawNet2.
Meanwhile, TDNN and LFCC spectrogram are explored by LCNN.
Then, the authors fused three results obtained from three individual models.
Similarly, multiple input features of raw audio, CQCC, and LFCC are explored by different models of LCNN, GMM, and RawNet2 in~\cite{m04}. Ensemble methods are widely adopted in many top-performing systems in DSD challenge competitions.

\textbf{Discussion:} Although many deep neural network architectures have been proposed for the DSD task and evaluated on various benchmark datasets, the best results have been obtained from ensemble methods with multiple inputs or/and different network architectures.
The statistics of ensemble models, as shown in Fig~\ref{F12}, indicate that multiple branches or models are the majority. 
However, ensemble models present the concern of large trainable parameters.
Moreover, none of the research has been analyzed to indicate the individual roles of input features or types of network architectures used in ensemble methods. 
To demonstrate a robust and general DSD model, the proposed model needs to be evaluated with multiple datasets, cross-datasets, or cross-languages.
However, only some recent research~\cite{m23, intwi_ds, m32, m37} evaluated the proposed models with multiple datasets such as ASVspoof 2019 (LA Task), ASVspoof 2021 (LA\&DF Task), In The Wild, etc. To the best of our knowledge, none of the research has proposed the evaluation on cross-languages. 

\textbf{Our contribution:} Given the analysis and the discussion above, we make the following main contributions:  
\begin{itemize}
    \item We evaluate various input features, indicating the effective input feature for DSD system performance.
    \item We also evaluate a wide range of network architectures leveraging the transfer learning technique, end-to-end training approach, and audio embeddings extracted from state-of-the-art pre-trained models.
    \item Given extensive experiments on different input features and various network architectures, we propose an ensemble model that is competitive to the state-of-the-art DSD systems. 
\end{itemize}
\subsection{Loss function and training strategy}
\label{loss}

From Table~\ref{S1},~\ref{S2},~\ref{S3}, it can be seen that most proposed models use a single loss function.
Statistics of the individual loss functions are also presented in Fig.~\ref{F10}.
As shown in Fig.~\ref{F10}, the cross entropy (CE) based losses (e.g., Binary Cross Entropy (BCE), Weight Cross Entropy (WCE), etc.) and Softmax-based losses (e.g., Additive-Margin-Based Softmax (AM-Softmax), Angular-Margin-Based Softmax (A-Softmax), etc.) present the most popular loss functions.
Some models combine different loss functions.
For example, CE and Contrastive loss were used in~\cite{m19}.
Similarly, authors in~\cite{m22} combined three loss functions of Cross Entropy, Triplet loss, and Adversarial loss.
Some papers such as~\cite{m10} and~\cite{m11} compared the DSD performance between large margin cosine loss (LMC loss), and A-Softmax loss functions or between OC-Softmax, MSE for P2SGrad loss functions, respectively. 

Generally, a single loss function is used in end-to-end based systems.
Meanwhile, the combination of multiple loss functions is related to different training strategies.
For example,~\cite{m23} proposed a teacher-student scheme in which the teacher was trained with contrastive loss and the student was trained by a combination of contrastive loss, Cross Entropy, and MSE loss.
Similarly, the student network in~\cite{m36, m52} was trained by a combination of Cosine Similarity/OC-Softmax and MSE loss functions.
It can be seen that muliple-loss functions used for teacher-student schemes help achieve a low-complexity model for the DSD task~\cite{m23, m36, m52}.
Additionally, using multiple-loss function in~\cite{m53} aims for multiple-task learning strategy.
Rather than focusing on loss functions, some researchers improve the DSD system by exploring the training strategy.
For example, authors in~\cite{x01} suggested to mix three datasets for the training process.
This enhances the generalization and stabilization of the authors' proposed DSD system.
Meanwhile, authors in~\cite{x02} generated more fake utterances by leveraging four types of Vocoders: HiFiGAN, MB-MelGAN, PWG, and WaveGlow, which helps to improve their DSD system performance.   

\section{Our Proposed Deepfake Speech Detection System and Extensive Evaluation}
\label{evidence}

\subsection{Our motivation}
Given the comprehensive analysis of the DSD systems in Section~\ref{systems}, we are motivated to conduct extensive experiments that address and evaluate the main concerns below.
\begin{itemize}
    \item We evaluate the role of offline data augmentation (codec) and compare this method with the conventional online data augmentation methods of Mixup~\cite{mixup2} and SpecAugment~\cite{spec_aug}. We also indicate if a combination of offline and online data augmentation methods is effective in enhancing the DSD system performance.  
    \item We conduct extensive experiments to evaluate different inputs and network architectures. Given the comparison, we indicate which input features, network architectures, combination of input features, and network architectures have the potential to be further explored. We then propose the best DSD ensemble system that is competitive to the state-of-the-art systems. 
    \item To deeply analyze the role of data augmentation methods, input features, and network architectures, we evaluated proposed DSD systems within cross-dataset and cross-language settings.
    \item To address the real-time ability, our proposed models are evaluated on two-second utterances and present low-complexity architectures.
\end{itemize}

\subsection{Selected datasets and evaluating metrics}
As the trade-off among the number of utterances, the deep-learning-based fake speech generation systems, the original/real human speech resource as shown in Fig.~\ref{F1} and the comprehensive analysis in Section~\ref{dataset}, we decide to use ASVspoof 2019 (LA Task) to evaluate the effect of data augmentations, different types of input features, and various network architectures. 
Given the results on ASVspoof 2019 (LA Task), we obtain the best DSD systems which are then evaluated with ASVspoof 2021 (LA \& DF Tasks) datasets for cross-dataset evaluation and with MLAAD dataset for cross-language evaluation.

%
We obey the ASVspoof 2019 (LA Task) and ASVspoof 2021 (LA \& DF Tasks) challenges, then use the Equal Error Rate (ERR) as the main metric for evaluating proposed models. 
We also report the Accuracy, F1 score, and AUC score to compare the performance among evaluating models.
 
\subsection{Proposed systems and experimental settings}
\begin{figure}[t]
    \centering
    \includegraphics[width =1.0\linewidth]{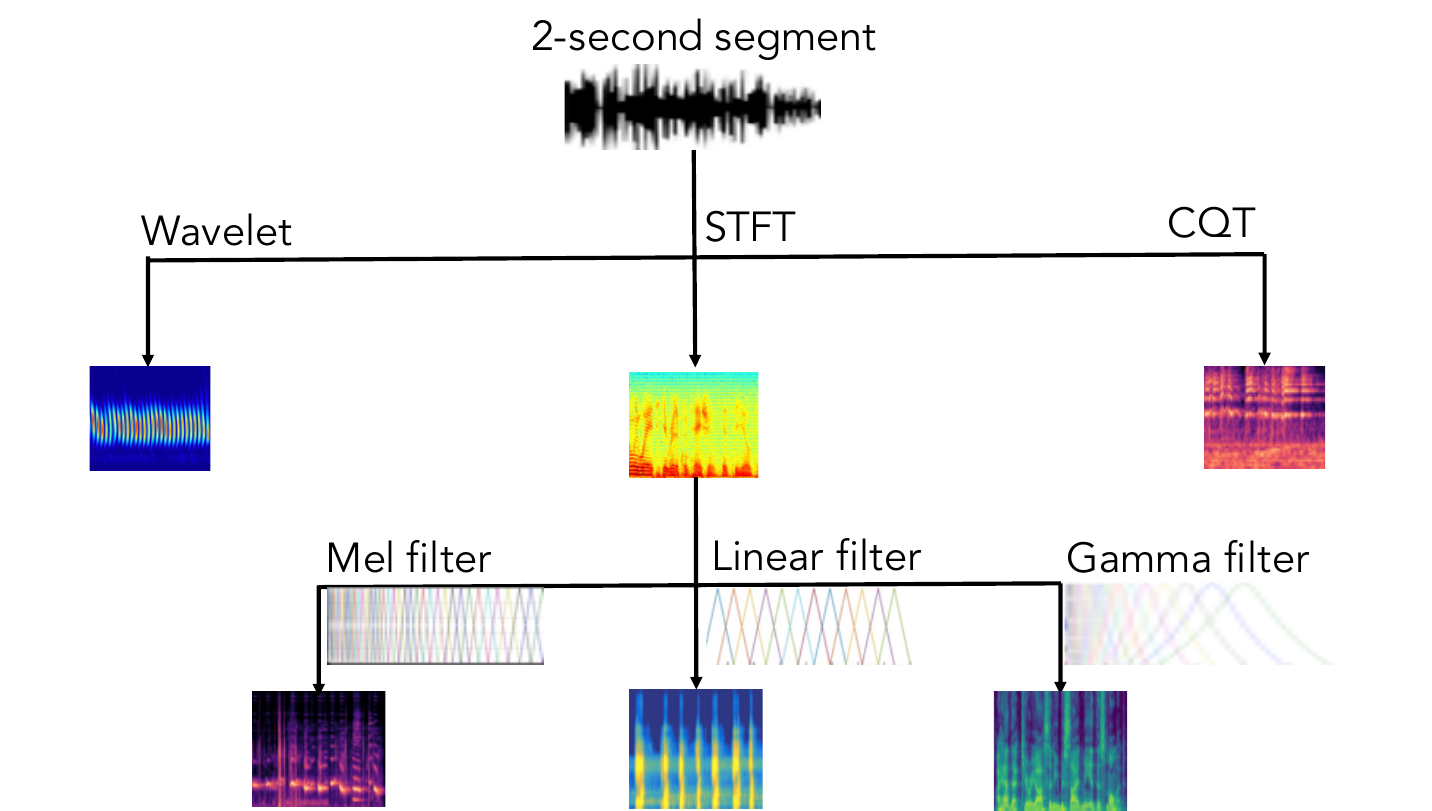}
           	\vspace{-0.1cm}
	\caption{Generate spectrograms using different spectrogram transformation methods and auditory filter models}
       	\vspace{-0.2cm}
    \label{F5}
\end{figure} 

\textbf{Data augmentations:} 
We evaluate the role of two data augmentation methods: offline data augmentation (codecs) and online data augmentation (Mixup and SpecAugment). Regarding offline data augmentation using codec-based methods, we use six popular codec formats MP3, OPUS, OGG, GSM, G722, and M4A.
While the codec-based methods compress and decompress raw audio before the training process, the online data augmentation methods of Mixup and SpecAugment work on batches of spectrograms during the training process.
By evaluating these two groups of data augmentation individually, we indicate if each of them presents a significant contribution and a combination of two data augmentation methods can help to enhance DSD task performance. 

\textbf{Multiple input features:} Fig.~\ref{F5} presents seven types of input features: raw audio and six different spectrograms, which are evaluated in this paper. 
In particular, we use three transformation methods of Short-time Fourier Transform (STFT), Constant-Q Transform (CQT), and Wavelet Transform. 
Presumably, each type of spectrogram focuses on different perspectives on frequency content and might catch different inconsistencies in the
audio signal. 
We then leverage different auditory-based filters: Mel and Gammatone filters focus on subtle variations relevant to human auditory perception and the linear filter (LF) isolates specific frequency bands. 
%

As we set the window length, the hop length, and the filter number with 1024, 512, and 64, we achieve the same spectrogram shape of 64×64. 
Then, we apply Discrete Cosine Transform (DCT) to spectrograms across the temporal dimension. 
Finally, the first and the second-order derivatives are applied to these spectrograms, generating a three-dimensional tensor of 64×64×3 (i.e., the original spectrogram, the first-order derivative, and the second-order derivative are concatenated across the third dimension).
\begin{table}[t]
\caption{The CNN, RNN, and C-RNN network architectures} 
       	\vspace{-0.2cm}
    \centering
    {
\scalebox{.95}{

\begin{tabular}{|c|c|}
\hline
\textbf{Models} & \textbf{Configuration} \\  
\hline
CNN-based model                  & \textbf{3} $\times$ \{Conv(32/64/128)-ReLU-AP-Dropout(0.2)\}   \\ 
                              & \textbf{1} $\times$ \{Dense(256)-ReLU-Dropout(0.2)\} \\
                              & \textbf{1} $\times$ \{Dense(2)-Softmax\}  \\
                              \hline
RNN-based model                  &\textbf{2} $\times$ \{BiLSTM(128/64)-ReLU-Dropout(0.2)\}   \\ 
                              & \textbf{1} $\times$ \{Dense(256)-ReLU-Dropout(0.2)\} \\
                              & \textbf{1} $\times$ \{Dense(2)-Softmax\}   \\
                              \hline
C-RNN-based model                &\textbf{3} $\times$ \{Conv(32/64/128)-ReLU-AP-Dropout(0.2)\}   \\ 
                              &\textbf{2} $\times$ \{BiLSTM(128/64)-ReLU-Dropout(0.2)\}   \\
                              &\textbf{1} $\times$ \{Dense(256)-ReLU-Dropout(0.2)\}\\
                              &\textbf{1} $\times$ \{Dense(2)-Softmax\} \\
                              \hline
\end{tabular}
}
    }
\label{ete_models}
\end{table}

\textbf{Back-end classification models:} 
This paper proposes three main approaches for back-end classification models: the end-to-end deep learning approach, the transfer learning approach, and the audio-embedding deep learning approach.
Regarding the end-to-end deep learning approach, four models of CNN-based model, SinC-CNN model (e.g., SinC-CNN architecture is a combination of SinC layer and CNN architecture. The CNN architecture component is reused from CNN-based model), CNN-based model, RNN-based model, and C-RNN-based model are evaluated with the detailed configuration in Table~\ref{ete_models}. 
The Sinc-CNN model proves powerful for raw audio input and has been widely used as the survey in Section~\ref{systems}
Meanwhile, CNN-based models are commonly used and effectively capture and learn spectral features. 
We also use RNNs to focus on detecting natural sequential patterns that can be disrupted in synthetic audio~\cite{rnn_deepfake} (e.g., temporal coherence, prosodic features such as rhythm, stress, and intonation).
Consequently, based on the idea of combining both spectral features and temporal features,  we use C-RNN-based model to distinguish characteristics of real and fake audio utterances.

With the transfer learning approach, a wide range of benchmark network architectures in the computer vision domain are evaluated. These networks are ResNet-18, MobileNet-V3, EfficientNet-B0, DenseNet-121, SuffleNet-V2, Swint, Convnext-Tiny, GoogLeNet, MNASnet, RegNet, which were trained on the ImageNet1K dataset~\cite{imagenet_ds} in advance.
Given the pre-trained networks,  trainable weights, which capture rich and generalized features of pattern recognition in images,have the potential to adapt patterns in spectrograms by the fine-tuning process. 
To adapt the DSD task, we modify the final dense layer of these mentioned networks to be compatible with the binary classification task.

For the audio-embedding deep learning approach, the state-of-the-art audio pre-trained models of Whisper~\cite{whisper}, Seamless~\cite{seamless}, Speechbrain~\cite{speechbrain}, and Pyannote~\cite{pyanote1, pyanote2} are leveraged. 

In particular, we feed the spectrogram inputs into these pre-trained models to obtain audio embeddings.
Given the audio embeddings, 
We then propose a Multilayer Perceptron (MLP) to classify the audio embeddings into fake or real classes.
The proposed MLP is shown in Table~\ref{mlp_be}, to detect real or fake audio.

\textbf{Ensemble method:}
As we train individual model works with two-second audio segment, the result on an entire audio recording is computed by averaging of results over all two-second segments.
Let consider $\boldsymbol{p}^{(n)} = [p_{1}^{(n)}, p_{2}^{(n)},..., p_{C}^{(n)}]$,  where $C$ is the category number of the n-th out of \(N\) two-second segments, as the predicted probability of one two-second segment. 
The predicted probability of an entire audio recording, as described by $\bar{\boldsymbol{p}} = [\bar{p}_{1},\bar{p}_{2},..., \bar{p}_{C}]$, is computed by:
\begin{equation}
    \label{eq:mean_stratergy_patch}
    \bar{p}_{c} = \frac{1}{N}\sum_{n=1}^{N}p_{c}^{(n)}  ~~~  for  ~~ 1 \leq c \leq C 
\end{equation}

Given the predicted probabilities from individual models, we propose a MEAN fusion for an ensemble of multiple models.
Let consider the predicted probability of one model as  \(\mathbf{\hat{p}}_{s}= (\bar{p}_{s_1}, \bar{p}_{s_2}, ..., \bar{p}_{s_C})\), where $C$ is the category number and the $s$-th out of \(S\) individual models. 
Next, the predicted probability after MEAN fusion \((\hat{p}_{1}, \hat{p}_{2}, ..., \hat{p}_{C}) \) is obtained by:
\begin{equation}
    \label{eq:mix_up_x1}
     \hat{p_{c}} = \frac{1}{S} \sum_{s=1}^{S} \hat{p}_{s_c} ~~~  for  ~~ 1 \leq c \leq C 
\end{equation}
Finally, the predicted label \(\hat{y}\) for an entire audio sample is computed by:

\begin{equation}
    \label{eq:label_determine}
    \hat{y} = \argmax (\hat{p}_{1}, \hat{p}_{2}, ...,\hat{p}_{C} )
\end{equation}

\subsection{Experimental results and discussion}
We first use ASVspoof 2019 (LA Task) to evaluate and indicate the best DSD systems.
The comprehensive result comparison is described in Table~\ref{table:R1}.

\textbf{Evaluation of data augmentation methods on ASVspoof 2019 (LA Task)}: Considering the performance of online and offline data augmentation methods as shown in systems A1 (no data augmentation), A2 (online data augmentation with codec), A3 (offline data augmentation with Mixup and SpecAugment), and A4 (both online and offline data augmentation), it can be seen that the offline data augmentations of Mixup and SpecAugment are appropriate for DSD task on ASVspoof 2019 (LA Task) dataset. 
Notably, the combination of online and offline data augmentations does not help to enhance the DSD task performance compared with only using offline data augmentation. 
\begin{table}[t]
\caption{The audio pre-trained models and the Multilayer Perceptron} 
       	\vspace{-0.2cm}
    \centering
    {
\scalebox{.98}{
\begin{tabular}{|l|c|c|}
\hline
\textbf{Models} & \textbf{Using License} & \textbf{Embedding size/} \\  
&& \textbf{Configuration}\\
\hline
Whisper~\cite{whisper} & MIT &512 \\
SpeechBrain~\cite{speechbrain} & Apache2-0 &192 \\
SeamLess~\cite{seamless} & MIT & 1024\\
Pyannote~\cite{pyanote1, pyanote2} & MIT & 512\\
\hline
MLP      & Our proposal    &\textbf{1} $\times$ \{Dense(128)-ReLU \} \\
         &                     &\textbf{1} $\times$ \{Dense(2)-Softmax \} \\
\hline
\end{tabular}
}
    }
\label{mlp_be}
\vspace{-0.3cm}
\end{table}

\textbf{Evaluation of input features on ASVspoof 2019 (LA Task)}: Considering the efficacy of raw audio and six types of spectrograms in systems from B1 to B7, STFT outperforms the raw audio and other spectrograms.
Models B2, B5, and B7 achieve the best ERR score of 0.08 while the combination of STFT \& LF obtains slightly better accuracy and F1 scores of 0.88 and 0.9, respectively. 
This indicates that STFT and applying filters such as Linear Filter or Gammatone filter are suitable for isolating specific frequency bands in classification algorithms.
\begin{table*}[t]
    \caption{Performance comparison among deep learning models and ensemble of high-performance models \\ on Logic Access evaluation subset in ASVspoofing 2019} 
        	\vspace{-0.2cm}
    \centering
    \scalebox{0.95}{
    \begin{tabular}{|l  |c|c|c|c|c|c|c|} 
        \hline 
        \textbf{Systems} &\textbf{Inputs}   &\textbf{Augmentations} &\textbf{Models}  &  \textbf{Acc$\uparrow$} &  \textbf{F1$\uparrow$} &\textbf{AUC$\uparrow$}  &\textbf{ERR $\downarrow$}\\  
        \hline  
        A1 & STFT \& LF  &None  &CNN   &0.82 &0.84   &0.91 &0.15\\
        A2 & STFT \& LF  &Codec  &CNN   &0.81 &0.84  &0.93 &0.13\\
        A3 & STFT \& LF  &Mixup, Spec.  &CNN   &\textbf{0.88} &\textbf{0.90} &\textbf{0.96} &\textbf{0.08}\\
        A4 & STFT \& LF  &Codec, Mixup, Spec. &CNN   &0.81 &0.84 &0.93 &0.13\\
        \hline
         B1 & Raw Audio  &None     &SinC-CNN   &0.84 &0.87 &0.96 &0.10\\
         B2 & STFT  &Mixup, Spec.     &CNN   &\textbf{0.87} &\textbf{0.89} &\textbf{0.96} &\textbf{0.08}\\
         B3 & CQT   &Mixup, Spec.    &CNN   &0.89 &0.90 &0.92 &0.14\\
         B4 & WT    &Mixup, Spec.    &CNN   &0.84 &0.86 &0.89 &0.17\\
         B5 & STFT \& LF  &Mixup, Spec.  &CNN   &\textbf{0.88} &\textbf{0.90} &\textbf{0.96} &\textbf{0.08}\\
         B6 & STFT \& MEL &Mixup, Spec.  &CNN   &0.86 &0.88 &0.95 &0.11\\
         B7 & STFT \& GAM &Mixup, Spec.  &CNN   &\textbf{0.85} &\textbf{0.87} &\textbf{0.96} &\textbf{0.08}\\         
         \hline
         \hline                 
         C1 & STFT \& LF &Mixup, Spec.  &RNN   &0.92 &0.91 &0.88 &0.17\\
         C2 & STFT \& LF &Mixup, Spec.  &CRNN  &0.88 &0.90 &0.96 &0.14\\
         \hline
         \hline
         D1 & STFT \& LF &Mixup, Spec.  &ResNet-18   &0.49 &0.58 &0.51 &0.47\\
         D2 & STFT \& LF &Mixup, Spec.  &MobileNet-V3    &0.59 &0.67 &0.52 &0.48\\
         D3 & STFT \& LF &Mixup, Spec.  &EfficientNet-B0     &0.52 &0.61 &0.51 &0.48\\
         D4 & STFT \& LF &Mixup, Spec.  &DenseNet-121     &0.58 &0.66 &0.51 &0.48\\
         D5 & STFT \& LF &Mixup, Spec.  &ShuffleNet-V2  &0.64 &0.71 &0.53 &0.48\\       
         D6 & STFT \& LF &Mixup, Spec.  &Swin\_T     &\textbf{0.84} &\textbf{0.87} &\textbf{0.94} &\textbf{0.09}\\         
         D7 & STFT \& LF &Mixup, Spec.  &ConvNeXt-Tiny     &\textbf{0.88} &\textbf{0.90} &\textbf{0.96} &\textbf{0.075}\\         
         D8 & STFT \& LF &Mixup, Spec.  &GoogLeNet     &0.53 &0.62 &0.51 &0.47\\                                    
         D9 & STFT \& LF &Mixup, Spec.  &MNASNet    &0.62 &0.70 &0.54 &0.47\\   
         D10 & STFT \& LF &Mixup, Spec.  &RegNet    &0.50 &0.60 &0.50 &0.48\\   
         \hline
         \hline
         E1 & Raw Audio  &None &Whisper+MLP   &\textbf{0.85} &\textbf{0.88} &\textbf{0.95} &\textbf{0.10}\\
         E2 & Raw Audio  &None &Speechbrain+MLP   &0.77 &0.81 &0.81 &0.25\\
         E3 & Raw Audio  &None &Seamless+MLP   &0.86 &0.88 &0.87 &0.20\\
         E4 & Raw Audio  &None &Pyannote+MLP   &0.64 &0.71 &0.78 &0.27\\
       \hline
       \hline             
       B2 + B3 &STFT, CQT &Mixup, Spec. &CNN  &\textbf{0.91} &\textbf{0.92} &\textbf{0.98} &\textbf{0.06}\\ 
       B2 + B4 &STFT, WT  &Mixup, Spec. &CNN  &0.88 &0.90 &0.96 &0.09\\ 
       B2 + B3 + B4 & STFT, CQT, WT  &Mixup, Spec. &CNN  &0.90 &0.92 &0.98 &0.07\\ 
       \hline                           
       B5 + B6 &STFT\&LF, STFT\&MEL &Mixup, Spec. &CNN  &0.88 &0.90 &0.97 &0.08\\ 
       B5 + B7 &STFT\&LF, STFT\&GAM &Mixup, Spec. &CNN  &0.87 &0.89 &\textbf{0.98} &\textbf{0.065}\\ 
       B5 + B6 + B7 &STFT\& LF, STFT\&MEL, STFT\&GAM &Mixup, Spec. &CNN  &0.88 &0.90 &0.98 &0.069\\ 
       \hline
       B5 + D6 & STFT\&LF &Mixup, Spec. & CNN, Swint\_T &0.87 &0.89 &0.96 &0.078 \\                                   
       B5 + D7 & STFT\&LF &Mixup, Spec. & CNN, ConvNeXt-Tiny &0.88 &0.90 &\textbf{0.97} &\textbf{0.07} \\   
       B5 + D6 + D7 & STFT\&LF &Mixup, Spec. & CNN, ConvNeXt-Tiny, Swint\_T &0.88 &0.89 &0.97 &0.072 \\                                                                 
       \hline
       \hline
       \textbf{B3 + B5 + B7} &\textbf{CQT, STFT\&LF, STFT\&GAM} &\textbf{Mixup, Spec.} &\textbf{CNN}  &\textbf{0.88} &\textbf{0.90} &\textbf{0.98} &\textbf{0.05}\\ 
       \textbf{D7 + E1} &\textbf{Raw Audio, STFT\&LF} &\textbf{Mixup, Spec.} &\textbf{Whisper, ConvNeXt-Tiny}  &\textbf{0.86} &\textbf{0.88} &\textbf{0.99} &\textbf{0.03}\\ 
       \textbf{D7 + B5} &\textbf{Raw Audio, STFT\&LF} &\textbf{Mixup, Spec.} &\textbf{Whisper, CNN}  &\textbf{0.87} &\textbf{0.89} &\textbf{0.99} &\textbf{0.03}\\

       \hline 
      
    \end{tabular}
    }
    \vspace{-0.1cm}
    \label{table:R1} 
\end{table*}

\textbf{Evaluate multiple deep learning approaches on ASVspoof 2019 (LA Task):} Regarding the end-to-end deep learning approach from A1 to C2, CNN systems outperform RNN or C-RNN systems.
Indeed, using the same input feature of STFT+LFCC, RNN and C-RNN approaches (C1 and C2 systems) obtain ERR scores of 0.14 and 0.17, which is significantly worse than CNN system (A3 or B2 or B7), with the best score of 0.08. 
This indicates that the specific patterns indicative of deepfake audio might not be primarily temporal but rather frequency in the spectrogram representation. 
Regarding the finetuning approach (D1 to D10), Convnext-Tiny stands out as the best system with competitive EER scores of 0.075.
Meanwhile, the embedding-based approach (E1 to E4) achieves the best EER scores of 0.10 using the pre-trained Whisper model.
This suggests the potential of these approaches when choosing the appropriate networks for further optimization.  

\textbf{Evaluate ensemble methods on ASVspoof 2019 (LA Task):} Given the performance of individual input features and network architecture, we conduct extensive experiments to evaluate a wide range of ensemble models.
First, ensembles of STFT, CQT, and WT spectrograms are evaluated, indicating the best EER score of 0.06 from the combination of STFT and CQT (B2+B3).
Then, ensembles of spectrogram with different filter banks (MEL, LF, GAM) are also evaluated, resulting in the best score of 0.065 from STFT+LF and STFT+GAM (B5+B7).
As a result, when an ensemble of CQT, STFT+LF, and STFT+GAM is conducted (B3+B5+B7), we can achieve the EER score of 0.05.
Regarding the ensemble of network architecture, CNN and ConvNeXt-Tiny (B5+D7) help obtain the EER score of 0.07.
Meanwhile, the combination of Whisper+MLP, ConvNeXt-Tiny (E1+D7) or  Whisper+MLP, CNN (E1+B5) achieves the best EER score of 0.03.

We continue evaluating cross-datasets on ASVspoof 2021 (LA \& DF Tasks)~\cite{ASV21} and cross-languages on MLAAD dataset~\cite{MAILABS_ds}.
For the cross-dataset evaluation, the evaluation sets of ASVspoof 2021 (LA \& DF Tasks)~\cite{ASV21} are tested with the DSD models which were trained and evaluated on ASVspoof 2019 (LA Task) in advance from Table~\ref{table:R1}. 
Regarding cross-language evaluation, we only select pairs of utterances from four languages (e.g. French, Spanish, Italian, and German).
A pair of utterances presents the original utterance and a deepfake utterance with the same transcription. 
Similar to the cross-dataset evaluation, pre-trained DSD systems on ASVspoof 2019 (LA Task) from Table~\ref{table:R1} are used to verify the cross-language evaluation.

\textbf{Data augmentation methods for cross-dataset evaluation on ASVspoof 2021 (LA \& DF Tasks):}
As experimental results on the B5 system are shown in Table~\ref{table:R2}, it indicates that using the offline data augmentation of codec helps improve the DSD system performance on both ASVspoof 2021 LA and DF tasks. 
Significantly, codec helps enhance by 0.11 in terms of EER score in the ASVspoof 2021 LA task.
The results also indicate that a combination of offline data augmenation (e.g., codec) and online data augmentation (e.g., Mixup and SpecAugment) are necessary to achieve a general DSD model to deal with the domain shift issue in cross-data evaluation.

\textbf{Input features for cross-dataset evaluation on ASVspoof 2021 (LA \& DF Tasks):}
Regarding the input features, three types of spectrograms (e.g., CQT, STFT+GAM, STFT+LF) which present the high performance on ASVspoof 2019 dataset are evaluated.
In particular, STFT+LF (B5 system) outperforms CQT (B3 system) and STFT+GAM (B7 system). 
This indicates that a combination of STFT and linear filter is suitable for DSD task.

\textbf{Network architecture for the cross-dataset evaluation on ASVspoof 2021 (LA \& DF Tasks):}
The experimental results from B5 (STFT+LF, CNN), D7 (STFT+LF, ConvNeXt-Tiny) and E1 (Raw Audio, Whisper+MLP) systems indicate that leveraging pre-trained model (E1) significantly outperforms the others. This again proves and explains why more Encoder-Decoder architectures have been recently proposed for the DSD task (i.e., Encoder architectures leveraging pre-trained models such as Whisper or Wave2vec2.0).
Regarding the ensemble methods, the combination of D7 and E1, which present CNN model trained from scratch and pre-trained Whisper model, achieves the best performance on both ASVspoof 2019 (LA Task) and ASVspoof 2021 (LA \& DF Tasks). This also proves that the ensemble of network architectures is more effective than the ensemble of input features.

The results obtained from the evaluation on ASVspoof 2019 (LA Task) and ASVspoof 2021 (LA \& DF Tasks) lead to some conclusions:
\begin{itemize}
    \item The results indicate a combination of offline data augmentation (codec) and online data augmentation (Mixup, SpecAugment) is essential for constructing a general DSD system.
    \item Not all network architectures are appropriate for the DSD task. As the good performance obtained from CNN-based network, ConvNeXt-Tiny, Whisper models, suggesting that CNN-based architectures are suitable for DSD task.
    \item The ensemble of network architectures is effective in enhancing the model performance on the DSD task rather than the ensemble of spectrograms. 
    \item Leveraging pre-trained models such as Whisper shows effectiveness, reinforcing the growing trend of using Encoder-Decoder architectures with pre-trained Encoders. This explains why these architectures have gained popularity in recent works.      
\end{itemize}

In the cross-language evaluation, as shown in Table~\ref{table:R3}, all proposed DSD systems exhibit poor performance. This suggests that training a model on a single language (e.g., English) and testing it on other languages (e.g., French, German, Spanish, Italian) is not effective. To develop a robust DSD model for multiple languages, training with multilingual datasets is essential. This highlights the need for the DSD research community to focus on creating and publishing more multilingual datasets for the task.

\begin{table*}[t]
    \caption{Performance comparison among deep learning models and ensemble of high-performance models \\ on ASVspoof 2021 (LA \&DF Tasks) for cross-dataset evaluation} 
        	\vspace{-0.2cm}
    \centering
    \scalebox{0.95}{
    \begin{tabular}{|l|l|c|c|c|c|c|c|c|} 
        \hline 
        \textbf{Systems} &\textbf{Inputs}   &\textbf{Augmentations} &\textbf{Models} &\textbf{Dataset} &  \textbf{Acc$\uparrow$} &  \textbf{F1$\uparrow$} &\textbf{AUC$\uparrow$}  &\textbf{ERR $\downarrow$}\\  
        \hline
        \hline
         B5 & STFT \& LF  &Codec                  &CNN &ASV21-LA    &0.84 &0.87 &0.89 &0.16\\
         B5 & STFT \& LF  &Mixup, Spec.           &CNN &ASV21-LA    &0.88 &0.88 &0.79 &0.27\\
         B5 & STFT \& LF  &Codec \& Mixup, Spec.  &CNN &ASV21-LA    &0.85 &0.87 &0.90 &\textbf{0.15}\\
         \hline
         B5 & STFT \& LF  &Codec                  &CNN &ASV21-DF    &0.88 &0.91 &0.80 &\textbf{0.25}\\
         B5 & STFT \& LF  &Mixup, Spec.           &CNN &ASV21-DF    &0.91 &0.88 &0.77 &0.28\\
         B5 & STFT \& LF  &Codec \& Mixup, Spec.  &CNN &ASV21-DF    &0.91  &0.93 &0.80 &0.27\\
         \hline
         \hline
         B3 & CQT         &Mixup, Spec.  &CNN &ASV21-LA    &0.89 &0.86 &0.49 &0.51\\
         B5 & STFT \& LF  &Mixup, Spec.           &CNN &ASV21-LA    &0.88 &0.88 &0.79 &\textbf{0.27}\\
         B7 & STFT \& GAM  &Mixup, Spec.  &CNN &ASV21-LA    &0.89 &0.87 &0.52 &0.49\\         
        \hline
         B3 & CQT         &Mixup, Spec.  &CNN &ASV21-DF    &0.95 &0.94 &0.51 &0.49\\
         B5 & STFT \& LF  &Mixup, Spec.           &CNN &ASV21-DF    &0.91 &0.88 &0.77 &\textbf{0.28}\\
         B7 & STFT \& GAM &Mixup, Spec.  &CNN &ASV21-DF    &0.96 &0.95 &0.61 &0.42\\ 
        \hline
        \hline
         D7 & STFT \& LF &Mixup, Spec.  &ConvNeXt-Tiny    &ASV21-LA &0.88 &0.88 &0.73 &0.33\\
         E1 & Raw Audio  &None &Whisper+MLP   &ASV21-LA &0.84 &0.86 &0.88 &\textbf{0.18}\\         
        \hline
         D7 & STFT \& LF &Mixup, Spec.  &ConvNeXt-Tiny    &ASV21-DF &0.93 &0.94 &0.76 &0.32\\

         E1 & Raw Audio  &None &Whisper+MLP   &ASV21-DF &0.84 &0.89 &0.92 &\textbf{0.14}\\
        \hline
        \hline
        B3 + B5 + B7 &CQT, STFT\&LF, STFT\&GAM &Mixup, Spec. &CNN  &ASV21-LA &0.90 &0.87 &0.75 &0.30\\ 
        D7 + E1 &Raw Audio, STFT\&LF &Mixup, Spec. &Whisper, CNN &ASV21-LA  &0.90 &0.91 &0.96 &\textbf{0.11}\\ 
        \hline
        B3 + B5 + B7 &CQT, STFT\&LF, STFT\&GAM &Mixup, Spec. &CNN  &ASV21-DF &0.96 &0.95 &0.77 &0.29\\         
        D7 + E1 &Raw Audio, STFT\&LF &Mixup, Spec. &Whisper, CNN &ASV21-DF  &0.94 &0.95 &0.95 &\textbf{0.13}\\ 

        \hline
   
\end{tabular}
    }
    \vspace{-0.1cm}
    \label{table:R2} 
\end{table*}

\begin{table*}[t]
    \caption{Performance comparison among deep learning models and ensemble of high-performance models \\ on MLAAD dataset for cross-language evaluation} 
        	\vspace{-0.2cm}
    \centering
    \scalebox{0.95}{
    \begin{tabular}{|l|l|c|c|c|c|c|c|c|} 
        \hline 
        \textbf{Systems} &\textbf{Inputs}   &\textbf{Augmentations} &\textbf{Models} &\textbf{Dataset-Language} &  \textbf{Acc$\uparrow$} &  \textbf{F1$\uparrow$} &\textbf{AUC$\uparrow$}  &\textbf{ERR $\downarrow$}\\  
        \hline
        \hline
         B5 & STFT \& LF  &Codec \& Mixup, Spec.           &CNN &MLAAD-DE    &0.45 &0.32 &0.53 &0.46\\
         B5 & STFT \& LF  &Codec \& Mixup, Spec.           &CNN &MLAAD-IT    &0.49 &0.34 &0.27 &0.69\\
         B5 & STFT \& LF  &Codec \& Mixup, Spec.           &CNN &MLAAD-FR    &0.49 &0.35 &0.48 &0.51\\        
         B5 & STFT \& LF  &Codec \& Mixup, Spec.           &CNN &MLAAD-ES    &0.48 &0.33 &0.45 &0.52\\         
        \hline
         E1 & Raw Audio  &None &Whisper+MLP   &MLAAD-DE &0.53 &0.52 &0.56 &0.45\\
         E1 & Raw Audio  &None &Whisper+MLP   &MLAAD-IT &0.52 &0.52 &0.54 &0.48\\       
         E1 & Raw Audio  &None &Whisper+MLP   &MLAAD-FR &0.59 &0.57 &0.62 &0.40\\
         E1 & Raw Audio  &None &Whisper+MLP   &MLAAD-ES &0.52 &0.52 &0.53 &0.48\\         
        \hline
         B5 + E1 & Raw Audio, STFT \& LF  &Codec \& Mixup, Spec. &CNN, Whisper+MLP   &MLAAD-DE &0.50 &0.38 &0.54 &0.47\\
         B5 + E1 & Raw Audio, STFT \& LF  &Codec \& Mixup, Spec. &CNN, Whisper+MLP   &MLAAD-IT &0.52 &0.38 &0.63 &0.40\\       
         B5 + E1 & Raw Audio, STFT \& LF  &Codec \& Mixup, Spec. &CNN, Whisper+MLP   &MLAAD-FR &0.50 &0.36 &0.59 &0.42\\
         B5 + E1 & Raw Audio, STFT \& LF  &Codec \& Mixup, Spec. &CNN, Whisper+MLP   &MLAAD-ES &0.50 &0.37 &0.49 &0.50\\
         \hline
         
\end{tabular}
    }
    \vspace{-0.1cm}
    \label{table:R3} 
\end{table*}

\section{Open Challenges and Potential Research Directions}
\label{potential}

\subsection{{Datasets for Deepfake Speech Detection}}
\subsubsection{Open challenges}
Building better datasets for audio deepfake detection is essential for improving the accuracy and robustness of detection systems. However, the current diversity of available datasets for audio deepfake detection remains limited, especially in terms of speaker identity, language, and deepfake generation methods. 

A large number of published datasets feature a narrow range of speaker identities, often focusing on a small group of speakers with limited gender, age, and accent diversity. For instance, datasets of ASVspoof and FakeAVCeleb include mainly English-speaking voices from certain groups of speakers (e.g., celebrity, predominantly synthesized voice) with a small number of speakers from different language backgrounds, resulting in biased models when applied to diverse populations. 

Many existing datasets are domain-specific, focusing on particular types of audio or speakers. For example, FakeAVCeleb primarily includes celebrity interviews, while LibriSpeech focuses on read recordings. These datasets often have limited variability in terms of recording conditions, speaker interactions, and speech styles, making it difficult to generalize detection models to new domains or unseen environments, such as detecting deepfakes in real-world scenarios with noisy or degraded audio, such as phone calls, public spaces, or online content.

The lack of language diversity is also a significant issue that limits the robustness of detection models. As shown at Table~\ref{table:T2}, most existing datasets support single languages (primarily English or Chinese). This imbalance raises challenges that hinder the development of robust, audio deepfake detection systems in multilingual settings. 

As deepfake generation techniques have been evolving rapidly, they produce fake audio that is increasingly difficult to detect.
This makes it difficult for existing datasets to stay up to date as they may be vulnerable to newer methods of audio synthesis. 
Therefore, datasets must be continuously updated to include samples produced by new techniques to ensure the robustness and adaptability of detection models.

\subsubsection{Future directions}
Given the open challenges discussed in the previous subsection, we highlight some potential future directions in dataset development for Deepfake Speech Detection:

\textbf{Multilingual and Multimodal Datasets:}
To address the issue of language diversity, future datasets should include a broader range of languages, accents, and dialects. This variety will enable detection models to better handle diverse linguistic and phonetic features across different languages, ensuring their stability in multilingual contexts and their effectiveness in developing global solutions.
Moreover, deepfake content in real-world scenarios often includes both audio and video elements, rather than just audio. Therefore, integrating multimodal datasets that combine both audio and video deepfakes is a crucial direction for future research. This integration enhances detection capabilities by allowing models to identify anomalies across multiple data types, improving their effectiveness in combating increasingly sophisticated forgeries

\textbf{Continuous Dataset Updates:} 
To stay updated, there needs to be ongoing collaboration between researchers developing deepfake generation methods and those working on the DSD task. 
Regular updates to datasets should include deepfake samples created by the latest synthesized generation techniques, allowing detection models to adapt to emerging threats.

\textbf{Cross-Domain and Real-World Dataset Adaptation:} One of the biggest challenges for DSD models is domain adaptation — the ability to generalize across different types of audio environments, speakers, and use cases. 
Future datasets should prioritize cross-domain generalization, including diverse data from various contexts (e.g., podcasts, phone calls, interviews, public speeches, and social media content). In addition, besides varied deepfake generation methods, future dataset development should include data from diverse online platforms (e.g., YouTube, TikTok, podcasts) and various speaker demographics that stimulate inclusive real-life scenarios.

\subsection{{The generalization and robustness of Deepfake Speech Detection models}}
\subsubsection{Open challenges}
A major challenge in developing deepfake detection systems is ensuring they can generalize to new samples that are not presented in the training data. 
While models may perform well on known attacks, they often struggle with novel manipulations and across different domains, such as varying languages, accents, or speaking styles. 
The limited size and diversity of training datasets hinder DSD models' ability to handle real-world variability without degraded performance. 
Some approaches have been adopted to address these challenges. 
For example, ensemble models, as discussed in Sections~\ref{challenges} and~\ref{systems}, have been effectively utilized to enhance DSD performance and generalization ability, often achieving top results in competition settings. 
They are also frequently employed in research papers to deliver competitive outcomes~\cite{m06, m15, m04}. 
While ensemble models are powerful and versatile, they often require significant computational costs during training. 
Additionally, detection systems leveraging pre-trained models have gained popularity~\cite{m23}. By fine-tuning models pre-trained on upstream audio tasks like speech-to-text~\cite{wav2vec20, whisper}, the training cost for DSD downstream tasks is greatly reduced. However, proving the generalization of these fine-tuned single models remains challenging. For instance, experiments on ASVspoof 2021 (DF Task) in~\cite{m23} achieved remarkable results, with an EER of 5.67 compared to 15.64 from the top-performing system in the challenge. In contrast, the performance on the ASVspoof 2021 (LA Task) was much lower, with an EER of 15.92, compared to 1.32 from the top-performing system.

In terms of improving the model's robustness to adversarial attacks, the majority of current methods for defending against adversarial attacks rely on adversarial training~\cite{survey_02}, which involves generating adversarial examples from known attacks to retrain the model. However, this approach incurs high computational costs.

\subsubsection{Future directions} To improve the generalization and robustness of detection systems, there has been much room for improving existing approaches as well as proposing new methods. 
For example, future directions can address challenges in ensemble methods by balancing the trade-off between cost and effectiveness using techniques such as pruning, quantization, and knowledge distillation or other efficient ensembling strategies to reduce model size. 
In the approach using transfer learning or fine-tuning, employing several strategies such as cross-dataset validation or an ensemble of fine-tuned models could address the challenges of proving generalization. 
Applying mechanisms to learn information from domain-invariant attacks could also enhance the robustness of models against different adversarial attacks.

\subsection{{Interpretability and Explainable AI (XAI) for Deepfake Speech Detection}}
\subsubsection{Open challenges}
\label{xai}
Improving interpretability and explainability in Deepfake Speech Detection remains a complex task due to the unique challenges posed by audio data and the black-box nature of deep learning methods. 
Although various explainable AI (XAI) techniques prove effectiveness in interpreting deep-learning-based models, applying XAI to DSD systems has not drawn much attention from the research community.
Indeed, only some recently published papers~\cite{xai_01, xai_05, xai_04, xai_03, xai_02} address the role of XAI, which mainly focus on the visualization-based XAI methods.  
For example, the conventional SHapley Additive exPlanations (SHAP)~\cite{shap_mt} and Local  Interpretable Model-agnostic Explanations (LIME)~\cite{lime_mt} methods were used to interpret the feature contribution in~\cite{xai_05, xai_03} and in~\cite{xai_04}, respectively.
Authors in~\cite{xai_01} applied Saliency Map~\cite{xai_01_sp01} and Smooth Grad~\cite{xai_01_sp02} techniques to visualize how their model processes audio in the frequency domain.
Similarly, layer-wise relevance propagation (LRP), a visualization-based XAI method, was leveraged in~\cite{xai_02} to indicate the difference of formants among fake and real audio utterances.
While more deep-learning-based models have been proposed to solve the DSD task, not many research papers focus on exploring XAI methods to interpret DSD systems.

\subsubsection{Future directions} Based on the above discussion, there is much room for applying XAI to improve transparency and trustworthiness within detection systems. 
Additionally, leveraging visualization tools for visualizing audio features or feature maps could also provide user-friendly platforms and valuable insights into the underlying decision-making process of detection models.
 
\subsection{{Real-time deepfake speech detection}}
\subsubsection{Open challenges}
Integrating DSD systems into real-world applications still presents several challenges. Key factors include the length of the audio utterance, the complexity of the model (e.g., the number of trainable parameters), computational costs (e.g., FLOPs), and the target edge devices (e.g., mobile phones, embedded systems, high-performance computers). 
These factors directly affect inference time and are carefully analyzed to ensure effective implementation.
For example, the trade-off between the performance and the model complexity was comprehensively analyzed in~\cite{lampham_01} and~\cite{ed_dv_01} concerning Acoustic Scene Classification (ASC) task and Acoustic Event Detection (AED) task, respectively.
Currently, most proposed DSD systems have been currently evaluated on high-performance computers with the advance of powerful GPUs without any computational constraints, while there is little research on real-time deepfake detection. Several studies, such as \cite{realtime-detection1} and \cite{realtime-detection2}, have proposed real-time deepfake audio detection systems. 
However, these systems often face significant limitations, such as being applicable to only a limited range of deepfake creation techniques (voice conversion) or domains (communication). 
These challenges highlight the need for further exploration and analysis of real-time DSD systems in future research.

\subsubsection{Future directions}
Future directions in developing real-time audio deepfake detection systems could rely on better handling the trade-off between model complexity and performance, facilitating model implementation in low-latency conditions. 
Some techniques such as quantization and pruning can be used to reduce model size, while other methods leverage edge computing or distributed computing to reduce inference time and handle large-scale data more efficiently.

\subsection{Ethical and legal considerations}
\subsubsection{Open challenges}
Training audio deepfake detection models requires large datasets, which may involve the collection and the use of personal voice recordings. For example, VoxCeleb and FakeAVCeleb corpora contain speech from thousands of celebrities in various environments. Personal data handling  raises threats of privacy and consent. 
Furthermore, there is also a risk of dual-use dilemma when some bad actors could manipulate detection technology and available individuals's speech for harmful purposes such as reinforcing disinformation narratives, defamation, and fraud, infringing on individuals' privacy rights.

\subsubsection{Future directions}
Future directions in addressing ethical and legal considerations for developing audio deepfake technologies focus on enhancing data privacy protection, fairness, and facilitating global regulatory frameworks. 
Developers will increasingly incorporate privacy-by-design principles in developing detection systems, ensuring that personal voice data is handled securely and with consent, minimizing the risk of misuse. 
Within DSD applications, access control mechanisms should be implemented to limit certain groups of people and the frequency of using detection technologies, reducing the potential risk of misuse by malicious actors. 
In terms of legal perspectives, legal frameworks may also evolve to introduce stricter penalties for misuse of both deepfake creation and detection technology. 

\subsection{The race between Deepfake Speech Generation and Detection}
\subsubsection{Open challenges} 
As mentioned and discussed in Section~\ref{dataset}, there is a tight relationship between Deepfake Speech Generation and Deepfake Speech Detection tasks. Deepfake Speech Generation systems (e.g., VC, TTS, and AT models) have been becoming more powerful and accessible, enabling the creation of hyper-realistic fake utterances that mimic normal speech patterns and produce fewer detectable flaws. This makes it hard for DSD systems to distinguish between real and manipulated content, presenting challenges to keep pace with these deepfake creation advancements.

\subsubsection{Future directions} 
As deepfakes have evolved rapidly, detection models must also adapt by learning from increasingly realistic fakes. By facilitating collaborative environments, researchers in both Deepfake Speech Generation and Detection can further explore and push boundaries of what is technically possible and ensure that detection methods keep pace with advances in deepfake generators. 
For example, ADD 2022~\cite{add22}, ADD 2023~\cite{add23}, and ASVspoof 2024~\cite{ASV24} challenge competitions were established to engage researchers in both Deepfake Speech Generation and Detection. 
This promotes innovations in addressing the race between creating and detecting deepfake, improving the robustness of detection systems in combating increasingly complicated deepfakes.

\subsection{Feature-free deepfake detection}
\subsubsection{Open challenges}
Deepfake detection faces the usual challenge of the cat-mouse logic of an attack-defense arms race, which is due to the fact that as soon as a feature is identified for detection, it can as quickly be neutralized in the next generation synthesis models. The only way to break this cycle is to develop feature-free detection approaches.

\subsubsection{Future directions}
For example, Bloom (\href{http://bloomsocialanalytics.com/}{bloomsocialanalytics.com}) proposed a feature-free approach that uses the very same synthesis technologies used to produce deepfakes for their own detection. The idea is based on the intuition that an AI model can reproduce speech produced by an AI more easily than by a human, because reality is always more complex than its model. In other words, real speech contains chaotic components that won’t be perfectly captured by AI models.
The proposed method consists of the training and detection phases. The training phase uses an advanced neural voice cloning system to synthesize voice samples based on the target speech files whose authenticity needs to be verified, and then computes a similarity metric between the target speech (authentic or synthetic) and the cloned speech. This distance distribution is used to find the optimal classification threshold, which is then applied to compute the likelihood of authenticity during the detection phase.

\subsection{The availability of Deepfake Speech Detection tools}
\subsubsection{Open challenges}
Deepfake speech detection tools still face challenges in increasing their quantity and quality due to the rapid development of deepfake speech generation techniques. 
Although DSD systems act as a critical function in Voice over Internet Protocol (VoIP) based platforms such as WhatsApp, Facebook, etc. or social media such as YouTube, Twister, etc. for a thread warning, very few VoIP platforms or social media have announced an available and independent DSD tool. 
Regarding non-commercial or commercial solutions, only some DSD tools or platforms such as Deepware, WeVerify, TrueMedia, and DeepFake-O-Meter are available as highlighted in the survey~\cite{aissd_tool_01}.
However, information on DSD models used in these tools has been not described in detail except TrueMeida and DeepFake-O-Meter with 3 and 5 systems replicated from published papers. 
Overall, the sufficiency of deepfake detection applications is primarily due to technical complexity in developing and updating models, resource demands such as computational costs and scalability, accuracy concerns, and privacy issues.

\subsubsection{Future directions} 
To address the mentioned challenges, future improvements in developing deepfake speech detection tools could rely on some approaches such as lightweight detection models that can operate on consumer devices such as smartphones, laptops, or cloud-based services. 
To ensure broader adaption, the development of open-source deepfake detection tools or libraries and established standards for their use could also be promoted by the collaboration between tech companies and academic institutions, making detection tools more accessible and reliable.

\section{Conclusion}
\label{conclusion}
This paper has provided a comprehensive survey for Deepfake Speech Detection (DSD) task by deeply analyzing the challenge competitions, the public and benchmark datasets, the main components in a deep-learning-based DSD system.
From the survey, we indicate exiting concerns and provide enhance solutions to motivate the research community for further contribution on this research topic.
More than a survey, we verified the role and the effect of data augmentation, feature extraction, and network architectures 
Given the comprehensive survey and extensive experiments, we indicate potential and promising research directions for Deepfake Speech Detection task.

\section*{ACKNOWLEDGMENTS}
The work described in this paper is performed in the H2020 project STARLIGHT (“Sustainable Autonomy and Resilience for LEAs using AI against High priority Threats”). This project has received funding from the European Union’s Horizon 2020 research and innovation program under grant agreement No 101021797.

The work in this paper has further received funding from the European Union - European Defence Fund under GA no. 101121418 (EUCINF). Views and opinions expressed are however those of the author(s) only and do not necessarily reflect those of the European Union or the European Commission. Neither the European Union nor the granting authority can be held responsible for them.




\begin{thebibliography}{100}

\bibitem{survey2021}
Zahra Khanjani, Gabrielle Watson, and Vandana~P Janeja,
\newblock ``How deep are the fakes? focusing on audio deepfake: A survey,''
\newblock {\em arXiv preprint arXiv:2111.14203}, 2021.

\bibitem{survey_11}
Momina Masood, Mariam Nawaz, Khalid~Mahmood Malik, Ali Javed, Aun Irtaza, and Hafiz Malik,
\newblock ``Deepfakes generation and detection: State-of-the-art, open challenges, countermeasures, and way forward,''
\newblock {\em Applied intelligence}, vol. 53, no. 4, pp. 3974--4026, 2023.

\bibitem{survey_12}
Rami Mubarak, Tariq Alsboui, Omar Alshaikh, Isa Inuwa-Dutse, Saad Khan, and Simon Parkinson,
\newblock ``A survey on the detection and impacts of deepfakes in visual, audio, and textual formats,''
\newblock {\em IEEE Access}, vol. 11, pp. 144497--144529, 2023.

\bibitem{survey_patel}
Yogesh Patel, Sudeep Tanwar, Rajesh Gupta, Pronaya Bhattacharya, Innocent~Ewean Davidson, Royi Nyameko, Srinivas Aluvala, and Vrince Vimal,
\newblock ``Deepfake generation and detection: Case study and challenges,''
\newblock {\em IEEE Access}, vol. 11, pp. 143296--143323, 2023.

\bibitem{survey_01}
Jiangyan Yi, Chenglong Wang, Jianhua Tao, Xiaohui Zhang, Chu~Yuan Zhang, and Yan Zhao,
\newblock ``Audio deepfake detection: A survey,''
\newblock {\em arXiv preprint arXiv:2308.14970}, 2023.

\bibitem{survey2023_usa}
Zahra Khanjani, Gabrielle Watson, and Vandana~P Janeja,
\newblock ``Audio deepfakes: A survey,''
\newblock {\em Frontiers in Big Data}, vol. 5, pp. 1001063, 2023.

\bibitem{survey2024video}
Zahid Akhtar, Thanvi~Lahari Pendyala, and Virinchi~Sai Athmakuri,
\newblock ``Video and audio deepfake datasets and open issues in deepfake technology: Being ahead of the curve,''
\newblock {\em Forensic Sciences}, vol. 4, no. 3, pp. 289--377, 2024.

\bibitem{survey_14}
Enes Altuncu, Virginia~NL Franqueira, and Shujun Li,
\newblock ``Deepfake: definitions, performance metrics and standards, datasets, and a meta-review,''
\newblock {\em Frontiers in Big Data}, vol. 7, pp. 1400024, 2024.

\bibitem{survey_02}
Menglu Li, Yasaman Ahmadiadli, and Xiao-Ping Zhang,
\newblock ``Audio anti-spoofing detection: A survey,''
\newblock {\em arXiv preprint arXiv:2404.13914}, 2024.

\bibitem{survey_03}
Jiayang Wu, Wensheng Gan, Zefeng Chen, Shicheng Wan, and Hong Lin,
\newblock ``Ai-generated content (aigc): A survey,''
\newblock {\em arXiv preprint arXiv:2304.06632}, 2023.

\bibitem{survey_04}
Burak Yeti{\c{s}}tiren, I{\c{s}}{\i}k {\"O}zsoy, Miray Ayerdem, and Eray T{\"u}z{\"u}n,
\newblock ``Evaluating the code quality of ai-assisted code generation tools: An empirical study on github copilot, amazon codewhisperer, and chatgpt,''
\newblock {\em arXiv preprint arXiv:2304.10778}, 2023.

\bibitem{survey_05_TTS}
Xu~Tan, Tao Qin, Frank Soong, and Tie-Yan Liu,
\newblock ``A survey on neural speech synthesis,''
\newblock {\em arXiv preprint arXiv:2106.15561}, 2021.

\bibitem{survey_06_VC}
Berrak Sisman, Junichi Yamagishi, Simon King, and Haizhou Li,
\newblock ``An overview of voice conversion and its challenges: From statistical modeling to deep learning,''
\newblock {\em IEEE/ACM Transactions on Audio, Speech, and Language Processing}, vol. 29, pp. 132--157, 2021.

\bibitem{survey_07_crime}
Fatima Dakalbab, Manar~Abu Talib, Omnia~Abu Waraga, Ali~Bou Nassif, Sohail Abbas, and Qassim Nasir,
\newblock ``Artificial intelligence \& crime prediction: A systematic literature review,''
\newblock {\em Social Sciences \& Humanities Open}, vol. 6, no. 1, pp. 100342, 2022.

\bibitem{dfdc_ds}
Brian Dolhansky, Joanna Bitton, Ben Pflaum, Jikuo Lu, Russ Howes, Menglin Wang, and Cristian~Canton Ferrer,
\newblock ``The deepfake detection challenge ({DFDC}) dataset,''
\newblock {\em arXiv preprint arXiv:2006.07397}, 2020.

\bibitem{wake_ds}
Joel Frank and Lea Sch{\"o}nherr,
\newblock ``Wavefake: A data set to facilitate audio deepfake detection,''
\newblock {\em NeurIPS}, 2024.

\bibitem{add22}
``Audio deep synthesis detection challenge ({ADD} 2022),'' \url{http://addchallenge.cn/add2022}, 2022.

\bibitem{MAILABS_ds}
``M-ailabs speech dataset,'' \url{https://github.com/imdatceleste/m-ailabs-dataset}, 2024.

\bibitem{svdd2024}
You Zhang, Yongyi Zang, Jiatong Shi, Ryuichi Yamamoto, Tomoki Toda, and Zhiyao Duan,
\newblock ``{SVDD} 2024: The inaugural singing voice deepfake detection challenge,''
\newblock {\em arXiv preprint arXiv:2408.16132}, 2024.

\bibitem{ASV15}
Zhizheng Wu, Tomi Kinnunen, Nicholas Evans, Junichi Yamagishi, Cemal Hanilçi, Md. Sahidullah, and Aleksandr Sizov,
\newblock ``Asvspoof 2015: the first automatic speaker verification spoofing and countermeasures challenge,''
\newblock in {\em Proc. INTERSPEECH}, 2015, pp. 2037--2041.

\bibitem{ASV19}
Xin Wang, Junichi Yamagishi, Massimiliano Todisco, H{\'e}ctor Delgado, Andreas Nautsch, Nicholas Evans, Md~Sahidullah, Ville Vestman, Tomi Kinnunen, Kong~Aik Lee, et~al.,
\newblock ``Asvspoof 2019: A large-scale public database of synthesized, converted and replayed speech,''
\newblock {\em Computer Speech \& Language}, vol. 64, pp. 101114, 2020.

\bibitem{FTC_ch}
``The ftc voice cloning challenge,'' \url{https://www.ftc.gov/news-events/contests/ftc-voice-cloning-challenge}, 2023.

\bibitem{ASV21}
Junichi Yamagishi, Xin Wang, Massimiliano Todisco, Md~Sahidullah, Jose Patino, Andreas Nautsch, Xuechen Liu, Kong~Aik Lee, Tomi Kinnunen, Nicholas Evans, et~al.,
\newblock ``Asvspoof 2021: accelerating progress in spoofed and deepfake speech detection,''
\newblock in {\em Workshop-Automatic Speaker Verification and Spoofing Coutermeasures Challenge (ASVspoof)}, 2021.

\bibitem{add23}
``Audio deep synthesis detection challenge ({ADD} 2023),'' \url{http://addchallenge.cn/add2023}, 2023.

\bibitem{cai2023av}
Zhixi Cai, Shreya Ghosh, Aman~Pankaj Adatia, Munawar Hayat, Abhinav Dhall, and Kalin Stefanov,
\newblock ``Av-deepfake1m: A large-scale llm-driven audio-visual deepfake dataset,''
\newblock {\em arXiv preprint arXiv:2311.15308}, 2023.

\bibitem{1mdeepfake_ch}
``1m-deepfakes detection challenge,'' \url{https://deepfakes1m.github.io/}, 2023.

\bibitem{ASV24}
``The asvspoof 2024 challenge,'' \url{https://www.asvspoof.org/}, 2024.

\bibitem{sing}
``The singing voice deepfake detection challenge (svdd),'' \url{https://challenge.singfake.org/}, 2024.

\bibitem{ASV17}
H{\'e}ctor Delgado, Massimiliano Todisco, Md~Sahidullah, Nicholas Evans, Tomi Kinnunen, Kong~Aik Lee, and Junichi Yamagishi,
\newblock ``Asvspoof 2017 version 2.0: meta-data analysis and baseline enhancements,''
\newblock in {\em The Speaker and Language Recognition Workshop}, 2018, pp. 296--303.

\bibitem{for_ds}
Ricardo Reimao and Vassilios Tzerpos,
\newblock ``For: A dataset for synthetic speech detection,''
\newblock in {\em International Conference on Speech Technology and Human-Computer Dialogue}, 2019, pp. 1--10.

\bibitem{for_source}
``Audio source used to generate for dataset,'' \url{https://www.kaggle.com/datasets/percevalw/englishfrench-translations}, 2018.

\bibitem{ljspeech_ds}
Nal Kalchbrenner, Erich Elsen, Karen Simonyan, Seb Noury, Norman Casagrande, Edward Lockhart, Florian Stimberg, Aaron Oord, Sander Dieleman, and Koray Kavukcuoglu,
\newblock ``Efficient neural audio synthesis,''
\newblock in {\em Proc. ICML}, 2018, pp. 2410--2419.

\bibitem{jsut_ds}
Ryosuke Sonobe, Shinnosuke Takamichi, and Hiroshi Saruwatari,
\newblock ``Jsut corpus: free large-scale japanese speech corpus for end-to-end speech synthesis,''
\newblock {\em arXiv preprint arXiv:1711.00354}, 2017.

\bibitem{kokd_ds}
Patrick Kwon, Jaeseong You, Gyuhyeon Nam, Sungwoo Park, and Gyeongsu Chae,
\newblock ``Kodf: A large-scale korean deepfake detection dataset,''
\newblock in {\em Proc. IEEE/CVF International Conference on Computer Vision}, 2021, pp. 10744--10753.

\bibitem{aishell-3}
Yao Shi, Hui Bu, Xin Xu, Shaoji Zhang, and Ming Li,
\newblock ``Aishell-3: A multi-speaker mandarin tts corpus,''
\newblock in {\em Proc. INTERSPEECH}, 2021, pp. 2756--2760.

\bibitem{khalid2021fakeavceleb}
Hasam Khalid, Shahroz Tariq, Minha Kim, and Simon~S Woo,
\newblock ``Fakeavceleb: A novel audio-video multimodal deepfake dataset,''
\newblock in {\em Thirty-fifth Conference on Neural Information Processing Systems Datasets and Benchmarks Track (Round 2)}, 2021.

\bibitem{voxceleb2}
Joon~Son Chung, Arsha Nagrani, and Andrew Zisserman,
\newblock ``{VoxCeleb2: Deep Speaker Recognition},''
\newblock in {\em Proc. INTERSPEECH}, 2018, pp. 1086--1090.

\bibitem{intwi_ds}
Nicolas Müller, Pavel Czempin, Franziska Diekmann, Adam Froghyar, and Konstantin Böttinger,
\newblock ``{Does Audio Deepfake Detection Generalize?},''
\newblock in {\em Proc. INTERSPEECH}, 2022, pp. 2783--2787.

\bibitem{cai2022you}
Zhixi Cai, Kalin Stefanov, Abhinav Dhall, and Munawar Hayat,
\newblock ``Do you really mean that? content driven audio-visual deepfake dataset and multimodal method for temporal forgery localization,''
\newblock in {\em International Conference on Digital Image Computing: Techniques and Applications}, 2022, pp. 1--10.

\bibitem{VoC_ds}
Xin Wang and Junichi Yamagishi,
\newblock ``Spoofed training data for speech spoofing countermeasure can be efficiently created using neural vocoders,''
\newblock in {\em Proc. ICASSP}, 2023, pp. 1--5.

\bibitem{pasp_ds}
Lin Zhang, Xin Wang, Erica Cooper, Nicholas Evans, and Junichi Yamagishi,
\newblock ``The partialspoof database and countermeasures for the detection of short fake speech segments embedded in an utterance,''
\newblock {\em IEEE/ACM Transactions on Audio, Speech, and Language Processing}, vol. 31, pp. 813--825, 2022.

\bibitem{f_ds_01}
Chengzhe Sun, Shan Jia, Shuwei Hou, and Siwei Lyu,
\newblock ``Ai-synthesized voice detection using neural vocoder artifacts,''
\newblock in {\em Proc. IEEE/CVF Conference on Computer Vision and Pattern Recognition}, 2023, pp. 904--912.

\bibitem{cfad2023}
Haoxin Ma, Jiangyan Yi, Chenglong Wang, Xinrui Yan, Jianhua Tao, Tao Wang, Shiming Wang, and Ruibo Fu,
\newblock ``{CFAD}: A chinese dataset for fake audio detection,''
\newblock {\em Speech Communication}, vol. 164, pp. 103122, 2024.

\bibitem{aishell1}
Hui Bu, Jiayu Du, Xingyu Na, Bengu Wu, and Hao Zheng,
\newblock ``{AISHELL-1}: An open-source mandarin speech corpus and a speech recognition baseline.,''
\newblock in {\em Proc. O-COCOSDA}, 2017, pp. 1--5.

\bibitem{aishell3}
Yao Shi, Hui Bu, Xin Xu, Shaoji Zhang, and Ming Li,
\newblock ``Aishell-3: A multi-speaker mandarin tts corpus,''
\newblock in {\em Proc. INTERSPEECH}, 2021, pp. 2756--2760.

\bibitem{magicdata}
Zehui Yang, Yifan Chen, Lei Luo, Runyan Yang, Lingxuan Ye, Gaofeng Cheng, Ji~Xu, Yaohui Jin, Qingqing Zhang, Pengyuan Zhang, Lei Xie, and Yonghong Yan,
\newblock ``Open source {MagicData-RAMC}: A rich annotated mandarin conversational({RAMC}) speech dataset,''
\newblock in {\em Proc. INTERSPEECH}, 2022, pp. 1736--1740.

\bibitem{muller2024mlaad}
Nicolas~M M{\"u}ller, Piotr Kawa, Wei~Herng Choong, Edresson Casanova, Eren G{\"o}lge, Thorsten M{\"u}ller, Piotr Syga, Philip Sperl, and Konstantin B{\"o}ttinger,
\newblock ``Mlaad: The multi-language audio anti-spoofing dataset,''
\newblock {\em International Joint Conference on Neural Networks (IJCNN)}, 2024.

\bibitem{mls}
Vineel Pratap, Qiantong Xu, Anuroop Sriram, Gabriel Synnaeve, and Ronan Collobert,
\newblock ``{MLS: A Large-Scale Multilingual Dataset for Speech Research},''
\newblock in {\em Proc. INTERSPEECH}, 2020, pp. 2757--2761.

\bibitem{dcase2022}
``{DCASE} 2022 challenge competition {Task 1A},'' \url{https://dcase.community/challenge2022/task-low-complexity-acoustic-scene-classification}, 2022.

\bibitem{kodf_s1}
Ran Yi, Zipeng Ye, Juyong Zhang, Hujun Bao, and Yong-Jin Liu,
\newblock ``Audio-driven talking face video generation with learning-based personalized head pose,''
\newblock {\em arXiv preprint arXiv:2002.10137}, 2020.

\bibitem{asv15_s1_01}
Thierry Dutoit, Andre Holzapfel, Matthieu Jottrand, Alexis Moinet, Javier Perez, and Yannis Stylianou,
\newblock ``Towards a voice conversion system based on frame selection,''
\newblock in {\em Proc. ICASSP}, 2007, vol.~4, pp. IV--513.

\bibitem{asv15_s1_02}
Zhizheng Wu, Tuomas Virtanen, Tomi Kinnunen, Engsiong Chng, and Haizhou Li,
\newblock ``Exemplar-based unit selection for voice conversion utilizing temporal information.,''
\newblock in {\em Proc. INTERSPEECH}, 2013, pp. 3057--3061.

\bibitem{asv15_s2}
Toshiaki Fukuda,
\newblock ``An adaptive algorithm for mel-cepstral analysis of speech,''
\newblock in {\em Proc. ICASSP}, 1992, pp. 137--140.

\bibitem{asv15_s3}
Junichi Yamagishi, Takao Kobayashi, Yuji Nakano, Katsumi Ogata, and Juri Isogai,
\newblock ``Analysis of speaker adaptation algorithms for hmm-based speech synthesis and a constrained smaplr adaptation algorithm,''
\newblock {\em IEEE Transactions on Audio, Speech, and Language Processing}, vol. 17, no. 1, pp. 66--83, 2009.

\bibitem{asv15_s5}
``Festvox voice conversion system,'' \url{http://www.festvox.org}, 2024.

\bibitem{asv15_s6}
Tomoki Toda, Alan~W Black, and Keiichi Tokuda,
\newblock ``Voice conversion based on maximum-likelihood estimation of spectral parameter trajectory,''
\newblock {\em IEEE Transactions on Audio, Speech, and Language Processing}, vol. 15, no. 8, pp. 2222--2235, 2007.

\bibitem{asv15_s8}
Daisuke Saito, Keisuke Yamamoto, Nobuaki Minematsu, and Keikichi Hirose,
\newblock ``One-to-many voice conversion based on tensor representation of speaker space,''
\newblock in {\em Proc. INTERSPEECH}, 2011, pp. 653--656.

\bibitem{asv15_s9}
Elina Helander, Hanna Sil{\'e}n, Tuomas Virtanen, and Moncef Gabbouj,
\newblock ``Voice conversion using dynamic kernel partial least squares regression,''
\newblock {\em IEEE transactions on audio, speech, and language processing}, vol. 20, no. 3, pp. 806--817, 2011.

\bibitem{asv15_s10}
``Marytts speech synthesis system,'' \url{http://mary.dfki.de}, 2024.

\bibitem{A1}
``Hts working group, the english tts system flite+hts engine,'' \url{http://hts-engine.sourceforge.net/}, 2014.

\bibitem{A2}
Masanori Morise, Fumiya Yokomori, and Kenji Ozawa,
\newblock ``World: a vocoder-based high-quality speech synthesis system for real-time applications,''
\newblock {\em IEICE Transactions on Information and Systems}, vol. 99, no. 7, pp. 1877--1884, 2016.

\bibitem{A3}
Zhizheng Wu, Oliver Watts, and Simon King,
\newblock ``Merlin: An open source neural network speech synthesis system,''
\newblock in {\em Speech Synthesis Workshop}, 2016, pp. 202--207.

\bibitem{A4}
Marc Schr{\"o}der, Marcela Charfuelan, Sathish Pammi, and Ingmar Steiner,
\newblock ``Open source voice creation toolkit for the mary tts platform,''
\newblock in {\em Proc. INTERSPEECH}, 2011, pp. 3253--3256.

\bibitem{A5}
Chin-Cheng Hsu, Hsin-Te Hwang, Yi-Chiao Wu, Yu~Tsao, and Hsin-Min Wang,
\newblock ``Voice conversion from non-parallel corpora using variational auto-encoder,''
\newblock in {\em Proc. APSIPA}, 2016, pp. 1--6.

\bibitem{A6}
Driss Matrouf, J-F Bonastre, and Corinne Fredouille,
\newblock ``Effect of speech transformation on impostor acceptance,''
\newblock in {\em Proc. ICASSP}, 2006, vol.~1, pp. I--I.

\bibitem{A7}
Kou Tanaka, Hirokazu Kameoka, Takuhiro Kaneko, and Nobukatsu Hojo,
\newblock ``Wavecyclegan2: Time-domain neural post-filter for speech waveform generation,''
\newblock {\em arXiv preprint arXiv:1904.02892}, 2019.

\bibitem{A8}
Xin Wang, Shinji Takaki, and Junichi Yamagishi,
\newblock ``Neural source-filter-based waveform model for statistical parametric speech synthesis,''
\newblock in {\em Proc. ICASSP}, 2019, pp. 5916--5920.

\bibitem{A9_1}
Heiga Zen, Yannis Agiomyrgiannakis, Niels Egberts, Fergus Henderson, and Przemys{\l}aw Szczepaniak,
\newblock ``Fast, compact, and high quality lstm-rnn based statistical parametric speech synthesizers for mobile devices,''
\newblock in {\em Proc. INTERSPEECH}, 2016, pp. 2273--2277.

\bibitem{A9_2}
Yannis Agiomyrgiannakis,
\newblock ``Vocaine the vocoder and applications in speech synthesis,''
\newblock in {\em Proc. ICASSP}, 2015, pp. 4230--4234.

\bibitem{A10_1}
Li~Wan, Quan Wang, Alan Papir, and Ignacio~Lopez Moreno,
\newblock ``Generalized end-to-end loss for speaker verification,''
\newblock in {\em Proc. ICASSP}, 2018, pp. 4879--4883.

\bibitem{A10_2}
Nal Kalchbrenner, Erich Elsen, Karen Simonyan, Seb Noury, Norman Casagrande, Edward Lockhart, Florian Stimberg, Aaron Oord, Sander Dieleman, and Koray Kavukcuoglu,
\newblock ``Efficient neural audio synthesis,''
\newblock in {\em Proc. ICML}, 2018, pp. 2410--2419.

\bibitem{A11}
Daniel Griffin and Jae Lim,
\newblock ``Signal estimation from modified short-time fourier transform,''
\newblock {\em IEEE Transactions on acoustics, speech, and signal processing}, vol. 32, no. 2, pp. 236--243, 1984.

\bibitem{A12}
Aäron {van den Oord}, Sander Dieleman, Heiga Zen, Karen Simonyan, Oriol Vinyals, Alex Graves, Nal Kalchbrenner, Andrew Senior, and Koray Kavukcuoglu,
\newblock ``{WaveNet: A Generative Model for Raw Audio},''
\newblock in {\em Proc. Workshop on Speech Synthesis}, 2016, p. 125.

\bibitem{A13_1}
``Voicetext,'' \url{http://dws2.voicetext.jp/tomcat/demonstration/top.html}, 2024.

\bibitem{A14_1}
Li-Juan Liu, Zhen-Hua Ling, Yuan Jiang, Ming Zhou, and Li-Rong Dai,
\newblock ``Wavenet vocoder with limited training data for voice conversion.,''
\newblock in {\em Proc. INTERSPEECH}, 2018, pp. 1983--1987.

\bibitem{A14_2}
Hideki Kawahara, Ikuyo Masuda-Katsuse, and Alain De~Cheveigne,
\newblock ``Restructuring speech representations using a pitch-adaptive time--frequency smoothing and an instantaneous-frequency-based f0 extraction: Possible role of a repetitive structure in sounds,''
\newblock {\em Speech communication}, vol. 27, no. 3-4, pp. 187--207, 1999.

\bibitem{A17_1}
Kazuhiro Kobayashi, Tomoki Toda, and Satoshi Nakamura,
\newblock ``Intra-gender statistical singing voice conversion with direct waveform modification using log-spectral differential,''
\newblock {\em Speech communication}, vol. 99, pp. 211--220, 2018.

\bibitem{A17_2}
Wen-Chin Huang, Yi-Chiao Wu, Kazuhiro Kobayashi, Yu-Huai Peng, Hsin-Te Hwang, Patrick~Lumban Tobing, Yu~Tsao, Hsin-Min Wang, and Tomoki Toda,
\newblock ``Generalization of spectrum differential based direct waveform modification for voice conversion,''
\newblock in {\em Proc. Workshop on Speech Synthesis}, 2019, pp. 57--62.

\bibitem{A18_1}
Najim Dehak, Patrick~J Kenny, R{\'e}da Dehak, Pierre Dumouchel, and Pierre Ouellet,
\newblock ``Front-end factor analysis for speaker verification,''
\newblock {\em IEEE Transactions on Audio, Speech, and Language Processing}, vol. 19, no. 4, pp. 788--798, 2010.

\bibitem{A18_2}
Patrick Kenny,
\newblock ``A small footprint i-vector extractor.,''
\newblock in {\em Odyssey}, 2012, vol. 2012, pp. 1--6.

\bibitem{A18_3}
Simon~JD Prince and James~H Elder,
\newblock ``Probabilistic linear discriminant analysis for inferences about identity,''
\newblock in {\em Proc. IEEE international conference on computer vision}, 2007, pp. 1--8.

\bibitem{dfdc_s1}
Adam Polyak, Lior Wolf, and Yaniv Taigman,
\newblock ``{TTS Skins: Speaker Conversion via ASR},''
\newblock in {\em Proc. INTERSPEECH}, 2020, pp. 786--790.

\bibitem{kodf_s2}
KR~Prajwal, Rudrabha Mukhopadhyay, Vinay~P Namboodiri, and CV~Jawahar,
\newblock ``A lip sync expert is all you need for speech to lip generation in the wild,''
\newblock in {\em Proceedings of the 28th ACM international conference on multimedia}, 2020, pp. 484--492.

\bibitem{asv_2021_journal}
Xuechen Liu, Xin Wang, Md~Sahidullah, Jose Patino, Héctor Delgado, Tomi Kinnunen, Massimiliano Todisco, Junichi Yamagishi, Nicholas Evans, Andreas Nautsch, and Kong~Aik Lee,
\newblock ``Asvspoof 2021: Towards spoofed and deepfake speech detection in the wild,''
\newblock {\em IEEE/ACM Transactions on Audio, Speech, and Language Processing}, vol. 31, pp. 2507--2522, 2023.

\bibitem{wake_s1}
Kundan Kumar, Rithesh Kumar, Thibault De~Boissiere, Lucas Gestin, Wei~Zhen Teoh, Jose Sotelo, Alexandre De~Brebisson, Yoshua Bengio, and Aaron~C Courville,
\newblock ``Melgan: Generative adversarial networks for conditional waveform synthesis,''
\newblock {\em Advances in neural information processing systems}, vol. 32, 2019.

\bibitem{wake_s3}
Jungil Kong, Jaehyeon Kim, and Jaekyoung Bae,
\newblock ``Hifi-gan: Generative adversarial networks for efficient and high fidelity speech synthesis,''
\newblock {\em Advances in neural information processing systems}, vol. 33, pp. 17022--17033, 2020.

\bibitem{wake_s4}
Durk~P Kingma and Prafulla Dhariwal,
\newblock ``Glow: Generative flow with invertible 1x1 convolutions,''
\newblock {\em Advances in neural information processing systems}, vol. 31, 2018.

\bibitem{wake_s2}
Ryuichi Yamamoto, Eunwoo Song, and Jae-Min Kim,
\newblock ``Parallel wavegan: A fast waveform generation model based on generative adversarial networks with multi-resolution spectrogram,''
\newblock in {\em Proc. ICASSP}, 2020, pp. 6199--6203.

\bibitem{FakeAVCeleb_s1}
Ye~Jia, Yu~Zhang, Ron Weiss, Quan Wang, Jonathan Shen, Fei Ren, Patrick Nguyen, Ruoming Pang, Ignacio Lopez~Moreno, Yonghui Wu, et~al.,
\newblock ``Transfer learning from speaker verification to multispeaker text-to-speech synthesis,''
\newblock {\em Advances in neural information processing systems}, vol. 31, 2018.

\bibitem{FakeAVCeleb_s2}
KR~Prajwal, Rudrabha Mukhopadhyay, Vinay~P Namboodiri, and CV~Jawahar,
\newblock ``A lip sync expert is all you need for speech to lip generation in the wild,''
\newblock in {\em Proc. ACM international conference on multimedia}, 2020, pp. 484--492.

\bibitem{la_vd_s1}
Ye~Jia, Yu~Zhang, Ron Weiss, Quan Wang, Jonathan Shen, Fei Ren, Patrick Nguyen, Ruoming Pang, Ignacio Lopez~Moreno, Yonghui Wu, et~al.,
\newblock ``Transfer learning from speaker verification to multispeaker text-to-speech synthesis,''
\newblock {\em Advances in neural information processing systems}, vol. 31, 2018.

\bibitem{voc_s1}
Xin Wang, Shinji Takaki, and Junichi Yamagishi,
\newblock ``Neural source-filter waveform models for statistical parametric speech synthesis,''
\newblock {\em IEEE/ACM Transactions on Audio, Speech, and Language Processing}, vol. 28, pp. 402--415, 2019.

\bibitem{wavernn}
Nal Kalchbrenner, Erich Elsen, Karen Simonyan, Seb Noury, Norman Casagrande, Edward Lockhart, Florian Stimberg, Aaron van~den Oord, Sander Dieleman, and Koray Kavukcuoglu,
\newblock ``Efficient neural audio synthesis,''
\newblock in {\em Proc. ICML}, 2018, pp. 2410--2419.

\bibitem{paralel_wavegan}
Ryuichi Yamamoto, Eunwoo Song, and Jae-Min Kim,
\newblock ``Parallel wavegan: A fast waveform generation model based on generative adversarial networks with multi-resolution spectrogram,''
\newblock in {\em Proc. ICASSP}, 2020, pp. 6199--6203.

\bibitem{wave_grad}
Nanxin Chen, Yu~Zhang, Heiga Zen, Ron~J. Weiss, Mohammad Norouzi, and William Chan,
\newblock ``Wavegrad: Estimating gradients for waveform generation,''
\newblock in {\em Proc. ICLR}, 2021.

\bibitem{diff_wave}
Zhifeng Kong, Wei Ping, Jiaji Huang, Kexin Zhao, and Bryan Catanzaro,
\newblock ``Diffwave: A versatile diffusion model for audio synthesis,''
\newblock in {\em Proc. ICLR}, 2021.

\bibitem{av_deepfake_1M_tts_s2}
Jaehyeon Kim, Jungil Kong, and Juhee Son,
\newblock ``Conditional variational autoencoder with adversarial learning for end-to-end text-to-speech,''
\newblock in {\em Proc. ICML}, 2021, pp. 5530--5540.

\bibitem{av_deepfake_1M_tts_s1}
Edresson Casanova, Julian Weber, Christopher~D Shulby, Arnaldo~Candido Junior, Eren G{\"o}lge, and Moacir~A Ponti,
\newblock ``Yourtts: Towards zero-shot multi-speaker tts and zero-shot voice conversion for everyone,''
\newblock in {\em Proc. ICML}, 2022, pp. 2709--2720.

\bibitem{straight}
Hideki Kawahara,
\newblock ``Straight, exploitation of the other aspect of vocoder: Perceptually isomorphic decomposition of speech sounds,''
\newblock {\em Acoustical Science and Technology}, vol. 27, no. 6, pp. 349--353, 2006.

\bibitem{griffin_lim}
Nathanaël Perraudin, Peter Balazs, and Peter~L. Søndergaard,
\newblock ``A fast griffin-lim algorithm,''
\newblock in {\em IEEE Workshop on Applications of Signal Processing to Audio and Acoustics}, 2013, pp. 1--4.

\bibitem{lpcnet}
Jean-Marc Valin and Jan Skoglund,
\newblock ``Lpcnet: Improving neural speech synthesis through linear prediction,''
\newblock in {\em Proc. ICASSP}, 2019, pp. 5891--5895.

\bibitem{hifigan}
Jungil Kong, Jaehyeon Kim, and Jaekyoung Bae,
\newblock ``Hifi-gan: generative adversarial networks for efficient and high fidelity speech synthesis,''
\newblock in {\em Proc. NeurIPS}, 2020.

\bibitem{world}
Masanori MORISE, Fumiya YOKOMORI, and Kenji OZAWA,
\newblock ``World: A vocoder-based high-quality speech synthesis system for real-time applications,''
\newblock {\em IEICE Transactions on Information and Systems}, vol. E99.D, no. 7, pp. 1877--1884, 2016.

\bibitem{fast_speech}
Yi~Ren, Chenxu Hu, Xu~Tan, Tao Qin, Sheng Zhao, Zhou Zhao, and Tie-Yan Liu,
\newblock ``Fastspeech 2: Fast and high-quality end-to-end text to speech,''
\newblock {\em arXiv preprint arXiv:2006.04558}, 2020.

\bibitem{tacotron}
Yuxuan Wang, RJ~Skerry-Ryan, Daisy Stanton, Yonghui Wu, Ron~J. Weiss, Navdeep Jaitly, Zongheng Yang, Ying Xiao, Zhifeng Chen, Samy Bengio, Quoc Le, Yannis Agiomyrgiannakis, Rob Clark, and Rif~A. Saurous,
\newblock ``Tacotron: Towards end-to-end speech synthesis,''
\newblock in {\em Proc. INTERSPEECH}, 2017, pp. 4006--4010.

\bibitem{asv24_tts1}
Jaehyeon Kim, Sungwon Kim, Jungil Kong, and Sungroh Yoon,
\newblock ``Glow-tts: A generative flow for text-to-speech via monotonic alignment search,''
\newblock {\em Advances in Neural Information Processing Systems}, vol. 33, pp. 8067--8077, 2020.

\bibitem{asv24_tts2}
Vadim Popov, Ivan Vovk, Vladimir Gogoryan, Tasnima Sadekova, and Mikhail Kudinov,
\newblock ``Grad-tts: A diffusion probabilistic model for text-to-speech,''
\newblock in {\em Proc. ICML}, 2021, pp. 8599--8608.

\bibitem{asv24_tts3}
Adrian {\L}a{\'n}cucki,
\newblock ``Fastpitch: Parallel text-to-speech with pitch prediction,''
\newblock in {\em Proc. ICASSP}, 2021, pp. 6588--6592.

\bibitem{asv24_tts4}
Jaehyeon Kim, Jungil Kong, and Juhee Son,
\newblock ``Conditional variational autoencoder with adversarial learning for end-to-end text-to-speech,''
\newblock in {\em Proc. ICML}, 2021, pp. 5530--5540.

\bibitem{asv24_tts5}
Florian Lux, Julia Koch, and Ngoc Thang~Vu,
\newblock ``Low-resource multilingual and zero-shot multispeaker tts,''
\newblock in {\em Proc. AACL}, 2022, pp. 741--751.

\bibitem{asv24_tts6}
Jungil Kong, Jaehyeon Kim, and Jaekyoung Bae,
\newblock ``Hifi-gan: Generative adversarial networks for efficient and high fidelity speech synthesis,''
\newblock {\em Advances in neural information processing systems}, vol. 33, pp. 17022--17033, 2020.

\bibitem{asv24_tts7}
Jonathan Shen, Ruoming Pang, Ron~J Weiss, Mike Schuster, Navdeep Jaitly, Zongheng Yang, Zhifeng Chen, Yu~Zhang, Yuxuan Wang, Rj~Skerrv-Ryan, et~al.,
\newblock ``Natural tts synthesis by conditioning wavenet on mel spectrogram predictions,''
\newblock in {\em Proc. ICASSP}, 2018, pp. 4779--4783.

\bibitem{asv24_vc1}
Yinghao~Aaron Li, Ali Zare, and Nima Mesgarani,
\newblock ``Starganv2-vc: A diverse, unsupervised, non-parallel framework for natural-sounding voice conversion,''
\newblock in {\em Proc. INTERSPEECH}, 2021, pp. 1349--1353.

\bibitem{asv24_tts9}
Edresson Casanova, Julian Weber, Christopher~D Shulby, Arnaldo~Candido Junior, Eren G{\"o}lge, and Moacir~A Ponti,
\newblock ``Yourtts: Towards zero-shot multi-speaker tts and zero-shot voice conversion for everyone,''
\newblock in {\em Proc. ICML}, 2022, pp. 2709--2720.

\bibitem{asv24_vc2}
Ehab~A AlBadawy and Siwei Lyu,
\newblock ``Voice conversion using speech-to-speech neuro-style transfer.,''
\newblock in {\em Proc. INTERSPEECH}, 2020, pp. 4726--4730.

\bibitem{asv24_tts10}
Cheng Gong, Xin Wang, Erica Cooper, Dan Wells, Longbiao Wang, Jianwu Dang, Korin Richmond, and Junichi Yamagishi,
\newblock ``Zmm-tts: Zero-shot multilingual and multispeaker speech synthesis conditioned on self-supervised discrete speech representations,''
\newblock {\em IEEE/ACM Transactions on Audio, Speech, and Language Processing}, 2024.

\bibitem{asv24_tts11}
Ingmar Steiner and S{\'e}bastien~Le Maguer,
\newblock ``Creating new language and voice components for the updated marytts text-to-speech synthesis platform,''
\newblock in {\em Proc. LREC}, 2018, pp. 1371--1375.

\bibitem{asv24_tts12}
Sang-gil Lee, Wei Ping, Boris Ginsburg, Bryan Catanzaro, and Sungroh Yoon,
\newblock ``Bigvgan: A universal neural vocoder with large-scale training,''
\newblock in {\em Proc. ICLR}, 2022.

\bibitem{asv24_tts13}
Florian Lux, Julia Koch, and Ngoc~Thang Vu,
\newblock ``Exact prosody cloning in zero-shot multispeaker text-to-speech,''
\newblock in {\em Proc. SLT}, 2023, pp. 962--969.

\bibitem{asv24_tts_xx}
Florian Lux, Julia Koch, and Ngoc~Thang Vu,
\newblock ``Low-resource multilingual and zero-shot multispeaker tts,''
\newblock in {\em Proc. AACL}, 2022.

\bibitem{asv24_vc5}
Vadim Popov, Ivan Vovk, Vladimir Gogoryan, Tasnima Sadekova, Mikhail Kudinov, and Jiansheng Wei,
\newblock ``Diffusion-based voice conversion with fast maximum likelihood sampling scheme,''
\newblock in {\em Proc. ICLR}, 2022.

\bibitem{asv24_tts14}
Edresson Casanova, Julian Weber, Christopher~D Shulby, Arnaldo~Candido Junior, Eren G{\"o}lge, and Moacir~A Ponti,
\newblock ``Yourtts: Towards zero-shot multi-speaker tts and zero-shot voice conversion for everyone,''
\newblock in {\em Proc. ICML}, 2022, pp. 2709--2720.

\bibitem{asv24_tts15}
Edresson Casanova, Kelly Davis, Eren G{\"o}lge, G{\"o}rkem G{\"o}knar, Iulian Gulea, Logan Hart, Aya Aljafari, Joshua Meyer, Reuben Morais, Samuel Olayemi, et~al.,
\newblock ``Xtts: a massively multilingual zero-shot text-to-speech model,''
\newblock in {\em Proc. INTERSPEECH}, 2024, pp. 4978--4982.

\bibitem{Malafide}
Michele Panariello, Wanying Ge, Hemlata Tak, Massimiliano Todisco, and Nicholas Evans,
\newblock ``Malafide: a novel adversarial convolutive noise attack against deepfake and spoofing detection systems,''
\newblock in {\em Proc. INTERSPEECH}, 2023, pp. 2868--2872.

\bibitem{Malacopula}
Massimiliano Todisco, Michele Panariello, Xin Wang, Hector Delgado, Kong~Aik Lee, and Nicholas Evans,
\newblock ``Malacopula: adversarial automatic speaker verification attacks using a neural-based generalised hammerstein model,''
\newblock {\em arXiv preprint arXiv:2408.09300}, 2024.

\bibitem{m01}
Joaqu{\i}n C{\'a}ceres, Roberto Font, Teresa Grau, Javier Molina, and Biometric~Vox SL,
\newblock ``The biometric vox system for the asvspoof 2021 challenge,''
\newblock in {\em Edition of the Automatic Speaker Verification and Spoofing Countermeasures Challenge}, 2021, pp. 68--74.

\bibitem{m04}
Rohan~Kumar Das,
\newblock ``Known-unknown data augmentation strategies for detection of logical access, physical access and speech deepfake attacks: Asvspoof 2021,''
\newblock in {\em Edition of the Automatic Speaker Verification and Spoofing Countermeasures Challenge}, 2021, pp. 29--36.

\bibitem{m05}
Wanying Ge, Jose Patino, Massimiliano Todisco, and Nicholas Evans,
\newblock ``Raw differentiable architecture search for speech deepfake and spoofing detection,''
\newblock in {\em Edition of the Automatic Speaker Verification and Spoofing Countermeasures Challenge}, 2021, pp. 22--28.

\bibitem{m06}
Woo~Hyun Kang, Jahangir Alam, and Abderrahim Fathan,
\newblock ``Crim’s system description for the asvspoof2021 challenge,''
\newblock in {\em Edition of the Automatic Speaker Verification and Spoofing Countermeasures Challenge}, 2021, pp. 100--106.

\bibitem{m07}
Nicolas~M M{\"u}ller, Franziska Dieckmann, Pavel Czempin, Roman Canals, Konstantin B{\"o}ttinger, and Jennifer Williams,
\newblock ``Speech is silver, silence is golden: What do asvspoof-trained models really learn?,''
\newblock in {\em Edition of the Automatic Speaker Verification and Spoofing Countermeasures Challenge}, 2021, pp. 55--60.

\bibitem{m08}
Hemlata Tak, Jee-weon Jung, Jose Patino, Madhu Kamble, Massimiliano Todisco, and Nicholas Evans,
\newblock ``End-to-end spectro-temporal graph attention networks for speaker verification anti-spoofing and speech deepfake detection,''
\newblock in {\em Edition of the Automatic Speaker Verification and Spoofing Countermeasures Challenge}, 2021, pp. 1--8.

\bibitem{m09}
Anton Tomilov, Aleksei Svishchev, Marina Volkova, Artem Chirkovskiy, Alexander Kondratev, and Galina Lavrentyeva,
\newblock ``Stc antispoofing systems for the asvspoof2021 challenge,''
\newblock in {\em Edition of the Automatic Speaker Verification and Spoofing Countermeasures Challenge}, 2021, pp. 61--67.

\bibitem{m12}
Xingming Wang, Xiaoyi Qin, Tinglong Zhu, Chao Wang, Shilei Zhang, and Ming Li,
\newblock ``The dku-cmri system for the asvspoof 2021 challenge: vocoder based replay channel response estimation,''
\newblock in {\em Edition of the Automatic Speaker Verification and Spoofing Countermeasures Challenge}, 2021, pp. 16--21.

\bibitem{m35}
Jee-weon Jung, Hee-Soo Heo, Hemlata Tak, Hye-jin Shim, Joon~Son Chung, Bong-Jin Lee, Ha-Jin Yu, and Nicholas Evans,
\newblock ``Aasist: Audio anti-spoofing using integrated spectro-temporal graph attention networks,''
\newblock in {\em Proc. ICASSP}, 2022, pp. 6367--6371.

\bibitem{m34}
Hemlata Tak, Massimiliano Todisco, Xin Wang, Jee-weon Jung, Junichi Yamagishi, and Nicholas Evans,
\newblock ``Automatic speaker verification spoofing and deepfake detection using wav2vec 2.0 and data augmentation,''
\newblock in {\em The Speaker and Language Recognition Workshop}, 2022.

\bibitem{rawboost}
Hemlata Tak, Madhu Kamble, Jose Patino, Massimiliano Todisco, and Nicholas Evans,
\newblock ``Rawboost: A raw data boosting and augmentation method applied to automatic speaker verification anti-spoofing,''
\newblock in {\em Proc. ICASSP}, 2022, pp. 6382--6386.

\bibitem{m17}
Rui Liu, Jinhua Zhang, Guanglai Gao, and Haizhou Li,
\newblock ``{Betray Oneself: A Novel Audio DeepFake Detection Model via Mono-to-Stereo Conversion},''
\newblock in {\em Proc. INTERSPEECH}, 2023, pp. 3999--4003.

\bibitem{m25}
Chenglong Wang, Jiangyan Yi, Jianhua Tao, Chu~Yuan Zhang, Shuai Zhang, and Xun Chen,
\newblock ``{Detection of Cross-Dataset Fake Audio Based on Prosodic and Pronunciation Features},''
\newblock in {\em Proc. INTERSPEECH}, 2023, pp. 3844--3848.

\bibitem{wav2vec20}
Alexis Conneau, Alexei Baevski, Ronan Collobert, Abdelrahman Mohamed, and Michael Auli,
\newblock ``{Unsupervised Cross-Lingual Representation Learning for Speech Recognition},''
\newblock in {\em Proc. INTERSPEECH}, 2021, pp. 2426--2430.

\bibitem{Hubert}
Wei-Ning Hsu, Benjamin Bolte, Yao-Hung~Hubert Tsai, Kushal Lakhotia, Ruslan Salakhutdinov, and Abdelrahman Mohamed,
\newblock ``Hubert: Self-supervised speech representation learning by masked prediction of hidden units,''
\newblock {\em IEEE/ACM transactions on audio, speech, and language processing}, vol. 29, pp. 3451--3460, 2021.

\bibitem{m43}
Xiao-Min Zeng, Jian-Tao Zhang, Kang Li, Zhuo-Li Liu, Wei-Lin Xie, and Yan Song,
\newblock ``Deepfake algorithm recognition system with augmented data for add 2023 challenge.,''
\newblock in {\em Proc. IJCAI}, 2023, pp. 31--36.

\bibitem{m53}
Zhongwei Teng, Quchen Fu, Jules White, Maria~E Powell, and Douglas~C Schmidt,
\newblock ``Sa-sasv: An end-to-end spoof-aggregated spoofing-aware speaker verification system,''
\newblock in {\em Proc. INTERSPEECH}, 2022, pp. 4391--4395.

\bibitem{m49}
Jun Xue, Cunhang Fan, Jiangyan Yi, Chenglong Wang, Zhengqi Wen, Dan Zhang, and Zhao Lv,
\newblock ``Learning from yourself: A self-distillation method for fake speech detection,''
\newblock in {\em Proc. ICASSP}, 2023, pp. 1--5.

\bibitem{m37}
Yuankun Xie, Haonan Cheng, Yutian Wang, and Long Ye,
\newblock ``Learning a self-supervised domain-invariant feature representation for generalized audio deepfake detection,''
\newblock in {\em Proc. INTERSPEECH}, 2023, pp. 2808--2812.

\bibitem{m15}
Yujie Yang, Haochen Qin, Hang Zhou, Chengcheng Wang, Tianyu Guo, Kai Han, and Yunhe Wang,
\newblock ``A robust audio deepfake detection system via multi-view feature,''
\newblock in {\em Proc. ICASSP}, 2024, pp. 13131--13135.

\bibitem{Encode_01}
Alexandre D{\'e}fossez, Jade Copet, Gabriel Synnaeve, and Yossi Adi,
\newblock ``High fidelity neural audio compression,''
\newblock {\em Transactions on Machine Learning Research}, 2023.

\bibitem{Encode_02}
Yi-Chiao Wu, Israel~D Gebru, Dejan Markovi{\'c}, and Alexander Richard,
\newblock ``Audiodec: An open-source streaming high-fidelity neural audio codec,''
\newblock in {\em Proc. ICASSP}, 2023, pp. 1--5.

\bibitem{Encode_03}
Po-Yao Huang, Hu~Xu, Juncheng Li, Alexei Baevski, Michael Auli, Wojciech Galuba, Florian Metze, and Christoph Feichtenhofer,
\newblock ``Masked autoencoders that listen,''
\newblock {\em Advances in Neural Information Processing Systems}, vol. 35, pp. 28708--28720, 2022.

\bibitem{Wavlm}
Sanyuan Chen, Chengyi Wang, Zhengyang Chen, Yu~Wu, Shujie Liu, Zhuo Chen, Jinyu Li, Naoyuki Kanda, Takuya Yoshioka, Xiong Xiao, et~al.,
\newblock ``Wavlm: Large-scale self-supervised pre-training for full stack speech processing,''
\newblock {\em IEEE Journal of Selected Topics in Signal Processing}, vol. 16, no. 6, pp. 1505--1518, 2022.

\bibitem{whisper}
Alec Radford et~al.,
\newblock ``Robust speech recognition via large-scale weak supervision,''
\newblock in {\em Proc. ICML}, 2023, pp. 28492--28518.

\bibitem{m20}
Yinlin Guo, Haofan Huang, Xi~Chen, He~Zhao, and Yuehai Wang,
\newblock ``Audio deepfake detection with self-supervised wavlm and multi-fusion attentive classifier,''
\newblock in {\em Proc. ICASSP}, 2024, pp. 12702--12706.

\bibitem{m32}
Alessandro Pianese, Davide Cozzolino, Giovanni Poggi, and Luisa Verdoliva,
\newblock ``Training-free deepfake voice recognition by leveraging large-scale pre-trained models,''
\newblock in {\em Proc. ACM Workshop on Information Hiding and Multimedia Security}, 2024, pp. 289--294.

\bibitem{beats}
Sanyuan Chen, Yu~Wu, Chengyi Wang, Shujie Liu, Daniel Tompkins, Zhuo Chen, Wanxiang Che, Xiangzhan Yu, and Furu Wei,
\newblock ``Beats: audio pre-training with acoustic tokenizers,''
\newblock in {\em Proc. ICML}, 2023, pp. 5178--5193.

\bibitem{clap}
Yusong Wu, Ke~Chen, Tianyu Zhang, Yuchen Hui, Taylor Berg-Kirkpatrick, and Shlomo Dubnov,
\newblock ``Large-scale contrastive language-audio pretraining with feature fusion and keyword-to-caption augmentation,''
\newblock in {\em Proc. ICASSP}, 2023, pp. 1--5.

\bibitem{audioClib}
Andrey Guzhov, Federico Raue, J{\"o}rn Hees, and Andreas Dengel,
\newblock ``Audioclip: Extending clip to image, text and audio,''
\newblock in {\em Proc. ICASSP}, 2022, pp. 976--980.

\bibitem{m48}
Tianxiang Chen, Avrosh Kumar, Parav Nagarsheth, Ganesh Sivaraman, and Elie Khoury,
\newblock ``Generalization of audio deepfake detection.,''
\newblock in {\em Odyssey}, 2020, pp. 132--137.

\bibitem{m50}
Yuankun Xie, Haonan Cheng, Yutian Wang, and Long Ye,
\newblock ``Single domain generalization for audio deepfake detection.,''
\newblock in {\em Proc. IJCAI}, 2023, pp. 58--63.

\bibitem{m02}
Xinhui Chen, You Zhang, Ge~Zhu, and Zhiyao Duan,
\newblock ``Ur channel-robust synthetic speech detection system for asvspoof 2021,''
\newblock in {\em Edition of the Automatic Speaker Verification and Spoofing Countermeasures Challenge}, 2021, pp. 75--82.

\bibitem{m10}
Zhor Benhafid, Sid~Ahmed Selouani, Mohammed~Sidi Yakoub, and Abderrahmane Amrouche,
\newblock ``Larihs assert reassessment for logical access asvspoof 2021 challenge,''
\newblock in {\em Edition of the Automatic Speaker Verification and Spoofing Countermeasures Challenge}, 2021, pp. 94--99.

\bibitem{m11}
You Zhang, Fei Jiang, and Zhiyao Duan,
\newblock ``One-class learning towards synthetic voice spoofing detection,''
\newblock {\em IEEE Signal Processing Letters}, vol. 28, pp. 937--941, 2021.

\bibitem{m31}
Woo~Hyun Kang, Jahangir Alam, and Abderrahim Fathan,
\newblock ``Investigation on activation functions for robust end-to-end spoofing attack detection system,''
\newblock in {\em Proc. INTERSPEECH}, 2021, pp. 83--88.

\bibitem{m33}
Lin Zhang, Xin Wang, Erica Cooper, and Junichi Yamagishi,
\newblock ``Multi-task learning in utterance-level and segmental-level spoof detection,''
\newblock in {\em Edition of the Automatic Speaker Verification and Spoofing Countermeasures Challenge}, 2021.

\bibitem{m46}
Yang Gao, Tyler Vuong, Mahsa Elyasi, Gaurav Bharaj, and Rita Singh,
\newblock ``Generalized spoofing detection inspired from audio generation artifacts,''
\newblock in {\em Proc. INTERSPEECH}, 2021, pp. 4184--4188.

\bibitem{m45}
Rui Yan, Cheng Wen, Shuran Zhou, Tingwei Guo, Wei Zou, and Xiangang Li,
\newblock ``Audio deepfake detection system with neural stitching for add 2022,''
\newblock in {\em Proc. ICASSP}, 2022, pp. 9226--9230.

\bibitem{m22}
Yuankun Xie, Haonan Cheng, Yutian Wang, and Long Ye,
\newblock ``Domain generalization via aggregation and separation for audio deepfake detection,''
\newblock {\em IEEE Transactions on Information Forensics and Security}, vol. 19, pp. 344--358, 2024.

\bibitem{m24}
Amit Kumar~Singh Yadav, Emily~R Bartusiak, Kratika Bhagtani, and Edward~J Delp,
\newblock ``Synthetic speech attribution using self supervised audio spectrogram transformer,''
\newblock {\em Electronic Imaging}, vol. 35, pp. 1--11, 2023.

\bibitem{m52}
Yeqing Ren, Haipeng Peng, Lixiang Li, and Yixian Yang,
\newblock ``Lightweight voice spoofing detection using improved one-class learning and knowledge distillation,''
\newblock {\em IEEE Transactions on Multimedia}, 2023.

\bibitem{m16}
Yuxiang Zhang, Zhuo Li, Jingze Lu, Wenchao Wang, and Pengyuan Zhang,
\newblock ``Synthetic speech detection based on the temporal consistency of speaker features,''
\newblock {\em IEEE Signal Processing Letters}, vol. 31, pp. 944--948, 2024.

\bibitem{m19}
Junlong Deng, Yanzhen Ren, Tong Zhang, Hongcheng Zhu, and Zongkun Sun,
\newblock ``Vfd-net: Vocoder fingerprints detection for fake audio,''
\newblock in {\em Proc. ICASSP}, 2024, pp. 12151--12155.

\bibitem{Decoder_01_GAN}
``Gan-based network decoders,'' \url{https://github.com/kan-bayashi/ParallelWaveGAN}, 2023.

\bibitem{m21}
Luca Cuccovillo, Milica Gerhardt, and Patrick Aichroth,
\newblock ``Audio transformer for synthetic speech detection via formant magnitude and phase analysis,''
\newblock in {\em Proc. ICASSP}, 2024, pp. 4805--4809.

\bibitem{m23}
Xin Wang and Junichi Yamagishi,
\newblock ``Can large-scale vocoded spoofed data improve speech spoofing countermeasure with a self-supervised front end?,''
\newblock in {\em Proc. ICASSP}, 2024, pp. 10311--10315.

\bibitem{m26}
Hyun-seo Shin, Jungwoo Heo, Ju-ho Kim, Chan-yeong Lim, Wonbin Kim, and Ha-Jin Yu,
\newblock ``Hm-conformer: A conformer-based audio deepfake detection system with hierarchical pooling and multi-level classification token aggregation methods,''
\newblock in {\em Proc. ICASSP}, 2024, pp. 10581--10585.

\bibitem{m39}
Galina Lavrentyeva, Sergey Novoselov, Andzhukaev Tseren, Marina Volkova, Artem Gorlanov, and Alexandr Kozlov,
\newblock ``Stc antispoofing systems for the asvspoof2019 challenge,''
\newblock in {\em Proc. INTERSPEECH}, 2019, pp. 1033--1037.

\bibitem{m40}
Guang Hua, {Andrew Beng Jin} Teoh, and Haijian Zhang,
\newblock ``Towards end-to-end synthetic speech detection,''
\newblock {\em IEEE Signal Processing Letters}, vol. 28, pp. 1265--1269, 2021.

\bibitem{m51}
Xin Wang and Junich Yamagishi,
\newblock ``A comparative study on recent neural spoofing countermeasures for synthetic speech detection,''
\newblock in {\em Proc. INTERSPEECH}, 2021, pp. 4259--4263.

\bibitem{m03}
Tianxiang Chen, Elie Khoury, Kedar Phatak, and Ganesh Sivaraman,
\newblock ``Pindrop labs’ submission to the asvspoof 2021 challenge,''
\newblock in {\em Edition of the Automatic Speaker Verification and Spoofing Countermeasures Challenge}, 2021, pp. 89--93.

\bibitem{m29}
Yan Wen, Zhenchun Lei, Yingen Yang, Changhong Liu, and Minglei Ma,
\newblock ``Multi-path gmm-mobilenet based on attack algorithms and codecs for synthetic speech and deepfake detection.,''
\newblock in {\em Proc. INTERSPEECH}, 2022, pp. 4795--4799.

\bibitem{m44}
Il-Youp Kwak et~al.,
\newblock ``Low-quality fake audio detection through frequency feature masking,''
\newblock in {\em Proceedings of the 1st International Workshop on Deepfake Detection for Audio Multimedia}, 2022, pp. 9--17.

\bibitem{m30}
Jiahui Pan, Shuai Nie, Hui Zhang, Shulin He, Kanghao Zhang, Shan Liang, Xueliang Zhang, and Jianhua Tao,
\newblock ``Speaker recognition-assisted robust audio deepfake detection.,''
\newblock in {\em Proc. INTERSPEECH}, 2022, pp. 4202--4206.

\bibitem{m47}
Alexander Alenin et~al.,
\newblock ``A subnetwork approach for spoofing aware speaker verification.,''
\newblock in {\em Proc. INTERSPEECH}, 2022, pp. 2888--2892.

\bibitem{m41}
Shunbo Dong, Jun Xue, Cunhang Fan, Kang Zhu, Yujie Chen, and Zhao Lv,
\newblock ``Multi-perspective information fusion res2net with randomspecmix for fake speech detection,''
\newblock in {\em Proc. IJCAI}, 2023.

\bibitem{m42}
Ziqian Wang, Qing Wang, Jixun Yao, and Lei Xie,
\newblock ``The npu-aslp system for deepfake algorithm recognition in add 2023 challenge.,''
\newblock in {\em Proc. IJCAI}, 2023, pp. 64--69.

\bibitem{m28}
Chenglong Wang, Jiayi He, Jiangyan Yi, Jianhua Tao, Chu~Yuan Zhang, and Xiaohui Zhang,
\newblock ``Multi-scale permutation entropy for audio deepfake detection,''
\newblock in {\em Proc. ICASSP}, 2024, pp. 1406--1410.

\bibitem{m38}
Yi~Zhu, Surya Koppisetti, Trang Tran, and Gaurav Bharaj,
\newblock ``Slim: Style-linguistics mismatch model for generalized audio deepfake detection,''
\newblock {\em arXiv preprint arXiv:2407.18517}, 2024.

\bibitem{hu2020device}
Hu~Hu et~al.,
\newblock ``Device-robust acoustic scene classification based on two-stage categorization and data augmentation,''
\newblock in {\em Proc. DCASE}, 2020.

\bibitem{PHAM2023120608}
Nhat~Truong Pham et~al.,
\newblock ``Hybrid data augmentation and deep attention-based dilated convolutional-recurrent neural networks for speech emotion recognition,''
\newblock {\em Expert Systems with Applications}, vol. 230, pp. 120608, 2023.

\bibitem{ALEX2023102949}
Ashish Alex, Lin Wang, Paolo Gastaldo, and Andrea Cavallaro,
\newblock ``Data augmentation for speech separation,''
\newblock {\em Speech Communication}, vol. 152, pp. 102949, 2023.

\bibitem{mixup2}
Y.~Tokozume, Y.~Ushiku, and T.~Harada,
\newblock ``Learning from between-class examples for deep sound recognition,''
\newblock {\em in ICLR}, 2018.

\bibitem{spec_aug}
Daniel~S Park, William Chan, Yu~Zhang, Chung-Cheng Chiu, Barret Zoph, Ekin~D Cubuk, and Quoc~V Le,
\newblock ``Specaugment: A simple data augmentation method for automatic speech recognition,''
\newblock in {\em Proc. INTERSPEECH}, 2019, pp. 2613--2617.

\bibitem{ASV15_result_sum}
Zhizheng Wu, Junichi Yamagishi, Tomi Kinnunen, Cemal Hanil{\c{c}}i, Mohammed Sahidullah, Aleksandr Sizov, Nicholas Evans, Massimiliano Todisco, and Hector Delgado,
\newblock ``Asvspoof: the automatic speaker verification spoofing and countermeasures challenge,''
\newblock {\em IEEE Journal of Selected Topics in Signal Processing}, vol. 11, no. 4, pp. 588--604, 2017.

\bibitem{xls-r}
Arun Babu et~al.,
\newblock ``{XLS-R: Self-supervised Cross-lingual Speech Representation Learning at Scale},''
\newblock in {\em Proc. INTERSPEECH}, 2022, pp. 2278--2282.

\bibitem{sincnet}
Mirco Ravanelli and Yoshua Bengio,
\newblock ``Speaker recognition from raw waveform with sincnet,''
\newblock in {\em IEEE spoken language technology workshop}, 2018, pp. 1021--1028.

\bibitem{leaf}
Neil Zeghidour, Olivier Teboul, F{\'e}lix de~Chaumont~Quitry, and Marco Tagliasacchi,
\newblock ``{LEAF}: A learnable frontend for audio classification,''
\newblock in {\em Proc. ICLR}, 2021.

\bibitem{librosa_tool}
McFee, Brian, R.~Colin, L.~Dawen, Daniel P., M.~Matt, B.~Eric, and N.~Oriol,
\newblock ``librosa: Audio and music signal analysis in python,''
\newblock in {\em Proc. Python in Science Conference}, 2015, pp. 18--25.

\bibitem{lampham_01}
Lam Pham, Dat Ngo, Dusan Salovic, Anahid Jalali, Alexander Schindler, Phu~X. Nguyen, Khoa Tran, and Hai~Canh Vu,
\newblock ``Lightweight deep neural networks for acoustic scene classification and an effective visualization for presenting sound scene contexts,''
\newblock {\em Applied Acoustics}, vol. 211, pp. 109489, 2023.

\bibitem{m36}
Jingze Lu, Yuxiang Zhang, Wenchao Wang, Zengqiang Shang, and Pengyuan Zhang,
\newblock ``One-class knowledge distillation for spoofing speech detection,''
\newblock in {\em Proc. ICASSP}, 2024, pp. 11251--11255.

\bibitem{x01}
Piotr Kawa, Marcin Plata, and Piotr Syga,
\newblock ``{Attack Agnostic Dataset: Towards Generalization and Stabilization of Audio DeepFake Detection},''
\newblock in {\em Proc. INTERSPEECH}, 2022, pp. 4023--4027.

\bibitem{x02}
Xin Wang and Junichi Yamagishi,
\newblock ``Spoofed training data for speech spoofing countermeasure can be efficiently created using neural vocoders,''
\newblock in {\em Proc. ICASSP}, 2023, pp. 1--5.

\bibitem{rnn_deepfake}
Akash Chintha et~al.,
\newblock ``Recurrent convolutional structures for audio spoof and video deepfake detection,''
\newblock {\em IEEE Journal of Selected Topics in Signal Processing}, vol. 14, no. 5, pp. 1024--1037, 2020.

\bibitem{imagenet_ds}
Jia Deng et~al.,
\newblock ``Imagenet: A large-scale hierarchical image database,''
\newblock in {\em Proc. CVPR}, 2009, pp. 248--255.

\bibitem{seamless}
Barrault Lo{\"\i}c et~al.,
\newblock ``Seamless: Multilingual expressive and streaming speech translation,''
\newblock {\em arXiv preprint arXiv:2312.05187}, 2023.

\bibitem{speechbrain}
Mirco Ravanelli et~al.,
\newblock ``{SpeechBrain}: A general-purpose speech toolkit,'' 2021,
\newblock arXiv:2106.04624.

\bibitem{pyanote1}
Alexis Plaquet and Hervé Bredin,
\newblock ``{Powerset multi-class cross entropy loss for neural speaker diarization},''
\newblock in {\em Proc. INTERSPEECH}, 2023, pp. 3222--3226.

\bibitem{pyanote2}
Hervé Bredin,
\newblock ``{pyannote.audio 2.1 speaker diarization pipeline: principle, benchmark, and recipe},''
\newblock in {\em Proc. INTERSPEECH}, 2023, pp. 1983--1987.

\bibitem{xai_01}
Nicolas~M. Müller, Philip Sperl, and Konstantin Böttinger,
\newblock ``{Complex-valued neural networks for voice anti-spoofing},''
\newblock in {\em Proc. INTERSPEECH}, 2023, pp. 3814--3818.

\bibitem{xai_05}
Wanying Ge, Jose Patino, Massimiliano Todisco, and Nicholas Evans,
\newblock ``Explaining deep learning models for spoofing and deepfake detection with shapley additive explanations,''
\newblock in {\em Proc. ICASSP}, 2022, pp. 6387--6391.

\bibitem{xai_04}
Davide Salvi, Paolo Bestagini, and Stefano Tubaro,
\newblock ``Towards frequency band explainability in synthetic speech detection,''
\newblock in {\em Proc. EUSIPCO}, 2023, pp. 620--624.

\bibitem{xai_03}
Ning Yu, Long Chen, Tao Leng, Zigang Chen, and Xiaoyin Yi,
\newblock ``An explainable deepfake of speech detection method with spectrograms and waveforms,''
\newblock {\em Journal of Information Security and Applications}, vol. 81, pp. 103720, 2024.

\bibitem{xai_02}
Suk-Young Lim, Dong-Kyu Chae, and Sang-Chul Lee,
\newblock ``Detecting deepfake voice using explainable deep learning techniques,''
\newblock {\em Applied Sciences}, vol. 12, no. 8, pp. 3926, 2022.

\bibitem{shap_mt}
Scott~M. Lundberg and Su-In Lee,
\newblock ``A unified approach to interpreting model predictions,''
\newblock in {\em Proc. International Conference on Neural Information Processing Systems}, 2017, p. 4768–4777.

\bibitem{lime_mt}
Marco~Tulio Ribeiro, Sameer Singh, and Carlos Guestrin,
\newblock ``"why should i trust you?": Explaining the predictions of any classifier,''
\newblock in {\em Proceedings of the 22nd ACM SIGKDD International Conference on Knowledge Discovery and Data Mining}, 2016, p. 1135–1144.

\bibitem{xai_01_sp01}
Karen Simonyan and Andrew Zisserman,
\newblock ``Very deep convolutional networks for large-scale image recognition,''
\newblock in {\em Proc. ICLR}, 2015.

\bibitem{xai_01_sp02}
Daniel Smilkov, Nikhil Thorat, Been Kim, Fernanda~B. Vi{\'{e}}gas, and Martin Wattenberg,
\newblock ``Smoothgrad: removing noise by adding noise,''
\newblock {\em CoRR}, vol. abs/1706.03825, 2017.

\bibitem{ed_dv_01}
Reza~Amini Gougeh, Zhang Nu, and Zeljko Zilic,
\newblock ``{Optimizing Auditory Immersion Safety on Edge Devices: An On-Device Sound Event Detection System},''
\newblock in {\em Proc. The Speaker and Language Recognition Workshop}, 2024, pp. 225--231.

\bibitem{realtime-detection1}
Jordan~J Bird and Ahmad Lotfi,
\newblock ``Real-time detection of ai-generated speech for deepfake voice conversion,''
\newblock {\em arXiv preprint arXiv:2308.12734}, 2023.

\bibitem{realtime-detection2}
Jonat~John Mathew, Rakin Ahsan, Sae Furukawa, Jagdish Gautham~Krishna Kumar, Huzaifa Pallan, Agamjeet~Singh Padda, Sara Adamski, Madhu Reddiboina, and Arjun Pankajakshan,
\newblock ``Towards the development of a real-time deepfake audio detection system in communication platforms,''
\newblock {\em arXiv preprint arXiv:2403.11778}, 2024.

\bibitem{aissd_tool_01}
Shuwei Hou, Yan Ju, Chengzhe Sun, Shan Jia, Lipeng Ke, Riky Zhou, Anita Nikolich, and Siwei Lyu,
\newblock ``Deepfake-o-meter v2. 0: An open platform for deepfake detection,''
\newblock {\em arXiv preprint arXiv:2404.13146}, 2024.

\end{thebibliography}

\end{document}